\ifdefined\CompletePaper
  \PassOptionsToClass{11pt,letterpaper}{article}
  \PassOptionsToPackage{margin=1in}{geometry}
\else
  \PassOptionsToClass{10pt,a4paper}{article}
  \PassOptionsToPackage{hmargin=.85in,vmargin=.75in}{geometry}
\fi
\documentclass{article}
\usepackage{geometry}
\def\ArxivPreprint{}
\usepackage[numbers,sort&compress]{natbib}
\usepackage[T1]{fontenc}
\usepackage{lmodern,amsmath,amssymb,amsthm,mathtools,microtype,needspace,booktabs}
\usepackage{xcolor,graphicx,placeins,tikz}
\definecolor{depth1}{HTML}{F5DE88}
\definecolor{depth2}{HTML}{F2AE55}
\definecolor{depth3}{HTML}{8A94C9}
\definecolor{depth4}{HTML}{AB4473}
\definecolor{depth5}{HTML}{58285F}

\colorlet{figRegion}{depth3!48!white}
\colorlet{figRim}{depth3!63!black}
\colorlet{figTrace}{depth5}

\usepackage{algorithm}
\usepackage[noend]{algpseudocode}
\algrenewcommand\algorithmicrequire{\textbf{Input:}}
\ifdefined\CompletePaper
  \usepackage{xr-hyper}
\fi
\usepackage{hyperref}
\usepackage{accsupp}
\usepackage{orcidlink}
\usepackage{tabularx}
\ifdefined\CompletePaper
  \newcommand{\ProofRef}[1]{\ref{full-#1}}
\else
  \newcommand{\ProofRef}[1]{\ref{#1}}
\fi
\newcommand{\CoverageTitle}[1]{On a unit ball placement problem:#1covering the most points in fixed dimension}
\newcommand{\CoverageGivenName}{Krišjānis}
\newcommand{\CoverageFamilyName}{Petručeņa}
\newcommand{\CoverageAuthors}{\CoverageGivenName\space\CoverageFamilyName}
\newcommand{\CoverageInstitution}{Institute of Mathematics and Computer Science of the University of Latvia}
\newcommand{\CoverageORCID}{0009-0008-5713-5914}
\newcommand{\CoverageORCIDLink}{\orcidlinkX{}{\CoverageORCID}{{\small\CoverageORCID}}}
\newcommand{\CoverageNamePart}[1]{%
  \BeginAccSupp{method=pdfstringdef,space=false,ActualText={#1}}#1\llap{\char32}\EndAccSupp{}}
\newcommand*{\CoverageBibLetter}[1]{#1}
\DeclareUnicodeCharacter{0101}{\CoverageBibLetter{\={a}}}
\DeclareUnicodeCharacter{0146}{\CoverageBibLetter{\c{n}}}
\AddToHook{env/thebibliography/begin}{\let\CoverageBibLetter\CoverageNamePart}
\newcommand{\CoverageKeywords}{Unit-ball placement, maximum depth, circle arrangements, geometric cuttings, affine degeneracy}
\newcommand{\CoverageDate}{November 2026%
  \ifdefined\CompletePaper\ \(\cdot\) Reading edition\fi}
\hypersetup{hidelinks,bookmarksdepth=2,pdftitle={\CoverageTitle{\space}},
 pdfauthor={\CoverageAuthors},pdfkeywords={\CoverageKeywords},
 pdfsubject={Deterministic geometric algorithms and rational hardness reductions}}
\newtheorem{theorem}{Theorem}[section]
\newtheorem{lemma}[theorem]{Lemma}
\newtheorem{proposition}[theorem]{Proposition}
\newtheorem{corollary}[theorem]{Corollary}
\theoremstyle{definition}
\newtheorem{example}[theorem]{Example}
\newcommand{\R}{\mathbb R}
\newcommand{\Z}{\mathbb Z}
\newcommand{\Q}{\mathbb Q}
\newcommand{\Sph}{\mathbb S}
\newcommand{\norm}[1]{\lVert #1\rVert}
\newcommand{\PlaneDepth}{\textsc{PlaneDepth}}
\newcommand{\SphereDepth}{\textsc{SphereDepth}}
\newcommand{\PlaceBall}{\textsc{PlaceBall}}
\newcommand*{\ProofPlateWidth}{\ifdim\linewidth<141mm \linewidth\else 141mm\fi}
\hypersetup{linktoc=all}
\newif\ifIllustrated
\ifdefined\JournalManuscript
  \Illustratedfalse
\else
  \Illustratedtrue
\fi
\AddToHook{cmd/bibsection/after}{\addcontentsline{toc}{section}{References}}
\makeatletter
\newcommand{\PaperContentsSize}{\small}
\newcommand{\PaperContents}{\begingroup\PaperContentsSize
  \renewcommand*{\l@section}{\@dottedtocline{1}{0pt}{1.7em}}%
  \tableofcontents\endgroup}
\makeatother
\newcommand*{\OpeningSpace}{\Needspace{5\baselineskip}}
\newcommand*{\LayoutFloatBarrier}{\ifdefined\ArxivPreprint\else\FloatBarrier\fi}
\newenvironment{walkthroughfigure}{\begin{figure}[H]}{\end{figure}}

\newcommand{\CoverageAbstract}{%
For \(N\) distinct sites in fixed dimension \(D\ge2\), we give a deterministic
algorithm that places a closed unit ball covering the most sites.
It returns the centre, count and all covered indices in
\(O_D(N^D)\) time and \(O_D(N)\) extra words on an exact real random-access
machine. The same bounds hold for maximizing the sum of nonnegative weights
of covered sites.
For \(D\ge3\), we search two-spheres of possible centres determined by
\(D-2\) sites on the ball boundary. Each site gives a cap of centres whose
ball covers it. To find a point where the covering caps have maximum total
weight, we use stereographic projection and at most two planar searches
over weighted closed disks and their closed exteriors.
For \(m\) circles of arbitrary positive radii, each planar search takes
\(O(m^2)\) time and \(O(m)\) extra words.
A linear-time reduction shows that any fixed saving in the exponent \(D\)
for deterministic placement would carry over to deciding whether \(D+1\)
of \(N\) distinct integer points lie on a common hyperplane.}

\ifdefined\CompletePaper\else
  \renewenvironment{walkthroughfigure}{\begin{figure}[!htb]}{\end{figure}}

\fi
\newcommand{\doi}[1]{doi: \href{https://doi.org/#1}{\nolinkurl{#1}}}
\title{\CoverageTitle{\\}}
\author{\CoverageNamePart{\CoverageAuthors}\enspace\CoverageORCIDLink\\[2pt]
  {\small\CoverageInstitution}}
\date{\CoverageDate}
\makeatletter
\renewcommand{\@maketitle}{\begin{center}
  {\LARGE\@title\par}\vspace{.65\baselineskip}
  \ifx\@author\@empty\else{\large\@author\par}\vspace{.45\baselineskip}\fi
  {\large\@date\par}
\end{center}}
\makeatother
\begin{document}
\maketitle
\begin{abstract}
\CoverageAbstract
\end{abstract}

\ifIllustrated
\begingroup
\par\smallskip\centering\small
\tikzset{
  site/.style={fill=black,draw=white,line width=.38pt},
  witness/.style={fill=white,draw=black,line width=.65pt},
}
\def\coverballradius{.085\linewidth}
\ifdim\linewidth<6in
  \def\spherevolumewidth{.20\linewidth}
  \def\spheredepthwidth{.26\linewidth}
\else
  \def\spherevolumewidth{.21\linewidth}
  \def\spheredepthwidth{.29\linewidth}
\fi
\def\coverpanel#1#2{\makebox[.245\linewidth][c]{\rlap{\kern#1\relax
  \makebox[0pt][c]{#2}}}}
\setlength{\tabcolsep}{0pt}
\begin{tabular}{@{}cccc@{}}
  \coverpanel{0pt}{\raisebox{-.5\height}{%
\begin{tikzpicture}[x=\coverballradius,y=\coverballradius]
\input{figures/ball-depth/cover-vectors}
  \foreach \i in {1,...,5}
    \draw[black!80,line width=.55pt] (0,0) -- (cover\i);
  \shade[ball color=depth1!75!white,opacity=.30] (0,0) circle[radius=1];
  \draw[rotate=12,depth1!52!black,opacity=.42,line width=.36pt]
    (-1,0) arc[start angle=180,end angle=360,x radius=1,y radius=.30];
  \draw[depth1!55!black,opacity=.67,line width=.45pt] (0,0) circle[radius=1];
  \foreach \i in {1,...,5}
    \path[site] (cover\i) circle[radius=1.55pt];
  \path[witness] (0,0) circle[radius=2pt];
\end{tikzpicture}%
  \unskip}}
  & \coverpanel{\dimexpr\spherevolumewidth*38/1280\relax}{\raisebox{-.5\height}{%
    \begin{tikzpicture}
      \node[anchor=south west,inner sep=0](im) at (0,0)
        {\includegraphics[width=\spherevolumewidth]{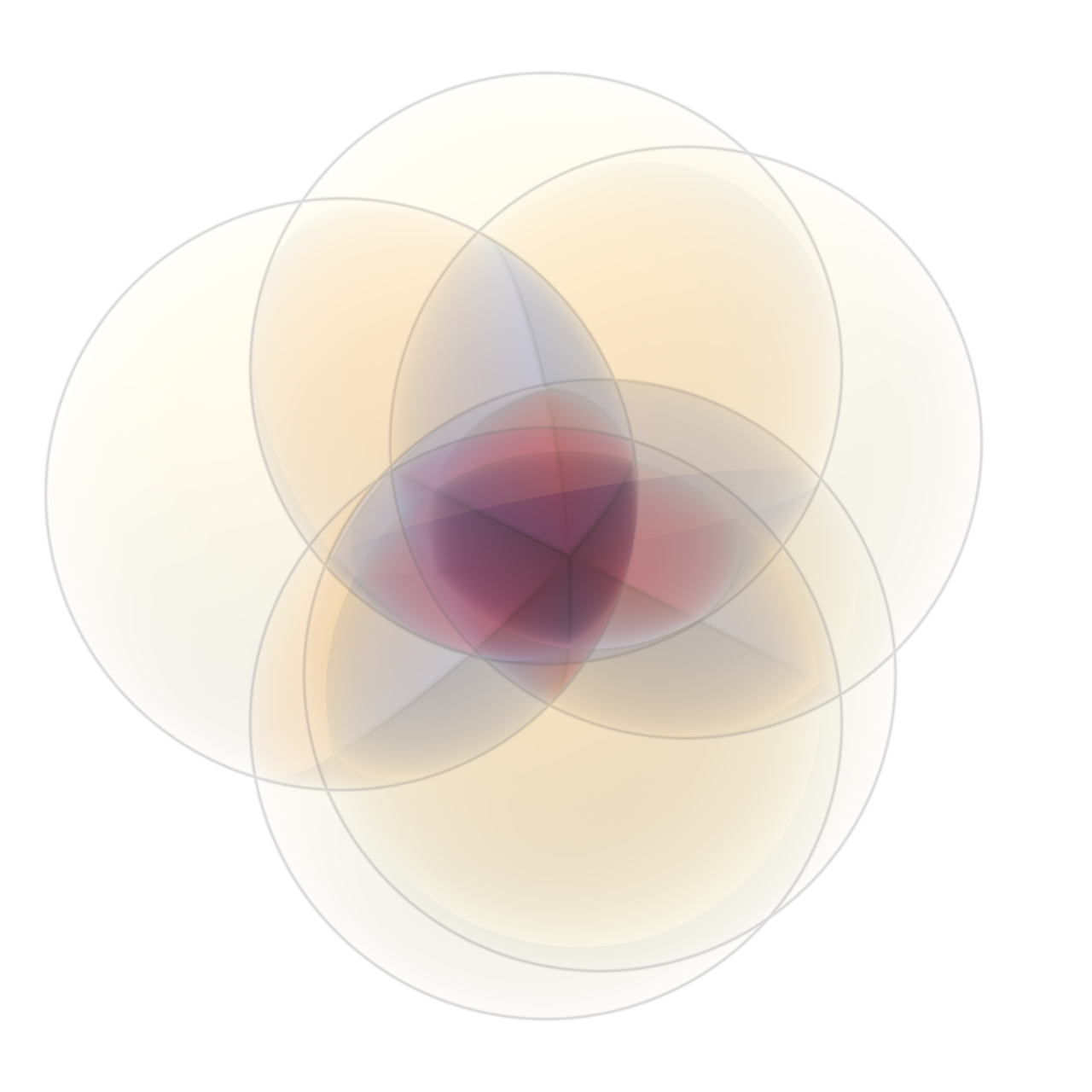}};
      \begin{scope}[x={(im.south east)},y={(im.north west)}]
\filldraw[site] (0.54926505,0.38157643) circle[radius=1.55pt];
\filldraw[site] (0.62808913,0.59473886) circle[radius=1.55pt];
\filldraw[site] (0.31279281,0.54736943) circle[radius=1.55pt];
\filldraw[site] (0.50000000,0.66235490) circle[radius=1.55pt];
\filldraw[site] (0.50000000,0.33764510) circle[radius=1.55pt];
\filldraw[witness] (0.48357832,0.53947452) circle[radius=2pt];

      \end{scope}
     \end{tikzpicture}}}
  & \coverpanel{\dimexpr-\spheredepthwidth*19/600\relax}{\raisebox{-.5\height}{%
      \includegraphics[width=\spheredepthwidth,trim=0 78bp 0 26bp,clip]{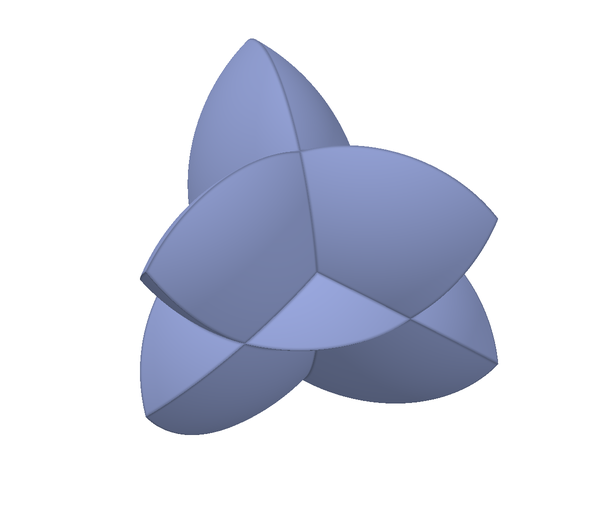}}}
  & \coverpanel{\dimexpr\spheredepthwidth*23/1200\relax}{\raisebox{-.5\height}{%
      \includegraphics[width=\spheredepthwidth,trim=0 78bp 0 26bp,clip]{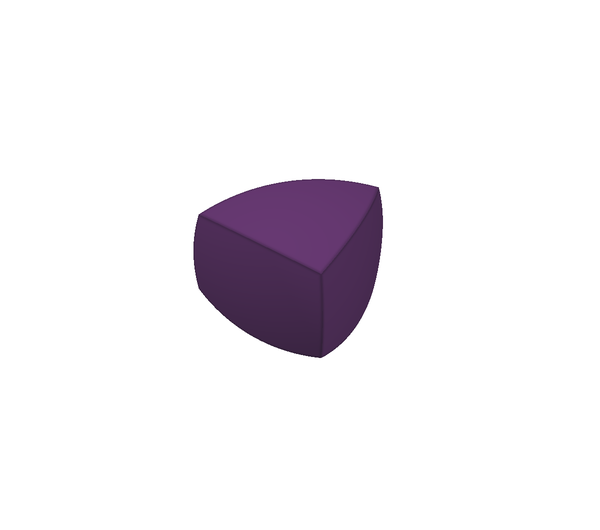}}}\\[2pt]
  Five sites in one ball & A ball at each site & At least three overlap & All five overlap
\end{tabular}\par
\endgroup

\PaperContents
\fi

\AddToHook{cmd/section/before}{\clearpage}
\section{Introduction}\label{sec:problem}
\ifIllustrated
The firefly orchestra has rehearsed everything except where to put the
conductor. The players hold their places in the evening air. The conductor
raises her baton (the small stick she uses to mark the beat). On the
downbeat, every player at most a metre from its tip lights up. Where
should she hover to bring in the most players?

Each player answers, ``Anywhere within a metre of me.''
\else
Where should a closed unit ball be centred to cover as many input
sites as possible?
\fi
The finite input \(P\subset\R^D\) consists of points called \emph{sites}. The unit ball about
\(c\) covers \(p\) exactly when \(c\in B(p,1)\), the unit ball about \(p\).
The \emph{depth} at \(c\) counts the site balls containing it; we seek the
maximum \(M(P)=\max_c|\{p\in P:\|p-c\|\le1\}|\).
This is unweighted maximum range sum (MaxRS) for a ball,
also studied in facility location~\citep{drezner1981modified}.

Two sites exactly two units apart fit in one unit ball only when its centre
is their midpoint. A maximum can therefore occur at a single contact.
We find an attaining centre without storing the arrangement of the
spheres that bound the site balls.
\begin{table}[!htb]
\centering
\caption{Time and space bounds on \(n\) inputs
(fixed \(D\)); \(e\) counts intersecting disk pairs, including containment.
The arrangement is the full map of regions, boundary pieces and intersections.
De et al.\ need \(O(n)\) words for a writable copy if the input is read-only.
\ifIllustrated\else The expected bound uses the arrangement labelling
in Section~\ref{sec:context} in general position.\fi}\label{tab:prior}
\small
\setlength{\tabcolsep}{1.7pt}
\begin{tabular*}{\linewidth}{@{\extracolsep{\fill}}lllll@{}}
\toprule
\ifIllustrated Authors\else Source\fi & Year & Regions & Time & Worst-case space\\
\midrule
Chazelle, Lee~\citep{chazelle1986circle} & \citeyear{chazelle1986circle}
 & Equal-radius disks & \(O(n^2)\) & \(\Theta(n^2)\) arcs\\
\ifIllustrated\else
Mulmuley~\citep{mulmuley1991fast} & \citeyear{mulmuley1991fast}
 & Disks/exteriors & \(O(n^2)\) exp. & Full arrangement\\
\fi
De, Nandy, Roy~\citep{de2014place} & \citeyear{de2014place}
 & Arbitrary disks & \(O(n^2+e\log n)\) & \(O(1)\) if writable\\
This paper, Thm.~\ref{thm:planar} & 2026 & Disks/exteriors & \(O(n^2)\) & \(O(n)\) words\\
This paper, Thm.~\ref{thm:main} & 2026 & Unit \(D\)-balls & \(O(n^D)\) & \(O(n)\) words\\
\bottomrule
\end{tabular*}
\end{table}

Our real random-access machine (real RAM) counts each arithmetic operation,
square root, comparison or indexed access as one step, without floor on
arbitrary reals or size-dependent advice. A word stores one real or index;
workspace is storage beyond the read-only input and the output. Constants may depend on \(D\).

\begin{theorem}\label{thm:main}
For \(N\) distinct points in fixed dimension \(D\ge2\), a deterministic
algorithm finds a closed unit ball of maximum coverage in \(O(N^D)\)
time using \(O(N)\) extra memory. It returns the centre, the count
\(M(P)\) and the indices of all points the ball covers.
\end{theorem}
\noindent\textit{Why stop at a sphere?} In 3D, fix one site on the ball
boundary. The possible \emph{centres} lie on a two-sphere about that site:
the surface, not the solid ball.
Each other site allows a patch of this sphere, its \emph{cap}. We project
the caps with circular rims to closed disks or closed exteriors. We map
the planar winners back to the sphere and check the one point the
projection omits (Theorem~\ref{thm:sphere}).
In dimension \(D\ge3\), fixing up to \(D-2\) boundary sites suffices:
\(O_D(N^{D-2})\) searches at \(O(N^2)\) time each give the
\hyperref[proof:main]{time bound of Theorem~\ref*{thm:main}}.
Fixing one more site and sorting events on each remaining circle
would cost an extra logarithm. We search one sphere at a time;
Theorem~\ref{thm:planar} bounds the planar workspace by \(O(N)\) words
even for unequal radii and weights on both closed sides.

\medskip\noindent\textit{Conditional lower bound.} Let \(H\) be the most of
\(s\) distinct integer points on one hyperplane. In linear time we build
\(2s\) rational sites with optimum \(s+H\) (Theorem~\ref{thm:lens}).
Subtracting \(s\) recovers \(H\): a deterministic exponent saving for coverage
would improve hyperplane detection as well (Corollary~\ref{cor:exponent}).
\ifIllustrated\else
Geometric terms are collected in the \hyperref[sec:glossary]{glossary
on p.~\pageref*{sec:glossary}}.
\fi

\ifIllustrated
\section{Placing one disk to cover the most points in the plane}\label{sec:plane}
Given \(N\) distinct points in the plane and a prescribed radius \(R>0\),
where should we centre one closed disk to cover as many points as possible?
We use \(R\) as the unit of length, so the disk has radius one.\footnote{Solve
on \(p_i/R\). For a returned centre \(c\), report \(Rc\) with the same covered indices.}
The centre need not be an input point. We first solve the problem by moving
the disk itself. Then we obtain a faster algorithm by looking at the set of
centres that cover each point.

\subsection{Where can we find an optimal centre?}\label{sec:unit-boundary}
A disk centred at \(c\) covers site \(p_i\) exactly when \(c\) is
in the unit disk \(B(p_i,1)\) about that site. Thus the centres that
cover all sites indexed by \(I\) form the intersection
\(\bigcap_{i\in I}B(p_i,1)\) (Figure~\ref{fig:planar-depth}).
A covered site is \emph{tight} at \(c\) if it lies on the rim:
its distance from \(c\) is exactly one.

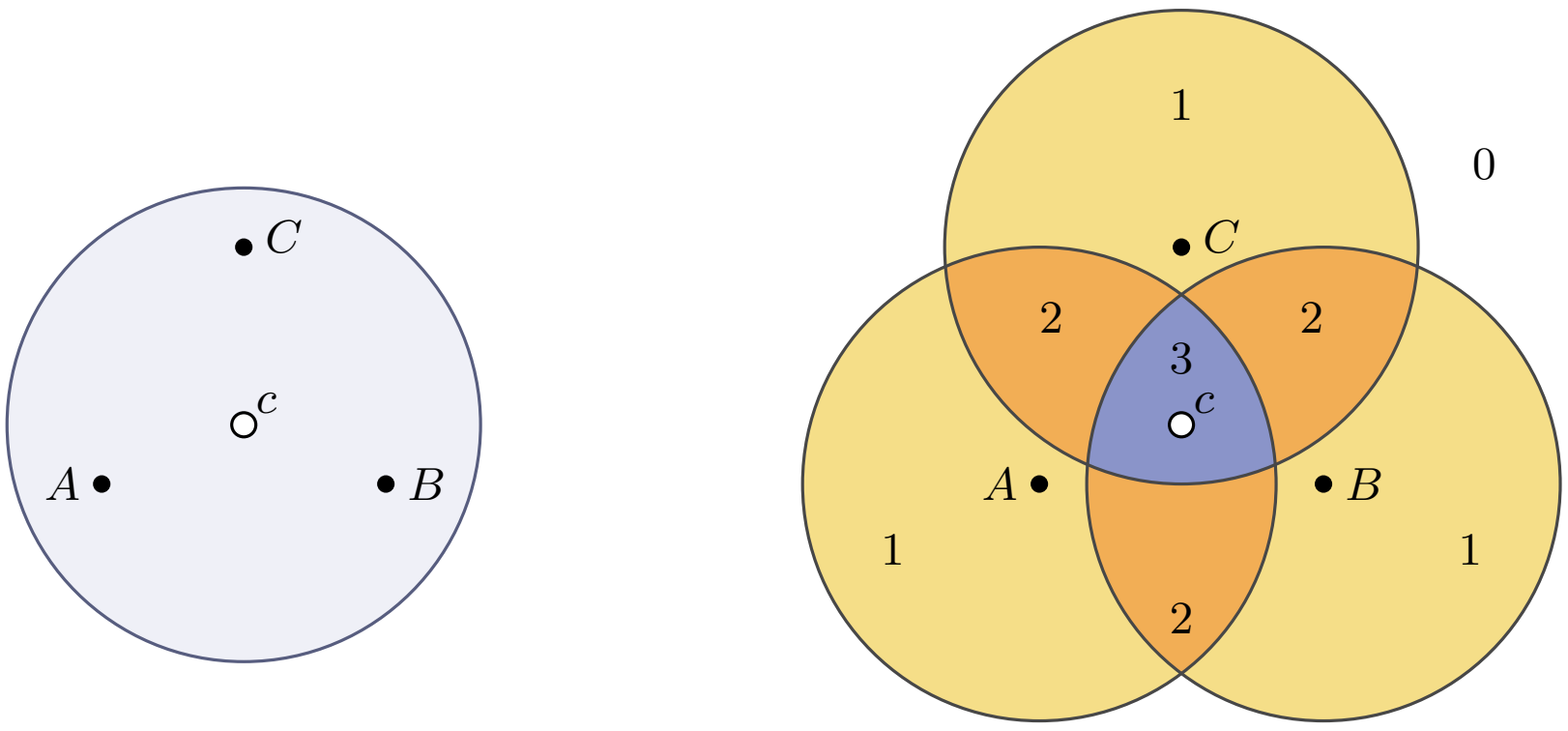
\begin{figure}[!htb]
\centering
\begin{tikzpicture}[scale=1.62,font=\small,
                    rim/.style={draw=black!72,semithick}]
  \def\marksites{%
    \foreach \x/\y/\site/\side/\lift in
      {-.6/-.25/A/east/0pt,.6/-.25/B/west/0pt,0/.75/C/west/2pt}{%
        \fill (\x,\y) circle[radius=1.05pt];
        \node[anchor=\side,inner sep=4pt,yshift=\lift] at (\x,\y) {\(\site\)};
    }%
    \draw[semithick,fill=white] (0,0) circle[radius=1.45pt];
    \node[above right,inner sep=2.2pt] at (0,0) {\(c\)};
  }
  \begin{scope}[shift={(-1.98,0)}]
    \draw[rim,draw=figRim,fill=depth3!14!white] (0,0) circle[radius=1cm];
    \marksites
  \end{scope}
  \begin{scope}[shift={(1.98,0)}]
    \coordinate (A) at (-.6,-.25);
    \coordinate (B) at (.6,-.25);
    \coordinate (C) at (0,.75);
    \foreach \s in {A,B,C} \fill[depth1] (\s) circle[radius=1cm];
    \foreach \s/\t in {A/B,B/C,C/A}{
      \begin{scope}
        \clip (\s) circle[radius=1cm];
        \fill[depth2] (\t) circle[radius=1cm];
      \end{scope}}
    \begin{scope}
      \clip (A) circle[radius=1cm];
      \clip (B) circle[radius=1cm];
      \fill[depth3] (C) circle[radius=1cm];
    \end{scope}
    \foreach \s in {A,B,C} \draw[rim] (\s) circle[radius=1cm];
    \marksites
    \foreach \x/\y/\depth in {-1.22/-.53/1,1.22/-.53/1,0/1.35/1,
                               0/-.82/2,-.55/.45/2,.55/.45/2,
                               0/.28/3,1.28/1.10/0}
      \node[inner sep=1pt] at (\x,\y) {\(\depth\)};
  \end{scope}
  \node[anchor=north] at (-1.98,-1.39) {(a) One disk around \(c\)};
  \node[anchor=north] at (1.98,-1.39) {(b) One disk around each site};
\end{tikzpicture}
\caption{Filled dots are sites; the open circle is the centre \(c\).
(a) One disk covers all three sites. (b) The numbers count site disks
in the open regions. The region marked \(3\), including its boundary,
is the set of all centres covering the three sites}\label{fig:planar-depth}
\end{figure}

\LayoutFloatBarrier
\else
\section{Planar placement and closed circle depth}\label{sec:plane}
A fixed input radius \(R>0\) can be scaled to one. A site \(p\) is
\emph{tight} when it lies on the boundary: \(\|c-p\|=1\). The following observation
works in every dimension.
\fi
\begin{lemma}[A tight site]\label{lem:tight}
For a nonempty finite input in any positive dimension, some optimal
closed unit ball has a site on its boundary.
\end{lemma}
\begin{proof}
An optimum exists because the count takes finitely many values, and
it covers at least one site since a site can serve as centre. Intersect
the closed balls of possible centres for its covered sites. The intersection
contains the optimal centre and is compact: it is closed and contained
in any one of these balls. Take its point of smallest first coordinate.
If all the distances there were strictly less than one, a small move to
the left would preserve every inequality, contradicting minimality.
Thus this optimal centre has a tight site.
\end{proof}

\ifIllustrated
Lemma~\ref{lem:tight} reduces the search from the whole plane to the
\(N\) circles about the sites. That is enough for an algorithm that
walks around each circle, but each circle still contains infinitely
many possible centres. Can we give a finite list instead?
Intersections are natural candidates, but there may be none.
For a single site at \((3,0)\), a unit disk centred at the origin
misses the site, whereas one centred at \((4,0)\), the rightmost
point of the site circle, covers it. Add one such mark per site
circle and the origin to the pair contacts.
\ifdefined\ArxivPreprint\else
To see why the list
suffices, follow the centre in
Figure~\ref{fig:candidate-walk}: it reaches a candidate without losing
any disk that contained it at the start.
\fi

\OpeningSpace
\fi
\begin{lemma}[Finite candidates]\label{lem:plane-candidates}
For a finite family \(F\) of distinct positive-radius circles in
the plane, let \(\operatorname{cand}(F)\) consist of the origin \(o\),
each circle's point \(R(C)\) of largest first coordinate, and all pairwise rim
intersections, including tangencies.
\ifIllustrated
\footnote{If a rightmost point is also a contact, testing it twice leaves
the maximum unchanged.}
\fi
For every \(x\in\R^2\) there is a \(y\in\operatorname{cand}(F)\)
such that, for each \(C\in F\), either \(y\) lies on \(C\) or
\(x\) and \(y\) are both strictly inside \(C\) or both strictly outside it.
\end{lemma}
\begin{proof}
Move along the segment from \(x\) to \(o\) until reaching \(o\) or
a rim, taking no step if \(x\) is already on a rim. At \(o\) or a
meeting of rims, stop. Otherwise exactly one rim \(C\) has been reached. Follow it towards
\(R(C)\), stopping there or at the first contact with another rim.
The endpoint is a candidate. Changing from inside a circle to outside
it, or vice versa, would require crossing its rim. Each move stops
at the first such contact, and a rim containing \(x\) is never left.
\end{proof}

\ifIllustrated
\noindent\textit{Corollary for unit disks.}
Start the walk of Lemma~\ref{lem:plane-candidates} at an optimal centre.
A site disk containing the starting centre still contains the endpoint: the
walk either stays inside or stops on its rim, which counts. Thus testing
just the candidates returns the exact maximum. There are at most
\(1+N+2\binom N2=1+N^2\) of them. Testing each against every site
takes \(O(N^3)\) time. The next algorithm uses Lemma~\ref{lem:tight}
to sweep circles instead; the quadratic algorithm will use
Lemma~\ref{lem:plane-candidates} to finish a region with few rims left.

\ifdefined\ArxivPreprint\else
\begin{figure}[!htb]
\centering

\caption{From \(x\), move toward the origin \(o\) until reaching \(C\)
at \(h\). Follow the upper arc toward \(R(C)\). The first
contact with \(B\), at \(y\), ends the walk: \(C\) still counts and \(B\)
now counts too}\label{fig:candidate-walk}
\end{figure}
\FloatBarrier
\fi

\OpeningSpace
\subsection{Finding the best placement by an angular sweep}\label{sec:walk}
Try each site \(p\) as the \emph{anchor}, keeping it on the rim of the
moving disk. The centre \(c=p+u\) traces a circle as the unit vector
\(u\) turns. Another site \(q=p+w\) counts exactly when
\else
Choose \(p\) as anchor, keeping it tight. Then the centre is \(p+u\)
with \(\|u\|=1\).
For a second site \(q=p+w\), expanding the squared distance gives
\fi
\begin{equation}\label{eq:anchor}
  \norm{u-w}^2=1-2w\cdot u+\norm w^2\le1
  \quad\Longleftrightarrow\quad w\cdot u\ge\norm w^2/2 .
\end{equation}
\ifIllustrated
Turn the axes so that \(w\) points to the right and write
\(d=\norm w>0\). Then \(w\cdot u=d\,u_x\): the allowed directions
are the part of the unit circle with \(u_x\ge d/2\). For \(d<2\)
this is a closed arc with two endpoints, for \(d=2\) one point, and for \(d>2\) the
empty set. No rotation is needed in the algorithm. Intersect the
unit circles about \(p\) and \(q\) to find the endpoints and keep
the arc facing \(q\), the one containing \(w/d\).

In Figure~\ref{fig:distance-walk}(a), \(p=(0,0)\) and \(q_+=(1,1)\).
The test \(u_1+u_2\ge1\) selects the quarter-circle from \((1,0)\) to \((0,1)\).

\begin{walkthroughfigure}
\centering
\ifdefined\ArxivPreprint
\begin{tabular}{@{}c@{\hspace{.55cm}}c@{\hspace{.55cm}}c@{}}
\begin{tikzpicture}[x=1.34cm,y=1.34cm,font=\small,
    baseline=(current bounding box.south)]
  \begin{scope}
    \clip (0,0) circle[radius=1];
    \fill[depth3!23!white] (0,1) -- (1,0) -- (2,0) -- (2,2) -- (0,2) -- cycle;
  \end{scope}
  \draw[black!55] (0,0) circle[radius=1];
  \draw[black!75,semithick,densely dashed] (0,1) -- (1,0);
  \draw[figRim,line width=1.6pt] (1,0) arc[start angle=0,end angle=90,radius=1];
  \fill[figRim] (1,0) circle[radius=2pt] (0,1) circle[radius=2pt];
  \fill (0,0) circle[radius=1.6pt] node[below left,inner sep=2pt] {\(p\)};
  \fill (1,1) circle[radius=1.6pt] node[above,inner sep=2pt] {\(q_+\)};
  \node[above left,inner sep=1.7pt] at (0,1) {\((0,1)\)};
  \node[below right,inner sep=2pt] at (1,0) {\((1,0)\)};
  \node[below,inner sep=1pt] at (0,-1.08) {\(u_1+u_2\ge1\)};
\end{tikzpicture}
&
\begin{tikzpicture}[x=1.43cm,y=1.43cm,font=\small,
    baseline=(current bounding box.south),
    centre/.style={draw=black,fill=white,semithick}]
  \fill[depth3!23!white] (0,1) circle[radius=1];
  \draw[draw=depth2!56!black,dashed] (.6,.8) circle[radius=1];
  \draw[draw=depth2!56!black,dashed] (-.6,.8) circle[radius=1];
  \draw[thick,draw=figRim] (0,1) circle[radius=1];
  \pgfmathsetmacro{\startangle}{atan2(4,3)}
  \draw[thick,draw=figTrace] (.6,.8)
    arc[start angle=\startangle,end angle={180-\startangle},radius=1];
  \draw[->,thick,draw=figTrace] (102:1) arc[start angle=102,end angle=113,radius=1];
  \fill (0,0) circle[radius=1.7pt] node[below,inner sep=2pt] {\(p\)};
  \fill (1,1) circle[radius=1.6pt] node[above right,inner sep=2pt] {\(q_+\)};
  \fill (-1,1) circle[radius=1.6pt] node[above left,inner sep=2pt] {\(q_-\)};
  \draw[centre] (.6,.8) circle[radius=1.7pt] node[below,inner sep=2pt] {\(c_+\)};
  \draw[centre,draw=figTrace] (0,1) circle[radius=1.7pt] node[above,inner sep=2pt] {\(c_0\)};
  \draw[centre] (-.6,.8) circle[radius=1.7pt] node[below,inner sep=2pt] {\(c_-\)};
\end{tikzpicture}
&
\begin{tikzpicture}[x=1.16cm,y=.65cm,font=\small,
    baseline=(current bounding box.south)]
  \foreach \y/\name in {3/p,2/q_+,1/q_-}
    \node[left,inner sep=3pt] at (0,\y) {\(\name\)};
  \draw[densely dashed,draw=figTrace,semithick] (1,.55) -- (1,3.48)
    node[above,inner sep=1pt] {\(c_0\)};
  \foreach \y/\l/\r in {3/0/4,2/0/1,1/1/2}{
    \draw[line width=2.15pt,draw=figRim] (\l,\y) -- (\r,\y);
    \fill[figRim] (\l,\y) circle[radius=1.9pt] (\r,\y) circle[radius=1.9pt];
  }
  \fill[figRim] (4,2) circle[radius=1.9pt];
  \draw[->] (0,.14) -- (4.13,.14);
  \foreach \x/\a in {0/0,1/90,2/180,3/270,4/360}
    \draw (\x,.22) -- (\x,.06) node[below,inner sep=1pt] {\(\a^\circ\)};
  \draw[black!75] (0,-1.05) -- (1,-1.05);
  \foreach \x in {0,1} \draw[black,fill=white] (\x,-1.05) circle[radius=1.8pt];
  \fill[figTrace] (.5,-1.05) circle[radius=2pt];
  \node[below,inner sep=3pt] at (.5,-1.16) {\(45^\circ\) start};
\end{tikzpicture}\\[.2cm]
{\small (a) Allowed directions} & {\small (b) Moving disk} & {\small (c) Covered angles}
\end{tabular}
\caption{One turn around the anchor \(p=(0,0)\), with \(q_\pm=(\pm1,1)\).
(a) The dashed chord bounds the closed arc for \(q_+\). (b) The moving disk
covers both \(q_+\) and \(q_-\) at \(c_0\). (c) Read the bars at
\(90^\circ\): one arc ends and another starts, so both sites count
along with \(p\). The \(360^\circ\) mark repeats \(0^\circ\); \(45^\circ\)
is a possible start between marks.}
\else

\caption{Three views of the anchor \(p=(0,0)\) and sites \(q_\pm=(\pm1,1)\).
(a) The chord is equality in~\eqref{eq:anchor} for \(q_+\); the heavy
arc consists of the allowed centre positions, endpoints included.
(b) The centre of the moving disk follows the arrow from \(c_+\)
through \(c_0\) to \(c_-\). (c) Unroll the same turn into a line: each
solid bar gives the angles covering that site. Three bars meet at \(90^\circ\),
but only two cover an angle just before or after it. On the bottom row,
\(45^\circ\) is a possible start, strictly between endpoint marks}
\fi\label{fig:distance-walk}
\end{walkthroughfigure}

\Needspace{6\baselineskip}
\begin{example}[A three-site walk]\label{ex:planar-walk}
Take the three sites in Figure~\ref{fig:distance-walk}. Measured
counterclockwise from \((1,0)\), the arc for \(q_+\) is
\([0^\circ,90^\circ]\). Reflecting it across the vertical axis gives \([90^\circ,180^\circ]\)
for \(q_-\). At \(c_+=(3/5,4/5)\) the disk covers \(p,q_+\);
at \(c_-=(-3/5,4/5)\) it covers \(p,q_-\). At \(c_0=(0,1)\) one
arc ends where the other begins, and all three count. Because the
two sites \(q_\pm\) are two units apart, a unit disk can cover both
only if it is centred at their midpoint \(c_0\). Section~\ref{sec:caps} uses
the same instance in three dimensions.
\end{example}

Sort the endpoint directions counterclockwise, grouping equal marks.
No angles are needed: sort first into the upper and lower half-circles,
assigning the positive horizontal ray to the upper half and the negative
ray to the lower. Within a half, \(a\) precedes \(b\) if
\(\det(a,b)=a_xb_y-a_yb_x>0\); zero means the same mark.
An \emph{open gap} is the arc strictly between consecutive marks,
or the rest of the circle if there is just one mark. Coverage need
not be zero there, but it is constant, so one count initializes the sweep.

Let \(s\) be the score in the gap \emph{before} a mark. It already
counts ending arcs. Add starts and one-point arcs to score the mark;
then remove ends and one-point arcs for the next gap. Every closed arc
covering a gap contains its endpoints, so no gap beats an adjacent mark.
At \(90^\circ\) the scores are \(2,3,2\): a net update alone misses the winner.

\begin{algorithm}[H]
\caption{\textsc{SweepCircle}: place one closed unit disk to cover the most sites}\label{alg:sweep}
\small
\begin{algorithmic}
\Require Distinct planar sites \(P=(p_1,\ldots,p_N)\)
\State Set the best centre to the origin and its count to \(|\{i:\norm{p_i}\le1\}|\).
\ForAll{anchors \(p\) in \(P\)}
  \State Form the arcs for the other sites on the unit circle about \(p\); sort and group endpoints.
  \State Choose \(z\) in an open gap; if there are no marks, use \(p+(1,0)\).
  \State \(s\gets|\{i:\norm{z-p_i}\le1\}|\); keep \((z,s)\) if its count improves the best.
  \ForAll{endpoint groups in cyclic order starting just after \(z\)}
    \State At \(x\), let \(a,e\) count starts and ends of two-endpoint arcs, and \(t\) count one-point arcs.
    \State Keep \((x,s+a+t)\) if its count improves the best.
    \State \(s\gets s+a-e\).
  \EndFor
\EndFor
\State For the best centre \(c\), collect \(L=\{i:\norm{c-p_i}\le1\}\); \Return \((c,|L|,L)\).
\end{algorithmic}
\end{algorithm}
To choose \(z\) in the counterclockwise gap from \(a\) to \(b\), the
unit direction \((a+b)/\norm{a+b}\) bisects the shorter arc. Use its
negative for the longer arc (\(\det(a,b)<0\)). For opposite directions,
turn \(a\) counterclockwise by \(90^\circ\); for a lone mark use \(-a\).
Add the chosen direction to \(p\).
Lemma~\ref{lem:tight} makes the best centre over all anchors optimal.
Sorting \(O(N)\) endpoints per anchor takes \(O(N^2\log N)\) total
time. Only one event list is live, so the extra memory is \(O(N)\).

\OpeningSpace
\subsection{A quadratic search for the best disk centre}\label{sec:unit-depth}
\ifdefined\ArxivPreprint
The angular sweep sorts a long list for every anchor. Instead, draw a
closed unit disk about each site: the score at a possible centre is the
number of disks containing it. We cut the plane into pieces so that later
sorts use shorter lists.

In Figure~\ref{fig:open-child}, the disks \(C,D,E\) have centres
\((0,0)\), \((2,0)\) and \((1/2,0)\), respectively. Cut the enclosing
rectangle along the rim of \(C\) and the vertical walls through its
leftmost and rightmost points. The other two rims are drawn for scoring.
With only three disks we could test the finite candidates directly;
this split illustrates what happens on a longer list.
\else
The angular sweep sorts a new list for every anchor. Can we avoid
sorting long lists over and over? We preview one split here;
Section~\ref{sec:depth} defines the cells and gives the algorithm.
Draw a closed unit disk about each site.
Any point \(c\) in the plane can be the centre of the disk we place;
its score is the number of drawn disks containing \(c\). In
Figure~\ref{fig:open-child}(a) the dot \(v\) marks a centre where two
rims touch. The unmarked point \((-3/4,0)\) is also a possible centre,
although it lies on no rim. Throughout the interior of \(C\), we can
count \(C\) once for the whole piece.

\medskip\noindent\textit{One split by hand.}
In Figure~\ref{fig:open-child}(a), enclose three unit disks in a
rectangle. Their centres are \((0,0)\) for \(C\), \((2,0)\) for \(D\)
and \((1/2,0)\) for \(E\). Choose the rim of \(C\) as a cut, and draw
vertical walls from its leftmost and rightmost points to the box.
The open pieces left by these cuts can be searched separately. The other
two rims are drawn so we can score positions; neither is a cut. We
make this split just to show the bookkeeping. With only three disks
the algorithm can test the finite candidates directly.
\fi

\begin{walkthroughfigure}
\centering
\ifdefined\ArxivPreprint
\begin{tikzpicture}[x=1.63cm,y=1.63cm,font=\small,line cap=round,line join=round,
  cut rim/.style={draw=figRim,line width=1pt},
  test rim/.style={draw=black!75,line width=.75pt,dash pattern=on 3.2pt off 2pt on .6pt off 2pt}]
  \draw[black!55,thin] (-1.38,-1.27) rectangle (3.38,1.3);
  \foreach \x in {-1,1} \draw[black!65,semithick] (\x,-1.27) -- (\x,1.3);
  \draw[black!65,dashed,semithick] (2,0) circle[radius=1];
  \draw[test rim] (.5,0) circle[radius=1];
  \draw[cut rim] (0,0) circle[radius=1];
  \node[below,inner sep=2pt] at (0,-1) {\(C\)};
  \node[above,inner sep=1.6pt] at (.53,1) {\(E\)};
  \node[above,inner sep=1.6pt] at (2,1) {\(D\)};
  \fill[figTrace] (1,0) circle[radius=2.4pt];
  \node[right,inner sep=4pt] at (1,0) {\(v\)};
  \node at (1,-1.57) {(a) On the cut: \(v\) scores 3};

  \begin{scope}[shift={(5.48,0)}]
    \fill[depth1!35!white] (0,0) circle[radius=1];
    \begin{scope}
      \clip (0,0) circle[radius=1];
      \fill[depth2!60!white] (.5,0) circle[radius=1];
      \draw[test rim] (.5,0) circle[radius=1];
    \end{scope}
    \draw[cut rim] (0,0) circle[radius=1];
    \node[font=\normalsize] at (-.75,0) {\(1\)};
    \node[font=\normalsize] at (.29,0) {\(2\)};
    \filldraw[fill=white,draw=figTrace,line width=.9pt] (1,0) circle[radius=2.4pt];
    \node[right,inner sep=4pt] at (1,0) {\(v\)};
    \node at (0,-1.57) {(b) Inside \(C\): test only \(E\)};
  \end{scope}
\end{tikzpicture}%
\caption{(a) The filled point \(v\) lies on the rims of \(C,D\) and
inside \(E\), so all three disks count. (b) The hollow mark shows the
same point excluded from the open child. Within this single child,
points in \(E\), including its dot-dashed rim, score two; the rest
score one.}
\else

\caption{The same split viewed twice. (a) On the solid rim near \(v\),
both \(C\) and \(E\) count; at \(v\), \(D\) counts too. (b) Inside \(C\)
only the rim of \(E\) still matters. Its two sides have scores one and
two. This entire open disk is one piece of the cut, even though its
score is not constant}
\fi\label{fig:open-child}
\end{walkthroughfigure}

\ifdefined\ArxivPreprint
Along the rim of \(C\), the scores just before, at and just after
\(v=(1,0)\) are \(2,3,2\): \(D\) counts only at the contact.
Strictly inside \(C\), start at one for \(C\) and test only \(E\).
Using that shorter list at \(v\) would lose the only centre of score three.

In general, score each new vertex. On each new edge, sort the contacts
with other rims and score both the contacts and the open gaps. Now search
each open piece: if a rim misses it, one test anywhere inside determines
the disk's contribution throughout. Once the list has constant size,
test the candidates of Lemma~\ref{lem:plane-candidates} strictly inside
the piece. Any winning candidate on a cut has already been scored.
Keep the best centre and scan the sites to report its covered indices.

\noindent\textit{What do all the sorts cost?}
If a level's limit on list length lies in \([k,2k)\), the cutting of
Lemma~\ref{lem:cut} has \(O((N/k)^2)\) pieces at that level. Each has
only a constant number of new edges, with \(O(k)\) contacts to sort per
edge. The limit falls by more than half at each step, so each row of
Table~\ref{tab:plane-sort-budget} pays for at most one level.
\else
\noindent\textit{Score the cut before setting any disk aside.}
Follow the solid rim towards the contact \(v=(1,0)\). Just before
\(v\), the score is two (\(C,E\)); at \(v\) it is three (\(C,D,E\));
just after \(v\) it is two again. In fact \(v\) is the only centre
that covers all three sites, because \(C\) and \(D\) touch only there.
On a general cut we check each new vertex. Along each new arc or wall
segment we sort the places where other rims meet it. Between those
places the score is constant; at a contact we update precisely the
disks that meet there, counting the contact itself.

\noindent\textit{Pass a count and a shorter list to the interior.}
Now look \emph{strictly inside} \(C\), as in Figure~\ref{fig:open-child}(b).
\(C\) always counts and \(D\) never does. The record for this piece
is therefore ``start at one; test only \(E\).'' At \((-3/4,0)\) the
answer is one; at \((1/4,0)\) it is two. Using that short list at
\(v\) would give two instead of three. This is why the cut was
scored first with the longer list.

Repeat inside each open piece. If a rim misses a piece, one test
anywhere inside tells us whether to add one or zero throughout it.
When only a constant number of rims remain, use
Lemma~\ref{lem:plane-candidates}: test the origin, their rightmost
marks and their pair contacts, keeping only points strictly inside the
piece. No winning candidate is lost: one on a cut was already scored;
one inside still has all its defining rims on the list. Keep the best
centre seen. A final scan reports the sites it covers.

\Needspace{9\baselineskip}
\noindent\textit{Why the sorts cost only quadratic time in total.}
Near the top of the search there are long lists but few regions.
Near the bottom there may be quadratically many regions, but their
lists are short. The one ingredient we use here without proof is how
to choose the cuts. Lemma~\ref{lem:cut} gives a deterministic choice
with two guarantees: each new level lowers the maximum allowed list
length by a fixed factor greater than two, and a level whose limit
lies in \([k,2k)\) has \(O((N/k)^2)\) regions. The construction is
proved in Appendix~\ProofRef{sec:construction}. In Table~\ref{tab:plane-sort-budget}
we group these limits by powers of two. In each region we sort \(O(k)\)
contacts per edge on only a constant number of new edges.
\fi

\begin{table}[!htb]
\centering
{\small
\newcommand{\budgetbar}[1]{\textcolor{figRim}{\rule{#1}{.62ex}}}
\setlength{\tabcolsep}{5pt}
\begin{tabular}{@{}rccc@{\hspace{4pt}}l@{}}
\toprule
List scale \(k\) & Regions & Work in one & \multicolumn{2}{c@{}}{Total divided by \(N^2\)}\\
\midrule
\(1\) & \(N^2\)    & \(1\)  & \(1\)     & \budgetbar{13mm}\\
\(2\) & \(N^2/4\)  & \(4\)  & \(1\)     & \budgetbar{13mm}\\
\(4\) & \(N^2/16\) & \(12\) & \(3/4\)   & \budgetbar{9.75mm}\\
\(8\) & \(N^2/64\) & \(32\) & \(1/2\)   & \budgetbar{6.5mm}\\
\addlinespace[2pt]
\(2^i\) & \(N^2/k^2\) & \(k\log_2(2k)\) & \multicolumn{2}{c@{}}{\((i+1)/2^i\)}\\
\bottomrule
\end{tabular}}

\caption{Sorting work, read from short lists to long ones. Each row
multiplies the region bound by the work per region; the last column
shows the result as a fraction of \(N^2\). Entries are upper bounds
up to constant factors, not counts from a particular run}\label{tab:plane-sort-budget}
\end{table}
\FloatBarrier

\ifdefined\ArxivPreprint
The last column sums to \(\sum_{i\ge0}(i+1)/2^i=4\), so all the sorts
together cost \(O(N^2)\). Constructing the cuts takes \(O(N^2)\) time
as well. A depth-first search keeps one branch at a time; its list
limits sum to \(O(N)\). Section~\ref{sec:depth} gives the full algorithm.
\else
Since the list limit drops by more than half, each row applies to at
most one level (some rows are skipped).
For \(k=2^i\) the last column is \((i+1)/2^i\); the geometric sums
\(\sum_{i\ge0} i/2^i=\sum_{i\ge0}1/2^i=2\) give four in total.
Thus all the sorts together cost \(O(N^2)\).
Constructing the cuts also takes \(O(N^2)\) time by Lemma~\ref{lem:cut}.
The search finishes one child before starting its siblings. Only the
lists along the current branch need to be kept, and their geometrically
decreasing sizes sum to \(O(N)\).
\fi

\LayoutFloatBarrier
\OpeningSpace
\subsection{Extending the solver to unequal radii and weights}\label{sec:plane-contract}
The overlap search also works for disks of different radii: two
distinct circles still meet at most twice. For weighted disks, add
their weights instead of counting them.

Why count exteriors? In Section~\ref{sec:caps}, the centres that cover
one site form a spherical cap. If the cap contains the projection
pole and its circular rim avoids the pole, its image is the
\emph{closed exterior} of a circle: the rim and everything outside it
(Figure~\ref{fig:pole-sides}f). Those points earn the site's weight;
the omitted pole is tested separately.
\else
The line in~\eqref{eq:anchor} is \(\norm w/2\) from the origin: it cuts
off a closed arc of unit directions for \(\norm w<2\), leaves one point
at equality, and none beyond. Sweeping those arcs in endpoint order
costs \(O(N^2\log N)\) over all anchors; at tied endpoints we count
both starting and ending arcs. Section~\ref{sec:depth} gives the
quadratic alternative: draw a unit disk about each site. On a connected
cell its rim misses, a disk either covers the whole cell or none of it.
Score the cell boundaries separately to retain contacts.
\subsection{Depth with both closed sides}\label{sec:plane-contract}
\fi

For a circle \(C\) of centre \(z_C\) and radius \(r_C>0\), give its
closed disk weight \(w_C^{\rm in}\) and its closed exterior weight
\(w_C^{\rm out}\), both nonnegative. A point on the rim belongs to
\emph{both} sets. The contribution used by the algorithm is therefore
\[
 w_C(x)=
 \begin{cases}
   w_C^{\rm in} & \norm{x-z_C}<r_C,\\
   w_C^{\rm in}+w_C^{\rm out} & \norm{x-z_C}=r_C,\\
   w_C^{\rm out} & \norm{x-z_C}>r_C .
 \end{cases}
\]
For distinct rims, Lemma~\ref{lem:plane-candidates} gives a candidate
\(y\) reachable from \(x\) without crossing a rim or leaving one that
contains \(x\). Moving from a strict side onto its rim only adds the
other side's nonnegative weight. Thus \(w_C(y)\ge w_C(x)\) for every
\(C\), and some candidate maximizes the total.

\Needspace{6\baselineskip}
\begin{theorem}[Closed planar depth]\label{thm:planar}
Given \(m\ge1\) positive-radius circles in the plane, each with
independent nonnegative inside and outside weights, one can find a point
of maximum total weight among their closed disks and closed exteriors in
\(O(m^2)\) time and \(O(m)\) extra memory on the real RAM with square roots.
Coincident rims, tangencies and meetings of any number of circles are allowed.
\end{theorem}
The \hyperref[proof:planar]{proof begins in Section~\ref*{sec:plane-scoring},
p.~\pageref*{proof:planar}}. It shows why the search finds a point of
maximum depth, then sums the work over the levels.
Appendix~\ProofRef{sec:construction} proves the splitting rule it uses.
Coincident rims are first grouped by centre and squared radius in
\(O(m\log(m+1))\) time and \(O(m)\) extra words, summing each
group's two weights separately. For empty input return the origin at
depth zero. Unit site disks with weights \((1,0)\) give \(M(P)\).

\section{Unit-ball placement in three and higher dimensions}\label{sec:lift}
Any site \(p\) can be tried as an \emph{anchor}. Covering \(p\) allows
any centre in the ball \(\norm{c-p}\le1\). Using \(p\) as anchor means
placing it on the boundary of our ball: the centres are then on the
sphere \(\norm{c-p}=1\) about \(p\). By Lemma~\ref{lem:tight}, some
such sphere contains an optimal centre whenever the input is nonempty.
Higher dimensions use more anchors; coplanar sites need only the
planar solver (Corollary~\ref{cor:affine-hull}).

\subsection{Three dimensions: one anchor, one sphere}\label{sec:three}
With \(p\) fixed, write \(c=p+u\). The direction \(u\) lies on
\(\Sph^2=\{u\in\R^3:\norm u=1\}\); the superscript counts the two
surface dimensions. The anchor always counts because \(\norm{c-p}=1\).
Each other site \(q=p+w\) contributes on the closed patch
\(w\cdot u\ge\norm w^2/2\), its \emph{cap}~\eqref{eq:anchor}.

\ifIllustrated
The sites in Figure~\ref{fig:spheres} are \((3/4,0,0)\),
\((-9/20,\pm3/5,0)\) and \((0,0,\pm3/4)\). All are \(3/4\) from the
origin, so every centre within \(1/4\) of it covers all five. At the
origin none is tight; at \(c=(-1/4,0,0)\) all five still count and
\(p=(3/4,0,0)\) is tight.

\begin{figure}[!htb]
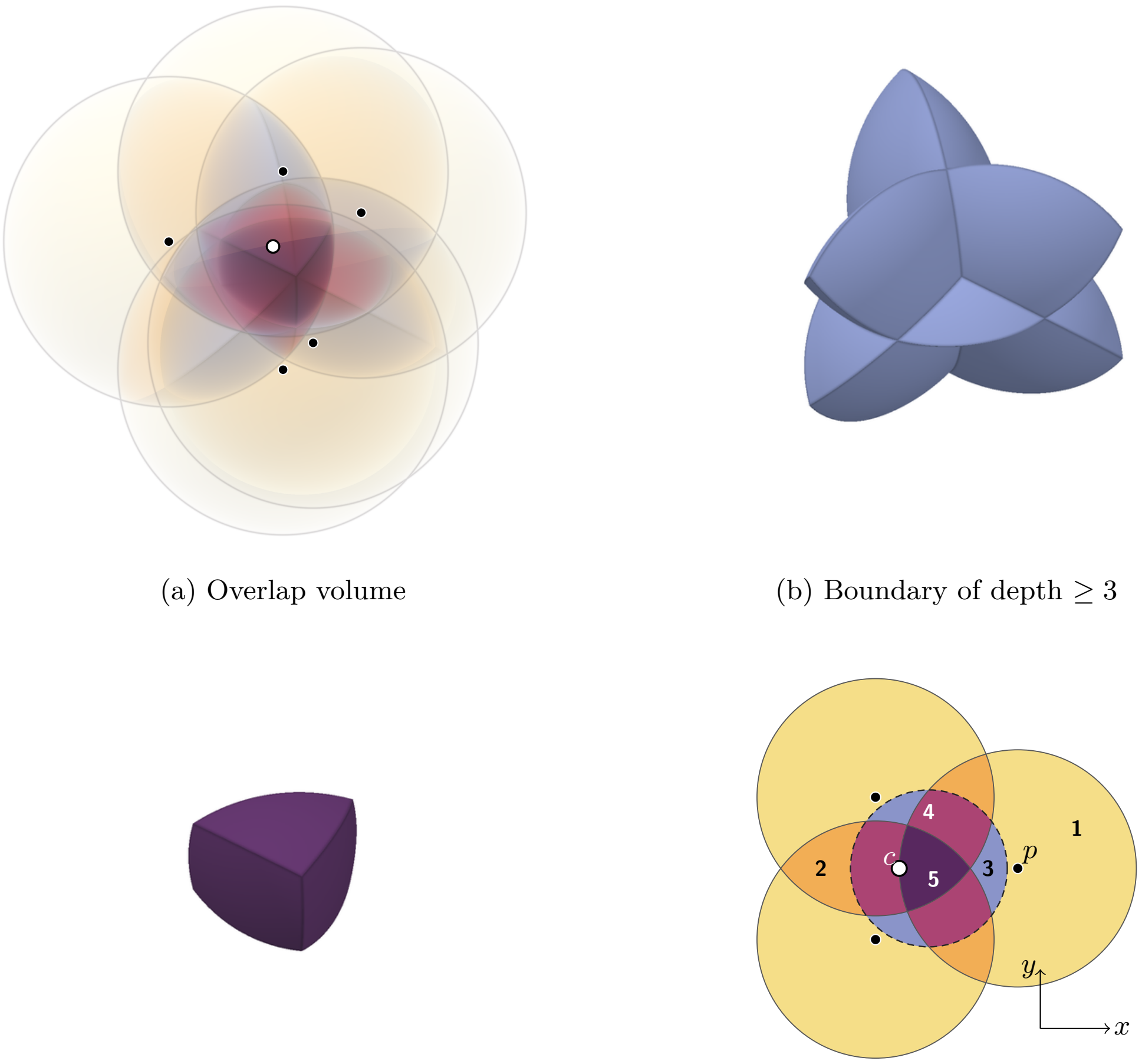

\centering
\begingroup
\centering\small
\tikzset{
  site/.style={fill=black,draw=white,line width=.38pt},
  witness/.style={fill=white,draw=black,line width=.65pt},
  ordinary rim/.style={draw=black!66,line width=.34pt},
  shared rim/.style={draw=black!85,line width=.44pt,dash pattern=on 2.2pt off 1.5pt},
  count/.style={inner sep=0pt,font=\sffamily\bfseries\scriptsize},
  ink1/.style={text=black},ink2/.style={text=black},
  ink3/.style={text=black},ink4/.style={text=white},ink5/.style={text=white},
}
\pgfmathsetmacro{\slicescalefour}{1.30*(\linewidth/(6.42in))}
\begin{tabular}{@{}c@{\hspace{.06231\linewidth}}c@{}}
  \begin{tikzpicture}[baseline=(im.center)]
    \node[anchor=south west,inner sep=0](im) at (0,0)
      {\includegraphics[width=.40966\linewidth]{figures/ball-depth/volume.png}};
    \begin{scope}[x={(im.south east)},y={(im.north west)}]
\filldraw[site] (0.54926505,0.38157643) circle[radius=1.55pt];
\filldraw[site] (0.62808913,0.59473886) circle[radius=1.55pt];
\filldraw[site] (0.31279281,0.54736943) circle[radius=1.55pt];
\filldraw[site] (0.50000000,0.66235490) circle[radius=1.55pt];
\filldraw[site] (0.50000000,0.33764510) circle[radius=1.55pt];
\filldraw[witness] (0.48357832,0.53947452) circle[radius=2pt];

    \end{scope}
  \end{tikzpicture}
  & \raisebox{-.5\height}{\includegraphics[width=.35826\linewidth]{figures/ball-depth/three.png}}\\[2pt]
  (a) Overlap volume & (b) Boundary of depth \(\ge3\)\\[.8em]
  \raisebox{-.5\height}{%
    \ifdim\linewidth<6in
      \includegraphics[width=.35826\linewidth,trim=0 80bp 0 80bp,clip]{figures/ball-depth/five.png}%
    \else
      \includegraphics[width=.35826\linewidth]{figures/ball-depth/five.png}%
    \fi}
  & \begin{tikzpicture}[scale=\slicescalefour,baseline=(current bounding box.center)]
\begin{scope}
\clip ({3/4},{0}) circle[radius={sqrt(1)}];
\fill[depth1] (-2,-2) rectangle (2,2);
\end{scope}
\begin{scope}
\clip ({-9/20},{3/5}) circle[radius={sqrt(1)}];
\fill[depth1] (-2,-2) rectangle (2,2);
\end{scope}
\begin{scope}
\clip ({-9/20},{-3/5}) circle[radius={sqrt(1)}];
\fill[depth1] (-2,-2) rectangle (2,2);
\end{scope}
\begin{scope}
\clip ({0},{0}) circle[radius={sqrt(7/16)}];
\fill[depth2] (-2,-2) rectangle (2,2);
\end{scope}
\begin{scope}
\clip ({3/4},{0}) circle[radius={sqrt(1)}];
\clip ({-9/20},{3/5}) circle[radius={sqrt(1)}];
\fill[depth2] (-2,-2) rectangle (2,2);
\end{scope}
\begin{scope}
\clip ({3/4},{0}) circle[radius={sqrt(1)}];
\clip ({-9/20},{-3/5}) circle[radius={sqrt(1)}];
\fill[depth2] (-2,-2) rectangle (2,2);
\end{scope}
\begin{scope}
\clip ({-9/20},{3/5}) circle[radius={sqrt(1)}];
\clip ({-9/20},{-3/5}) circle[radius={sqrt(1)}];
\fill[depth2] (-2,-2) rectangle (2,2);
\end{scope}
\begin{scope}
\clip ({3/4},{0}) circle[radius={sqrt(1)}];
\clip ({0},{0}) circle[radius={sqrt(7/16)}];
\fill[depth3] (-2,-2) rectangle (2,2);
\end{scope}
\begin{scope}
\clip ({-9/20},{3/5}) circle[radius={sqrt(1)}];
\clip ({0},{0}) circle[radius={sqrt(7/16)}];
\fill[depth3] (-2,-2) rectangle (2,2);
\end{scope}
\begin{scope}
\clip ({-9/20},{-3/5}) circle[radius={sqrt(1)}];
\clip ({0},{0}) circle[radius={sqrt(7/16)}];
\fill[depth3] (-2,-2) rectangle (2,2);
\end{scope}
\begin{scope}
\clip ({3/4},{0}) circle[radius={sqrt(1)}];
\clip ({-9/20},{3/5}) circle[radius={sqrt(1)}];
\clip ({-9/20},{-3/5}) circle[radius={sqrt(1)}];
\fill[depth3] (-2,-2) rectangle (2,2);
\end{scope}
\begin{scope}
\clip ({3/4},{0}) circle[radius={sqrt(1)}];
\clip ({-9/20},{3/5}) circle[radius={sqrt(1)}];
\clip ({0},{0}) circle[radius={sqrt(7/16)}];
\fill[depth4] (-2,-2) rectangle (2,2);
\end{scope}
\begin{scope}
\clip ({3/4},{0}) circle[radius={sqrt(1)}];
\clip ({-9/20},{-3/5}) circle[radius={sqrt(1)}];
\clip ({0},{0}) circle[radius={sqrt(7/16)}];
\fill[depth4] (-2,-2) rectangle (2,2);
\end{scope}
\begin{scope}
\clip ({-9/20},{3/5}) circle[radius={sqrt(1)}];
\clip ({-9/20},{-3/5}) circle[radius={sqrt(1)}];
\clip ({0},{0}) circle[radius={sqrt(7/16)}];
\fill[depth4] (-2,-2) rectangle (2,2);
\end{scope}
\begin{scope}
\clip ({3/4},{0}) circle[radius={sqrt(1)}];
\clip ({-9/20},{3/5}) circle[radius={sqrt(1)}];
\clip ({-9/20},{-3/5}) circle[radius={sqrt(1)}];
\clip ({0},{0}) circle[radius={sqrt(7/16)}];
\fill[depth5] (-2,-2) rectangle (2,2);
\end{scope}
\draw[ordinary rim] ({3/4},{0}) circle[radius={sqrt(1)}];
\draw[ordinary rim] ({-9/20},{3/5}) circle[radius={sqrt(1)}];
\draw[ordinary rim] ({-9/20},{-3/5}) circle[radius={sqrt(1)}];
\draw[shared rim] ({0},{0}) circle[radius={sqrt(7/16)}];
\filldraw[site] ({3/4},{0}) circle[radius=1.25pt];
\filldraw[site] ({-9/20},{3/5}) circle[radius=1.25pt];
\filldraw[site] ({-9/20},{-3/5}) circle[radius=1.25pt];
\node[count,ink1] at (1.25,.34) {1};
\node[count,ink2] at (-.91,0) {2};
\node[count,ink3] at (.50,0) {3};
\node[count,ink4] at (0,.48) {4};
\node[count,ink5] at (.04,-.09) {5};

      \filldraw[witness] (-.25,0) circle[radius=1.8pt];
      \node[above left,text=white,inner sep=1pt] at (-.25,0) {\(c\)};
      \node[above right,inner sep=1.4pt] at (.75,0) {\(p\)};
      \draw[->] (.94,-1.35) -- (1.53,-1.35) node[right,inner sep=1pt] {\(x\)};
      \draw[->] (.94,-1.35) -- (.94,-.85) node[left,inner sep=1pt] {\(y\)};
    \end{tikzpicture}\\[2pt]
  (c) Depth \(5\) (\(1.25\times\) scale of b) & (d) Slice \(z=0\)\\
\end{tabular}\par
\endgroup

\caption{Five unit balls viewed along \(-(4,-1,3.1)\). In (a) the dots
mark sites; the open circle marks an optimal centre \(c\). Shading
blends along each ray. For exact counts see the \(z=0\) slice (d).
The two off-plane balls share the dashed rim; \(c\) lies on the solid
rim about the anchor \(p\)}\label{fig:spheres}
\end{figure}
\fi

Write \(d=\norm{q-p}>0\) for the distance between the sites. On the
boundary of the cap, \(u\cdot(w/d)=d/2\). For \(d<2\) the plane is
\(d/2\) from the origin, so it cuts the unit sphere in a circle of
radius \(\sqrt{1-(d/2)^2}\), the \emph{rim}. The corresponding
centres are one unit from both sites, in their perpendicular bisector plane. At \(d=2\)
the only direction is \(u=(q-p)/2\), giving the
centre \(c=(p+q)/2\); for \(d>2\) no direction covers \(q\).

\LayoutFloatBarrier
\subsection{Calling the planar solver from a sphere}\label{sec:caps}
A general cap is the part of \(\Sph^2\) in a closed halfspace
\(a\cdot u\ge\beta\): one side of a plane, including the plane itself.
For \(a\ne0\) we call the cap \emph{regular} if \(|\beta|<\norm a\);
its rim is a circle of positive radius. At \(\beta=\norm a\) only the
\emph{tip} \(a/\norm a\) counts. A larger threshold gives an empty
cap, and \(\beta\le-\norm a\) gives a full cap. A zero normal gives a
constant test. Empty caps earn nothing, full caps add the same weight
everywhere, and a singleton changes only its tip's score.

\begin{figure}[!htb]
\centering
\begingroup
\input{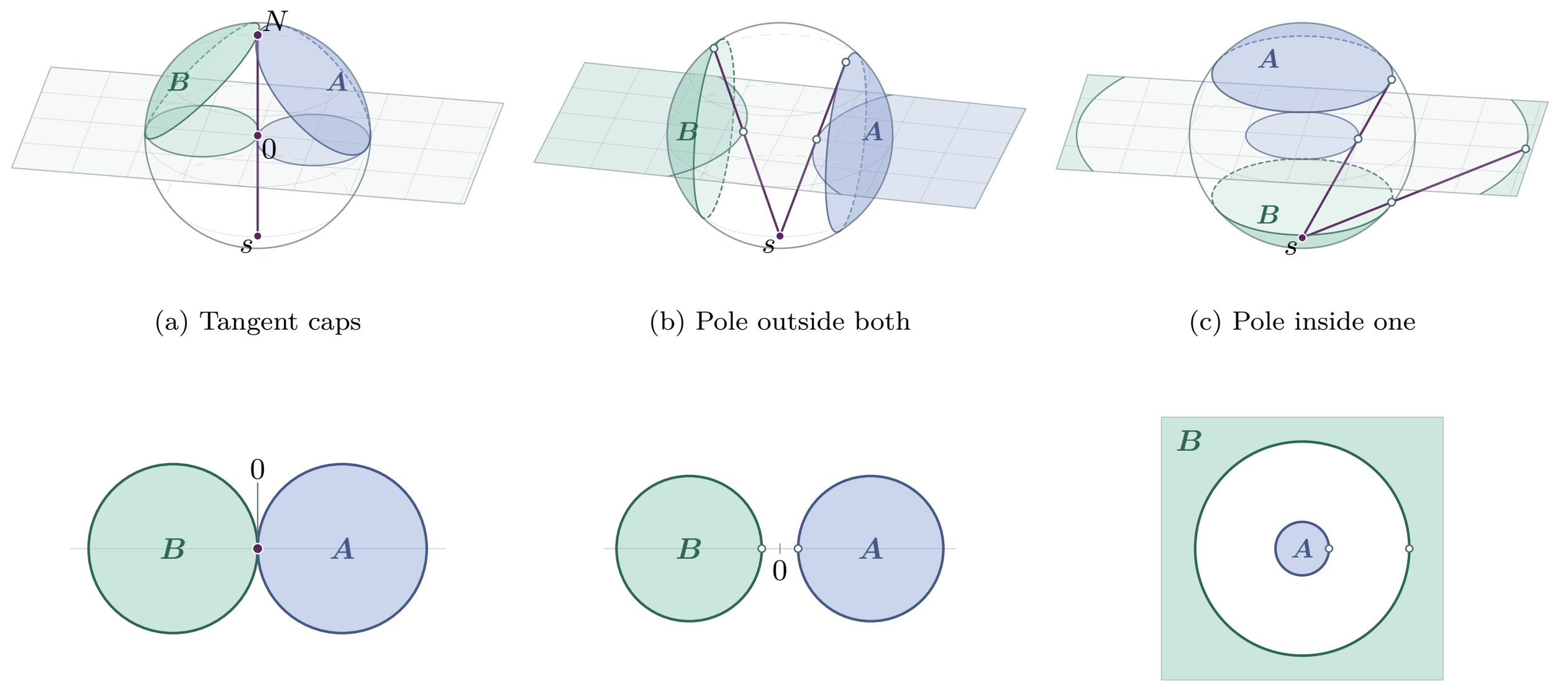}
\pgfmathsetmacro{\colC}{\linewidth/1mm/3}
\pgfmathsetmacro{\colA}{-\colC}\def\colB{0}
\pgfmathsetlengthmacro{\sphereunit}{.072*\linewidth}
\begin{tikzpicture}[x=1mm,y=1mm,font=\small,line join=round,line cap=round,
    every node/.style={inner sep=1pt},
    raymark/.style={fill=white,draw=glassInk,line width=.47pt},
    witness/.style={fill=glassRay,draw=white,line width=.4pt}]
  \path ({-1.5*\colC},45) -- ({1.5*\colC},45);
  \begin{scope}[shift={(\colA,65)},x=\sphereunit,y=\sphereunit]
    \glassE
    \node[text=capB,font=\footnotesize] at (-.70,.48) {\(\boldsymbol{B}\)};
    \node[text=capA,font=\footnotesize] at (.70,.48) {\(\boldsymbol{A}\)};
    \filldraw[witness] (eS) circle[radius=1.25pt] node[below left,text=black] {\(s\)};
    \filldraw[witness] (eN) circle[radius=1.4pt] node[above right,text=black] {\(N\)};
    \filldraw[witness] (eO) circle[radius=1.35pt] node[below right,text=black] {\(0\)};
  \end{scope}
  \begin{scope}[shift={(\colB,65)},x=\sphereunit,y=\sphereunit]
    \glassF
    \node[text=capB,font=\footnotesize] at (-.82,.04) {\(\boldsymbol{B}\)};
    \node[text=capA,font=\footnotesize] at (.82,.04) {\(\boldsymbol{A}\)};
    \filldraw[witness] (fS) circle[radius=1.25pt] node[below left,text=black] {\(s\)};
    \foreach \p in {fUPlus,fUMinus,fXPlus,fXMinus}
      \filldraw[raymark] (\p) circle[radius=1.1pt];
  \end{scope}
  \begin{scope}[shift={(\colC,65)},x=\sphereunit,y=\sphereunit]
    \glassG
    \node[text=capA,font=\footnotesize] at (-.30,.68) {\(\boldsymbol{A}\)};
    \node[text=capB,font=\footnotesize] at (-.30,-.70) {\(\boldsymbol{B}\)};
    \filldraw[witness] (gS) circle[radius=1.25pt] node[below left,text=black] {\(s\)};
    \foreach \p in {gUBlue,gURed,gXBlue,gXRed}
      \filldraw[raymark] (\p) circle[radius=1.1pt];
  \end{scope}
  \node[font=\footnotesize] at (\colA,45) {(a) Tangent caps};
  \node[font=\footnotesize] at (\colB,45) {(b) Pole outside both};
  \node[font=\footnotesize] at (\colC,45) {(c) Pole inside one};
  \begin{scope}[shift={(\colA,21)}]
    \draw[glassInk!28,line width=.3pt] (-20,0) -- (20,0);
    \filldraw[fill=capBFill,fill opacity=.52,draw=capB,line width=.78pt] (-9,0) circle[radius=9mm];
    \filldraw[fill=capAFill,fill opacity=.52,draw=capA,line width=.78pt] (9,0) circle[radius=9mm];
    \node[text=capB] at (-9,0) {\(\boldsymbol{B}\)};
    \node[text=capA] at (9,0) {\(\boldsymbol{A}\)};
    \draw[glassInk,line width=.37pt] (0,7) -- (0,1);
    \node[above] at (0,7) {\(0\)};
    \filldraw[witness] (0,0) circle[radius=1.55pt];
  \end{scope}
  \begin{scope}[shift={(\colB,21)},x=5.8mm,y=5.8mm]
    \draw[glassInk!32,line width=.31pt] (-3.22,0) -- (3.22,0);
    \filldraw[fill=capBFill,fill opacity=.52,draw=capB,line width=.78pt] ({-5/3},0) circle[radius={4/3}];
    \filldraw[fill=capAFill,fill opacity=.52,draw=capA,line width=.78pt] ({5/3},0) circle[radius={4/3}];
    \node[text=capB] at ({-5/3},0) {\(\boldsymbol{B}\)};
    \node[text=capA] at ({5/3},0) {\(\boldsymbol{A}\)};
    \foreach \p in {-1,1} \filldraw[raymark] ({\p/3},0) circle[radius=1.1pt];
    \draw[glassInk!75,line width=.35pt] (0,-.09) -- (0,.09);
    \node[below=1pt] at (0,-.09) {\(0\)};
  \end{scope}
  \begin{scope}[shift={(\colC,21)}]
    \fill[capBFill,fill opacity=.52,even odd rule] (-15,-14) rectangle (15,14) (0,0) circle[radius=11.4mm];
    \draw[glassInk!42,line width=.30pt] (-15,-14) rectangle (15,14);
    \draw[capB,line width=.78pt] (0,0) circle[radius=11.4mm];
    \filldraw[fill=capAFill,fill opacity=.52,draw=capA,line width=.75pt] (0,0) circle[radius=2.85mm];
    \node[text=capA,font=\footnotesize] at (0,0) {\(\boldsymbol{A}\)};
    \node[text=capB] at (-12,11.5) {\(\boldsymbol{B}\)};
    \filldraw[raymark] (2.85,0) circle[radius=1.05pt];
    \filldraw[raymark] (11.4,0) circle[radius=1.05pt];
  \end{scope}
  \node[font=\footnotesize] at (\colA,2) {(d) Tangent disks};
  \node[font=\footnotesize] at (\colB,2) {(e) Disjoint disks};
  \node[font=\footnotesize] at (\colC,2) {(f) Disk and exterior};
\end{tikzpicture}%
\endgroup%
\caption{Each column pairs two spherical caps with their images in the
equatorial plane below. The rays project from the south pole \(s\);
the planar views use different scales. Left: the caps touch
at north \(N\); their disks touch at the origin. Middle: the opposite
caps \(u_1\ge3/5\) and \(u_1\le-3/5\) both miss \(s\). Their disks have
centres \((\pm5/3,0)\) and radius \(4/3\). Right: rotate the same pair
to north and south, keeping \(s\) fixed. The north cap maps to the disk
\(\norm{x}\le1/2\); the south cap, omitting \(s\), maps to the exterior
\(\norm{x}\ge2\). The exterior extends beyond the frame. The open
ring between the two circles belongs to neither image}\label{fig:pole-sides}
\end{figure}

The \emph{pole} of stereographic projection is its viewpoint. In
coordinates with the pole at \(s=(0,0,-1)\), the ray from \(s\)
through \(u\ne s\) meets the equatorial plane at
\(x=\pi_s(u)=(u_1,u_2)/(1+u_3)\). Conversely, each \(x=(X,Y)\)
gives exactly one direction
\(\pi_s^{-1}(x)=(2X,2Y,1-\lambda)/(1+\lambda)\), where
\(\lambda=X^2+Y^2\). In the placement problem the corresponding centre
is \(p+\pi_s^{-1}(x)\).
\ifIllustrated
Embed the planar walk by \((a,b)\mapsto(a,0,b)\). The rim of cap
\(A\), where \(u_1+u_3=1\), looks elliptical from above
(Figure~\ref{fig:pole-sides}a). Its stereographic image is circular:
\(\|u\|=1\) gives \(X^2+Y^2=(1-u_3)/(1+u_3)\), so off the pole
\[
 u_1\ge1-u_3\iff X\ge X^2+Y^2
                \iff (X-\tfrac12)^2+Y^2\le\tfrac14 .
\]
\else
Figure~\ref{fig:pole-sides} shows what happens as a cap moves over the pole.
\fi

\ifIllustrated
Reflection across \(u_1=0\) gives cap \(B\) and mirrors the disk
across \(X=0\). Lengths change; membership, including equality, does not.
\fi
To decide which side to keep, substitute the inverse map in
\(a\cdot u\ge\beta\) and multiply by \(1+\lambda>0\):
\[
   (a_3+\beta)(X^2+Y^2)\ \le\ 2a_1X+2a_2Y+a_3-\beta .
\]
Completing the square gives a closed disk when \(a_3+\beta>0\) and
a closed exterior when \(a_3+\beta<0\). The sign has a geometric
meaning: \(a\cdot s=-a_3\), so the exterior occurs exactly when
the pole is strictly inside the cap. If \(a_3+\beta=0\), the rim
passes through the pole and projects to a line. We avoid lines by
choosing the pole off all regular rims: Appendix~\ProofRef{sec:chart-details}
finds a point off their projected circles and lines in two scans.
It uses the reflected chart, with north in place of south.
We score the pole directly because no planar point represents it.

Opposite closed sides of the same rim give one disk and its exterior.
With unit weights their circle record is \((1,1)\): a rim point
belongs to both caps and earns two.

No choice of pole always avoids exteriors. For example, fix the anchor at
the origin and put six other sites at the positive and negative coordinate
unit vectors. Their caps are \(\pm u_i\ge 1/2\) for \(i=1,2,3\).
Every unit vector has some \(|u_i|\ge1/\sqrt3>1/2\), so every pole lies
strictly inside at least one cap. Thus a solver for disks alone would
miss a required exterior.\footnote{Parra Bustos, Chin and
Suter~\citep[Fig.~3 and Section~3.3]{bustos2014fast} use the same images
to test patch intersection. Here the objective is maximum total weight.}

Cap \(K_i\) keeps its input index \(i\) as its \emph{label}, even if
caps coincide. A 3D anchor gives \(N-1\) other sites' caps, or \(N\)
records including the anchor as a full cap.

\ifdefined\ArxivPreprint\Needspace{8\baselineskip}\fi
\begin{theorem}[Spherical and planar depth]\label{thm:sphere}
Given \(m\ge1\) labelled closed halfspace caps on a two-sphere, with
nonnegative weights \(w_i\), we can return a maximizing point, its total weight
and the indices of all caps containing it. This takes at most two calls
to the planar problem of Theorem~\ref{thm:planar}, each on at most \(2m\)
positive-radius circles, plus \(O(m\log(2+m))\) time and \(O(m)\) extra
words. Conversely, \(m\) planar circle records reduce to one weighted
cap-depth call on at most \(2m\) caps, plus \(O(m)\) time and extra words.
Each call need only return an attaining point. Empty inputs take constant time.
\end{theorem}
\begin{proof}[Proof idea]
The first planar call sees exactly the regular caps:
\[
 \text{planar score at }x=R(x):=
 \sum_{\substack{i\ \mathrm{regular}\\ \pi_s^{-1}(x)\in K_i}} w_i .
\]
Its winner maximizes \(R\); away from the pole and the tips, only the
constant full-cap weight is missing. For the second call keep these
circles. At each distinct projected tip of total singleton weight \(g\),
add two disks touching there, with different pairs disjoint. Each
disk pays \(A+g/2\) inside or on its rim and zero outside, where \(A\)
exceeds the sum of all original weights. Only a tip earns two copies
of \(A\), and every tip scores \(2A+R(x)+g\). Thus the second winner
is a tip with the greatest original score among tips.
With \(r\) regular caps and \(t\) distinct projected tips there are at
most \(r+2t\le2m\) circles. Compare the lifted winners and the pole
on the original caps. Appendix~\ProofRef{app:cap-transfer} proves
the score gap and the converse: one cap for each closed side of a
planar circle. The converse establishes equivalence; placement uses
only the forward direction.
\end{proof}

For the quadratic solver of Theorem~\ref{thm:planar}, we can test
at most \(m\) tips against \(m\) caps each instead of making the second
call. These scans fit the quadratic budget but would hide a saving
from a faster planar solver. For unit weights the procedure is as follows.
\ifIllustrated
\begin{algorithm}[htb]
\caption{\SphereDepth: quadratic search for unweighted caps}\label{alg:sphere}
\small
\begin{algorithmic}
\Require Caps \(a_i\cdot u\ge\beta_i\) on the unit two-sphere
\State Test any sphere point; keep it and its full score.
\State Classify caps as empty, full, singleton or regular.
\State Test every singleton point against all original caps; keep the best so far.
\If{there are regular caps}
  \State Choose a pole off their rims; test the pole against all caps.
  \State Project the regular caps; form one record per distinct circle.
  \State In each record count disk images and exterior images separately.
  \State \((x,d)\gets\Call{PlaneDepth}{\text{grouped circles}}\)
  \State Test the inverse image of \(x\) against all original caps.
\EndIf
\State \Return the best tested sphere point and its full score.
\end{algorithmic}
\end{algorithm}

\LayoutFloatBarrier
\else
Test an arbitrary sphere point and every tip. If regular caps remain,
test the pole, group coincident image circles by summing their two
sides separately, and test the inverse image of the planar winner.
Keep the best score on the original caps.
\fi
If an optimum is a tip or the omitted pole, its direct test suffices.
At any other optimum the score is \(F+R(x)\), where \(F\) counts full
caps. The planar winner maximizes \(R\); after lifting it still earns
\(F\), and any tip there can only add weight. If no regular caps remain,
every non-tip earns just \(F\). The scans, grouping and at most one
quadratic call take \(O((m+1)^2)\) time and \(O(m+1)\) extra words.

\ifIllustrated
Return to Example~\ref{ex:planar-walk}, now with anchor \(p=(0,0,0)\)
and sites \(q_\pm=(\pm1,0,1)\). Adding the two cap tests
\(u_3+u_1\ge1\) and \(u_3-u_1\ge1\) gives \(u_3\ge1\): only north
covers both. In the south chart, disk \(A\) and its mirror touch only
at \(x=(0,0)\) (Figure~\ref{fig:pole-sides}(d)). That planar winner lifts
to \(c=(0,0,1)\), the midpoint of \(q_+q_-\). The two disk labels
and the anchor give all three indices.
\fi

\ifIllustrated\FloatBarrier\fi
\subsection{When an optimal ball misses few sites}\label{sec:missed-anchors}
Trying every site as the anchor costs \(N\) sphere searches in 3D.
We can shorten that list by first finding a ball that misses few sites
whenever the optimum does. The sites to try as anchors are the ones
the first ball \emph{misses}.

\begin{lemma}[A short list of anchors]\label{lem:missed-first}
Let \(P\) consist of \(N\ge1\) distinct sites in \(\R^3\), and put \(k=N-M(P)\).
Without knowing \(k\), we can find a reference centre \(c_0\) whose unit
ball misses at most \(4k\) sites, in \(O((k+1)N)\) time and \(O(N)\)
extra words. Either \(c_0\) is optimal, or a site missed by \(c_0\) lies
on the boundary of some optimal ball.
\end{lemma}
\begin{proof}
First see why the missed sites suffice, for \emph{any} reference centre.
If \(c_0\) is not optimal, imagine moving an optimal centre \(c^*\)
along the segment towards \(c_0\). A site covered at both ends stays
covered: its ball of possible centres is convex. Now watch the sites
covered at \(c^*\) but missed at \(c_0\). There is at least one.
Each starts at distance at most one and ends at distance greater than
one. Stop the first time any of them is exactly one unit away, even
if that is at \(c^*\) itself. None has yet left the closed ball, so
the centre is still optimal and one of the missed sites is tight.
The motion proves the list suffices; the algorithm does not need \(c^*\).

To obtain a short list, start with all the sites. If no unit ball
covers them together, their closed balls of possible centres have empty
intersection. By Helly's theorem, a finite family of convex sets in
\(\R^3\) with empty intersection has a subfamily of at most four with
empty intersection. Set those sites aside and repeat.
On \(m\) remaining sites, finding such a group or finding a centre
that covers the remainder takes \(O(m+1)\) time
(Appendix~\ProofRef{app:missed-anchors}). Use the latter centre as
\(c_0\); if no sites remain, any centre will do.

Fix one optimal ball. It misses at least one site in each group set
aside. Those groups are disjoint, so there are at most \(k\) of them.
All misses at \(c_0\) come from those groups, hence there are at most
\(4k\). The construction makes at most \(k+1\) passes and reuses the
same linear workspace.
\end{proof}

Write \(b\) for the actual number of misses at \(c_0\). If \(b=0\), we
are done. Otherwise keep \(c_0\) as a candidate. If it is not optimal,
Lemma~\ref{lem:missed-first} puts an optimal centre on the sphere of
one of its \(b\) misses. Run \SphereDepth{} on each of those spheres,
counting \emph{all} original sites. This gives
\(O((k+1)N^2)\) time by Theorem~\ref{thm:sphere} and Lemma~\ref{lem:missed-first}.
The next theorem gives a cheaper option when \(b\) is small.

\begin{theorem}[3D placement with few missed sites]\label{thm:missed}
For \(N\ge1\) distinct sites in \(\R^3\), let \(k=N-M(P)\). A deterministic
algorithm returns an optimal closed unit ball, its count and all covered
indices, without being given \(k\), in
\[
    O\!\left(N(k+1)\min\{N,(k+1)\log(2+N)\}\right)
\]
time and \(O(N)\) extra words.
\end{theorem}
On each anchor's sphere, the second option again sets aside groups of
at most four sites that cannot be covered together there. Finding
\(b+1\) disjoint groups forces \(b+1\) distinct misses, worse than
\(c_0\). Otherwise a reference centre covers the remainder and can
miss only the sites set aside, at most \(4b\). An arc on the sphere
plays the role of the segment in Lemma~\ref{lem:missed-first}
(Appendix~\ProofRef{app:missed-anchors} handles opposite directions).
Only the reference's misses need be tried as a \emph{second} anchor:
at most \(b\cdot4b\) pairs in total. Two tight sites leave a circle of possible centres,
a single point, or none. On a circle one angular sweep suffices.

Before searching any sphere, use \(b\) to choose between \SphereDepth{}
at \(O(N^2)\) per sphere and the pair searches at
\(O((b+1)N\log(2+N))\) per sphere. Thus if the optimum misses
\(O(\sqrt N)\) sites, the total time is \(O(N^2\log(2+N))\).
Appendix~\ProofRef{app:missed-anchors} proves the second-anchor rule
and gives the full cost.

\subsection{Fixing more than one boundary site}\label{sec:high}
A \emph{section} is the set of centres exactly one unit from every
chosen site. The \emph{cap} for another site keeps those centres on
the spherical section that are \emph{at most} one unit from it.
For the sites \((\pm a,0,0,0)\) in \(\R^4\), \(a>0\), keeping both tight
requires \(c_1=0\) and \(c_2^2+c_3^2+c_4^2=1-a^2\).
The section is a two-sphere for \(a<1\), a point for \(a=1\), and empty for \(a>1\).

\ifIllustrated
For \(a=3/5\), Figure~\ref{fig:section-equator} compares the same
two tight sites in three and four dimensions. The circle of centres
in three dimensions becomes the equator \(c_4=0\) in four.

\begin{walkthroughfigure}
\centering
\includegraphics[width=\ProofPlateWidth]{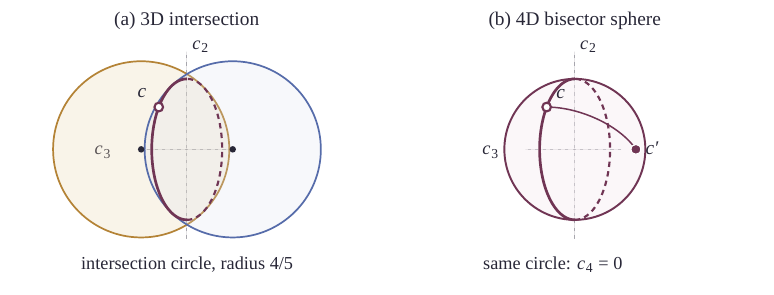}
\caption{Keeping both sites at distance one. (a) The dots are the sites
\(\pm(3/5)e_1\); their unit spheres intersect in a circle of possible centres
of radius \(4/5\). (b) In the three coordinates left after imposing \(c_1=0\),
that circle has the same scale and orientation. The marked centre \(c\)
is still feasible; moving along the indicated meridian reaches \(c'\)
with \(c_4=4/5\) while both sites remain tight}\label{fig:section-equator}
\end{walkthroughfigure}
\fi

For a nonempty set \(I\) of proposed tight sites, write its section as
\(S_I=\{c:\norm{c-p}=1\text{ for all }p\in I\}\).
We choose \emph{affinely independent} sites: from one \(p\in I\), the
vectors \(q-p\) to the other sites must be linearly independent.
Equivalently, they are the vertices of a simplex of dimension \(|I|-1\);
three such sites do not lie on a line.
Subtracting the squared-distance equation for \(p\)
from the others then leaves
\(|I|-1\) independent linear constraints. In the remaining
\(D-|I|+1\) coordinates the sphere equation cuts out an empty set,
a point, or a sphere of dimension \(D-|I|\). At \(|I|=D-2\)
a positive-radius section is a two-sphere. Use unit-sphere coordinates
\(z\) on this section. Write \(L_I(z)\) for the corresponding centre
in the input coordinates in \(\R^D\): the map scales, embeds and translates the unit
sphere. Appendix~\ProofRef{sec:section-coordinates} gives \(L_I\) and
the cap inequality for each site. In 3D, Section~\ref{sec:missed-anchors}
replaces the loop over all anchors in Algorithm~\ref{alg:place}.

\begin{algorithm}[htb]
\caption{\PlaceBall: return an attaining centre, count and indices}\label{alg:place}
\small
\begin{algorithmic}
\Require Distinct sites \(P\subset\R^D\), \(D\ge2\)
\If{\(P\) is empty} \State \Return \((0,0,\varnothing)\), with centre the origin. \EndIf
\If{\(D=2\)}
  \State \((c,d)\gets\Call{PlaneDepth}{\text{unit site circles with weights }(1,0)}\)
\Else
  \State Set \(c\) to any site and record its coverage count.
  \ForAll{affinely independent \(I\subseteq P\) with \(1\le |I|\le D-2\)}
    \State Form \(S_I\); skip it if empty.
    \If{\(S_I\) is a point} \State Let \(y\) be that point.
    \ElsIf{\(S_I\) is a two-sphere}
      \State Form the site caps in unit coordinates on \(S_I\).
      \State \((z,d)\gets\Call{SphereDepth}{\text{these caps}}\);\quad \(y\gets L_I(z)\).
    \Else \State Let \(y\) be one fixed representative of \(S_I\).
    \EndIf
    \State Score \(y\) on \(P\); if its count improves the best, keep \(y\) as \(c\) and store its count.
  \EndFor
\EndIf
\State \(L\gets\{i:\norm{c-p_i}\le1\}\);\quad \Return \((c,|L|,L)\).
\end{algorithmic}
\end{algorithm}
\LayoutFloatBarrier

\begin{lemma}[Sufficient sections]\label{lem:sections}
For finite nonempty \(P\subset\R^D\), \(D\ge3\), the best candidate
over the sections tested by Algorithm~\ref{alg:place} attains \(M(P)\).
Only the proposed tight sites in \(I\) must be affinely independent.
Other sites may lie on the same ball boundary, and the full input may
lie in a lower-dimensional affine subspace. The order of \(I\) does not matter.
\end{lemma}
\noindent\textit{Proof sketch.}
\ifdefined\ArxivPreprint
Suppose a section of dimension greater than two contains an optimum
but its representative is not optimal. The centres retaining all sites
covered by that optimum form a nonempty proper closed part of the sphere.
On its relative boundary some site's cap is tight. A site already in
\(\operatorname{aff}(I)\) cannot supply that boundary: its coverage test
is constant on the section, since its difference from \(p\in I\) is a linear
combination of the fixed-site differences in~\eqref{eq:anchor}.
Adding a site that supplies the boundary preserves an optimum on a smaller
section. At \(|I|=D-2\) it is a point or a two-sphere; both are searched
(Appendix~\ProofRef{sec:section-coordinates}).
\else
On a sphere of dimension greater than two, the representative need
not maximize depth. It suffices that either it is optimal or a
lower-dimensional section still contains an optimum. To see why,
keep all sites covered by a winning centre on the sphere. If the
representative misses one, the portion covering
them all is a nonempty proper closed subset of the section. At its
boundary \emph{within the section} another site becomes tight. Why must it enlarge the
affine span? On \(S_I\), every \((r-p)\cdot(c-p)\), \(r\in I\), is
fixed by~\eqref{eq:anchor}. If \(q-p=\sum_{r\in I}\alpha_r(r-p)\), then
\((q-p)\cdot(c-p)\) is the same linear combination of those fixed values.
Thus the coverage test for any \(q\in\operatorname{aff}(I)\) is constant
on \(S_I\). Such a site cannot supply a relative
boundary. Any site \(q\) supplying it must therefore lie outside
\(\operatorname{aff}(I)\); the algorithm enumerates \(I\cup\{q\}\)
and its smaller section. Appendix~\ProofRef{sec:section-coordinates}
gives the induction, including point sections.
\fi

\phantomsection\label{proof:main}
\begin{proof}[Completing the proof of Theorem~\ref{thm:main}]
In the plane the result is Theorem~\ref{thm:planar} with inside
weights one and outside weights zero. In \(D\ge3\), Lemma~\ref{lem:sections}
ensures that at least one tested candidate attains \(M(P)\).
Enumerate combinations of the indices and test affine independence;
there are \(O(N^{D-2})\) combinations. For fixed \(D\), forming the
section takes constant time. Scoring a representative or forming all
\(N\) caps takes \(O(N)\). The costliest case uses
Theorems~\ref{thm:sphere} and~\ref{thm:planar}, hence \(O(N^2)\) time per section and
\(O(N^{D-2}N^2)=O(N^D)\) overall. In contrast, continuing to at most
\(D-1\) tight sites and sorting events on each remaining circle uses
\(O(N^{D-1})\) sections and up to \(O(N\log N)\) time per section.
We store only the current section, the data for its search, and the
best centre and score so far. This takes \(O(N)\) extra memory. The
final scan returns the covered indices.
\end{proof}

\paragraph{What a faster planar solver would buy.}
Let \(T(m)\) and \(S(m)\) bound the time and extra words needed to
return a maximizing point for \emph{at most} \(m\) weighted planar
circles as in Theorem~\ref{thm:planar}. For fixed \(D\ge3\) and
\(N\ge1\), each of the \(O_D(N^{D-2})\) sections needs at most two
calls costing \(2T(2N)\) in total, plus \(O(N\log(2+N))\) for grouping
and scans (Theorem~\ref{thm:sphere}). The calls are sequential, so we get
\[
 O_D\!\left(N^{D-2}\bigl(T(2N)+N\log(2+N)\bigr)\right)
 \ \text{time},\qquad
 O_D\!\left(N+S(2N)\right)\ \text{extra words}.
\]
Thus a linear-workspace planar
solver saving a logarithmic factor would save the same factor for
unit-ball placement in every fixed dimension. This is an implication,
not a faster worst-case bound for the present algorithm.
\OpeningSpace
\subsection{When the sites use fewer dimensions}\label{sec:affine-hull}
If every 3D site has last coordinate zero, solve the planar problem
on the first two coordinates and append a zero to the winning centre.
A tilted plane works just as well once we choose coordinates in it.

\begin{corollary}[Projection onto the affine hull]\label{cor:affine-hull}
Let \(H\) be the smallest affine subspace containing the \(N\ge1\)
sites in \(\R^D\). Some optimal centre lies in \(H\). For fixed
\(r=\dim H\ge2\), we can return an optimal centre, its count and all
covered indices in \(O_r(ND+N^r)\) time and \(O_r(N+D)\) extra words.
In particular, coplanar sites take \(O(ND+N^2)\) time even when
the plane is unknown and \(D\) is part of the input.
\end{corollary}
\begin{proof}
Project any centre \(c\) orthogonally onto \(H\), obtaining \(h\).
For every site \(p\), \(p-h\) lies along \(H\) and \(c-h\) is
perpendicular to it. Hence
\[
   \norm{p-c}^2=\norm{p-h}^2+\norm{c-h}^2.
\]
Moving the centre to \(h\) keeps every site it covered, including
boundary sites. Conversely, a centre chosen in \(H\) is already a
legal centre in \(\R^D\). The two maximum counts therefore agree.

Take one site as origin. For each difference from it to another site,
subtract its components along the basis directions found so far.
If the remainder is nonzero, normalize it and add it to the basis.
A second scan gives every site's coordinates in this orthonormal
basis. The two scans take \(O(ND(r+1))\) time.
Apply Theorem~\ref{thm:main} in \(r\) dimensions and use the basis
and origin to return its centre to the original coordinates. This
preserves distances and hence the covered indices. We store \(r\)
basis vectors and \(r\) coordinates per site, in addition to the
solver's workspace. On a line, a centre at \(t\) covers exactly the
sites in \([t-1,t+1]\). Sort the sites and slide
such an interval to cover the most in \(O(N\log(2+N))\) further time;
its midpoint is a winning centre.
At rank zero the common site suffices.
\end{proof}

\paragraph{What if the plane does not contain the sites?}
Orthogonal projection cannot increase the distance from a site to a
centre when both are projected. The projected optimum, counting coincident images
with multiplicity, is therefore an upper bound. If a ball in the
original space attains that count, it is optimal.

The bound can be strict for \emph{every} plane. Take the regular
tetrahedron with vertices \((3/5)(\pm1,\pm1,\pm1)\) having an even
number of minus signs. Each face vertex is at squared distance
\(24/25\) from that face's centroid. No unit ball covers all four
vertices: their mean squared distance from \(c\) is \(27/25+\norm{c}^2\).
Project a face centroid along with its three vertices. Distances can
only decrease, so the projected centroid is in all three corresponding
unit disks. This holds for each triple. By Helly's
theorem for convex sets in the plane, a common point for each triple
implies a common point for all four.
Every planar answer is thus four, although the 3D answer is three.

\subsection{When sites have weights}\label{sec:weighted-sites}
\begin{corollary}[Nonnegative site weights]\label{cor:weighted-sites}
Give each of the \(N\ge1\) distinct sites in fixed dimension \(D\ge2\)
a nonnegative real weight. In deterministic \(O_D(N^D)\) time and
\(O_D(N)\) extra words one can return a closed unit ball of maximum
total weight, that weight, and all covered site indices.
\end{corollary}
\begin{proof}
Retaining every site covered by an optimum cannot lower its weight.
Thus the tight-site and section arguments (Lemmas~\ref{lem:tight}
and~\ref{lem:sections}) still apply; if all weights vanish, start at any
site. A planar site of weight \(w\) gives a unit disk with side weights
\((w,0)\). On a two-sphere each cap inherits its site's weight; the
anchors contribute everywhere. Theorem~\ref{thm:sphere} gives the same
cost per section. A final scan includes covered indices of weight zero.
\end{proof}

The bound can also be attained by a \emph{weight-oblivious} algorithm:
the coordinates fix the sequence of arithmetic operations on the
weights; score comparisons may still select a winner. Use the full
section search. Omit the pruning test of Section~\ref{sec:plane-scoring},
since pruning by the best score would change which sums are evaluated.
Score fixed
representatives, poles and singleton tips directly. If regular caps
remain, add the full-cap weight to both the planar root offset and the
initial corner score; otherwise the direct tests suffice.
Compare candidate scores as computed, without resumming
weights at a selected winner. At a singleton tip of total weight \(g\)
the regular search omits \(g\), but the direct tip test includes it;
at other finite points the score is already exact. Thus a maximum
computed score attains the true optimum. The resulting time matches
the \(\Omega(N^D)\) lower bound of de Figueiredo
and da Fonseca~\citep[Theorem~2.2]{defigueiredo2009enclosing} for this
restricted class of weighted algorithms.

\section{The PlaneDepth algorithm for disks and exteriors}\label{sec:depth}
For planar unit-disk placement, give
Algorithm~\ref{alg:plane} one unit circle about each site. Set its
inside and outside weights to \((1,0)\): a centre on or inside the
rim counts that site once, and a centre outside does not count it.
The returned point is the centre of the disk to place. A scan of the
original sites gives the covered indices.

The same search allows arbitrary positive radii and nonnegative
weights. Group coincident rims by adding their inside weights and
their outside weights separately. Write \(\mathcal C\) for the
resulting family of \(m\) distinct circles.
A point earns the inside weight in the disk's interior, the outside
weight beyond the rim, and both on the rim. \ifIllustrated Even with ordinary unit disks,
a single contact can beat all the open regions around it
(Figure~\ref{fig:open-child}).
\else A tangency can beat both adjacent open cells.\fi

One could score the \(O(m^2)\) candidates of
Lemma~\ref{lem:plane-candidates} against all \(m\) circles in
\(O(m^3)\) time. Sweeping whole rims is faster: test the origin once,
then on each rim score its leftmost and rightmost points and sweep
the two open semicircles using Lemma~\ref{lem:edge}. This covers all
candidates, but sorting contacts afresh on every rim takes
\(O(m^2\log(m+1))\) time and \(O(m)\) extra memory.

\PlaneDepth{} searches one region at a time. When it cuts a region
into open pieces, it scores the new edges and vertices before
shortening the circle list. Once the list is short, the algorithm
tests the finite candidates directly.

\subsection{From one rectangle to a tree of cells}\label{sec:cell-tree}
\ifIllustrated
\noindent\begin{minipage}{\linewidth}
Enclose all the disks strictly inside an axis-aligned rectangle.
On its boundary and everywhere outside it, the depth is the sum of
the outside weights. Its open interior \(W\) is the first \emph{cell}:
one connected region to search. When we split a cell, we draw circle
arcs and vertical segments inside it. The cell we split is the
\emph{parent}. Its \emph{children} are the open pieces left by the
cuts: two points are in the same child if they can be joined by a
path inside the parent without touching a cut. The new edges and
vertices belong to no child; the parent scores them separately.
A child can itself be split. Joining each parent to its children gives
a tree rooted at \(W\), the \emph{cutting}. A cell we finish without
another split is a \emph{leaf}.

\begin{figure}[H]
\centering
\begingroup%
\input{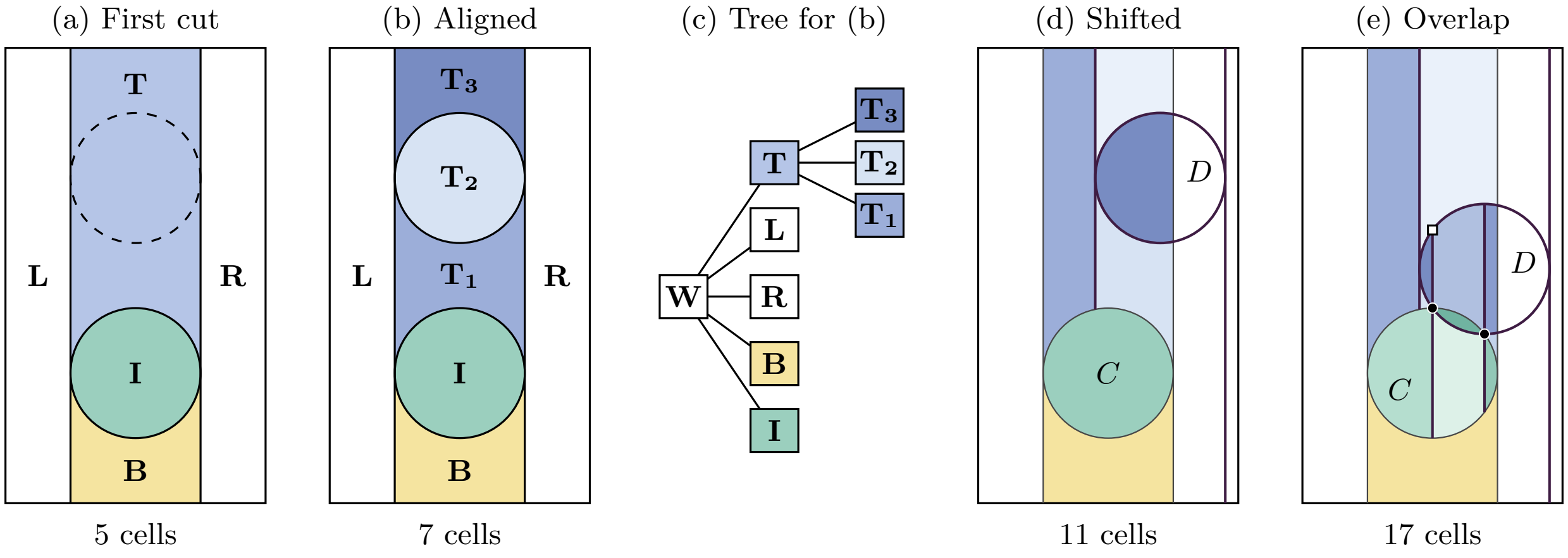}
\unskip%
\ifdim\linewidth<15cm
  \def\mapunit{.58cm}%
\else
  \def\mapunit{.68cm}%
\fi
\pgfmathsetlengthmacro{\panelstep}{(\linewidth-4*\mapunit-.4cm)/4}%
\begin{tikzpicture}[font=\small,oldcut/.style={semithick},
  newcut/.style={draw=figTrace!70!black,line width=.76pt}]
  \foreach \n/\variant in {0/a,1/b,3/d,4/e}{
    \pgfmathsetlengthmacro{\panelx}{\n*\panelstep}
    \begin{scope}[xshift=\panelx,x=\mapunit,y=\mapunit]
      \cellPaint{\variant}
      \ifnum\n>2
        \tikzset{oldcut/.style={draw=black!72,line width=.42pt}}
      \fi
      \draw[semithick] (-2,-2) rectangle (2,5);
      \foreach \x in {-1,1}
        \draw[oldcut] (\x,-2) -- (\x,5);
      \draw[oldcut] (0,0) circle[radius=1];
      \if\variant a
        \draw[semithick,dashed] (0,3) circle[radius=1];
        \node at (0,4.45) {\(\mathbf T\)};
      \fi
      \if\variant b
        \draw[semithick] (0,3) circle[radius=1];
        \node at (0,1.5) {\(\mathbf{T_1}\)};
        \node at (0,3) {\(\mathbf{T_2}\)};
        \node at (0,4.5) {\(\mathbf{T_3}\)};
      \fi
      \ifnum\n<2
        \node at (-1.5,1.5) {\(\mathbf L\)};
        \node at (1.5,1.5) {\(\mathbf R\)};
        \node at (0,-1.5) {\(\mathbf B\)};
        \node at (0,0) {\(\mathbf I\)};
      \fi
      \if\variant d
        \draw[newcut] (.8,3) circle[radius=1];
        \draw[newcut] (-.2,{sqrt(24/25)}) -- (-.2,5)
          (1.8,-2) -- (1.8,5);
        \node at (0,0) {\(C\)};
        \node at (1.4,3.1) {\(D\)};
      \fi
      \if\variant e
        \draw[newcut] (.8,1.6) circle[radius=1];
        \draw[newcut] (-.2,{sqrt(24/25)}) -- (-.2,5)
          (1.8,-2) -- (1.8,5)
          (0,-1) -- (0,1) (0,1) -- (0,2.2)
          (.8,-.6) -- (.8,.6) (.8,.6) -- (.8,2.6);
        \node[inner sep=1pt] at (-.5,-.25) {\(C\)};
        \node[inner sep=1pt] at (1.4,1.7) {\(D\)};
        \filldraw[fill=black,draw=white,line width=.36pt]
          (0,1) circle[radius=1.5pt] (.8,.6) circle[radius=1.5pt];
        \draw[fill=white,semithick] (0,2.2) ++(-1.35pt,-1.35pt)
          rectangle ++(2.7pt,2.7pt);
      \fi
    \end{scope}
  }
  \node at (0,{5*\mapunit+8pt}) {(a) First cut};
  \node at (\panelstep,{5*\mapunit+8pt}) {(b) Aligned};
  \node at ({2*\panelstep},{5*\mapunit+8pt}) {(c) Tree for (b)};
  \node at ({3*\panelstep},{5*\mapunit+8pt}) {(d) Shifted};
  \node at ({4*\panelstep},{5*\mapunit+8pt}) {(e) Overlap};
  \node at (0,{-2*\mapunit-9pt}) {5 cells};
  \node at (\panelstep,{-2*\mapunit-9pt}) {7 cells};
  \node at ({3*\panelstep},{-2*\mapunit-9pt}) {11 cells};
  \node at ({4*\panelstep},{-2*\mapunit-9pt}) {17 cells};

  \begin{scope}[xshift={2*\panelstep},shift={(-1.05cm,-.3cm)}]
    \draw[semithick] (0,1.1) -- (.95,2.5)
      (0,1.1) -- (.95,1.8) (0,1.1) -- (.95,1.1)
      (0,1.1) -- (.95,.4) (0,1.1) -- (.95,-.3);
    \draw[semithick] (.95,2.5) -- (2.05,3.05)
      (.95,2.5) -- (2.05,2.5) (.95,2.5) -- (2.05,1.95);
    \tikzset{hiernode/.style={draw,semithick,fill=white,inner sep=1pt,
      minimum width=5mm,minimum height=4.5mm}}
    \node[hiernode,fill=\cellNode{W}] at (0,1.1) {\(\mathbf W\)};
    \node[hiernode,fill=\cellNode{T}] at (.95,2.5) {\(\mathbf T\)};
    \node[hiernode,fill=\cellNode{L}] at (.95,1.8) {\(\mathbf L\)};
    \node[hiernode,fill=\cellNode{R}] at (.95,1.1) {\(\mathbf R\)};
    \node[hiernode,fill=\cellNode{B}] at (.95,.4) {\(\mathbf B\)};
    \node[hiernode,fill=\cellNode{I}] at (.95,-.3) {\(\mathbf I\)};
    \node[hiernode,fill=\cellNode{T3}] at (2.05,3.05) {\(\mathbf{T_3}\)};
    \node[hiernode,fill=\cellNode{T2}] at (2.05,2.5) {\(\mathbf{T_2}\)};
    \node[hiernode,fill=\cellNode{T1}] at (2.05,1.95) {\(\mathbf{T_1}\)};
  \end{scope}
\end{tikzpicture}%
\endgroup
\caption{One first cut and three positions of the second disk.
The three blue cells in (b) are the children of \(T\) shown in (c).
The heavier lines in (d) and (e) are new cuts. With \(C\) centred at
the origin, \(D\) is centred at \((0,3)\) in (a,b), \((4/5,3)\) in (d) and
\((4/5,8/5)\) in (e). The dots mark the two rim crossings in (e).
The wall above the left dot stops at the hollow square on the upper
rim of \(D\). We prescribe the cuts here; with only two circles the
algorithm would finish at \(W\)}\label{fig:cell-hierarchy}
\end{figure}
\end{minipage}\par\addvspace{\intextsep}

\Needspace{3\baselineskip}
\ifdefined\ArxivPreprint
In Figure~\ref{fig:cell-hierarchy}(a), the rim of \(C\) and its two
vertical walls leave five children: the side strips \(L,R\), the region
\(B\) below \(C\), its interior \(I\), and the region \(T\) above it.
The dashed rim of \(D\) is not yet a cut: we can cross it without leaving \(T\).
Cutting along it in (b) splits only \(T\), into the gap \(T_1\), the disk
interior \(T_2\) and the band \(T_3\). These are the three descendants
directly below \(T\) in (c). In (d) the shifted \(D\) splits \(T\) and
\(R\); in (e) it also splits \(I\). The three placements leave 7, 11
and 17 open cells, respectively.
\else
In Figure~\ref{fig:cell-hierarchy}(a), draw the rim of the lower
unit disk \(C\). From its leftmost and rightmost points run vertical
walls to the top and bottom of the rectangle. Count the children of
\(W\): a strip on the left (\(L\)), one on the right (\(R\)), and
three pieces between the walls: below the disk (\(B\)), inside it
(\(I\)), and above it (\(T\)). We can cross the dashed rim of the
upper unit disk \(D\) and stay in the same blue cell \(T\): that rim
is not yet a cut.

Panels (b), (d) and (e) start from this same first cut. In each,
cut along the rim of \(D\) in every open cell it enters. In (b) its
leftmost and rightmost points already lie on the old walls. Only \(T\)
splits: \(T_1\) is the gap between the disks, \(T_2\) the interior of
\(D\), and \(T_3\) the band above it. The four unchanged cells and
these three give seven open cells. In the tree (c), \(W\) still has
five children; the three new nodes belong to \(T\).

In (d) the rim of \(D\) crosses the old wall between \(T\) and \(R\)
but misses \(C\). Each of \(T\) and \(R\) splits into four children;
the four in the right strip are unshaded. This gives \(4+4+3=11\)
open cells. In (e) we lower \(D\) so its rim also enters \(I\).
Now \(T\) has seven children, \(I\) and \(R\) have four each, and
\(L\) and \(B\) stay whole: seventeen open cells.
\fi
\else
Enclose all disks strictly inside an axis-aligned rectangle. A corner
represents every point on or outside it: all earn just the exterior
weights. Its open interior \(W\) is the root cell. Cuts consist of sampled circle arcs
and vertical segments, called \emph{walls}. The \emph{children} are the
connected components left in the open parent after removing the cuts.
The parent owns the new edges and vertices. The resulting tree of
cells is the \emph{cutting}; its final cells are the \emph{leaves}.
\fi

Each cell has a simple shape. Fix a horizontal position \(x\)
strictly between its left and right sides. The vertical line there
meets the cell in one open segment, from a bottom height \(g_-(x)\)
to a top height \(g_+(x)\). A point on the line is in the cell
exactly when \(g_-(x)<y<g_+(x)\). Each boundary curve is part of a
horizontal side of \(W\) or a circle \emph{branch}, the graph of
\(y=z_y\pm\sqrt{r^2-(x-z_x)^2}\) for one fixed sign on an interval.
\ifIllustrated In \(T\) the bottom curve is the upper semicircle
of the lower disk, and the top is straight. In \(I\) the two curves
meet at the left and right ends, so each vertical side is just a point.
\else Where the two graphs meet at an end, the vertical side is a point.\fi

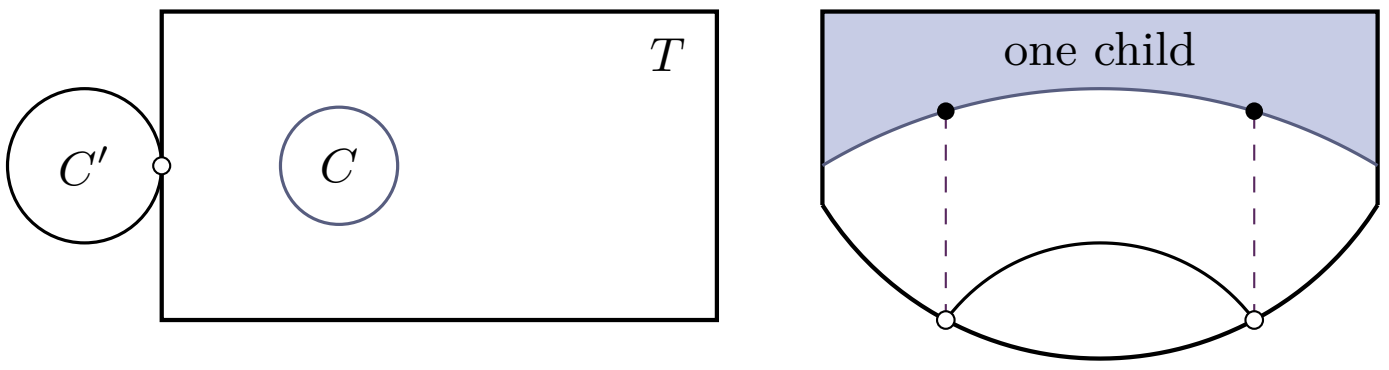
\begin{figure}[!htb]
\centering
\begin{tikzpicture}[x=1.02cm,y=1.02cm,font=\small]
  \begin{scope}[xshift=-2.35cm]
    \draw[thick] (-1.8,0) rectangle (1.8,2);
    \draw[semithick,draw=figRim] (-.65,1) circle (.38);
    \draw[semithick] (-2.3,1) circle (.5);
    \node at (-.65,1) {\(C\)};
    \node at (-2.3,1) {\(C'\)};
    \node at (1.48,1.72) {\(T\)};
    \filldraw[fill=white,draw=black] (-1.8,1) circle (1.6pt);
    \node at (-.4,-.56) {(a) Which rims meet \(T\)?};
  \end{scope}
  \begin{scope}[xshift=2.02cm]
    \fill[figRegion] (-1.8,2) -- (1.8,2) --
      plot[domain=1.8:-1.8,samples=57] (\x,{sqrt(12.25-\x*\x)-2}) -- cycle;
    \draw[thick] (-1.8,{15/8-sqrt((17/8)^2-1.8^2)}) -- (-1.8,2)
      -- (1.8,2) -- (1.8,{15/8-sqrt((17/8)^2-1.8^2)});
    \draw[thick,domain=-1.8:1.8,samples=57] plot (\x,{15/8-sqrt((17/8)^2-\x*\x)});
    \draw[semithick,domain=-1:1,samples=51] plot (\x,{sqrt(1.5625-\x*\x)-.75});
    \draw[semithick,draw=figRim,domain=-1.8:1.8,samples=57] plot (\x,{sqrt(12.25-\x*\x)-2});
    \foreach \x in {-1,1}{
      \draw[dashed,draw=figTrace] (\x,0) -- (\x,{sqrt(11.25)-2});
      \filldraw[fill=white,draw=black] (\x,0) circle (1.65pt);
      \fill (\x,{sqrt(11.25)-2}) circle (1.65pt);
    }
    \node at (0,1.76) {one child};
    \node at (0,-.56) {(b) Where do walls stop?};
  \end{scope}
\end{tikzpicture}%
\caption{(a) \(C\) belongs to the list for the open cell \(T\);
\(C'\) touches only its boundary. (b) The hollow endpoints lie on
the parent's curved bottom side, a boundary already searched. Each dashed
wall stops on the upper arc at a filled vertex. That vertex is scored
once, separately from its incident edges. Above the stops, the shaded
band remains one child}\label{fig:wall}
\end{figure}

The purpose of a split is to shorten the list of circles still to
test. The \emph{conflict list} of an open cell contains exactly the
circles whose rims have a point inside the cell. If the rim misses
the open cell, the continuous function \(\|x-z_C\|-r_C\) has no zero
there. It has one sign throughout the connected cell, so the same
weight applies everywhere. A rim wholly inside the cell is a conflict;
one tangent only to its boundary is not (Figure~\ref{fig:wall}(a)).

To split the cell, choose a subset of its list, called the
\emph{sample}. Appendix~\ProofRef{sec:localcut} explains the two
requirements on this choice; Lemma~\ProofRef{lem:split} constructs it
deterministically from a compressed list.
Keep the portions of the sampled rims inside the parent. Mark where
these arcs meet one another and where their ends lie on the parent
boundary. Also mark each retained leftmost or rightmost point of a
sampled circle, where its tangent is vertical. From each mark
draw vertical segments called \emph{walls} in each direction that
enters the parent. A wall
stops at the next sampled arc or parent boundary; an unselected rim
does not stop it. Split each incident arc at these events and wall
endpoints, merging coincident vertices. Each circular edge now lies
on a single upper or lower branch. We follow Wang's adaptation of
Chazelle's method to circular arcs \citep[Appendix~A]{wang2023unit};
Appendix~\ProofRef{sec:construction} gives the construction for
tangencies and points shared by several rims as well.

\Needspace{7\baselineskip}
\begin{lemma}[Dense circle cutting]\label{lem:cut}
For a sufficiently large fixed integer \(\rho>2\), one can generate
and visit a cutting rooted at \(W\) in \(O(m^2)\) time and
\(O(m)\) extra memory for \(m\ge1\). The levels run from
\(0\) to \(J=\lfloor\log_\rho m\rfloor\). Level \(j\) has \(O(\rho^{2j})\) cells,
each with an explicit list of at most \(m/\rho^j\) conflicts. A parent
is carried forward unchanged or split. Its open children, new open edges
and new vertices inside the parent together number \(O_\rho(1)\).
Final conflict lists have
size less than \(\rho\).
\end{lemma}
Choose \(\rho\) once, independently of the input. The bound on the
number of children depends only on \(\rho\); the actual number depends on the geometry.
The list length determines whether to split.

\paragraph{Why a level has few cells.}
\ifdefined\ArxivPreprint
If a parent's list already meets the next threshold, carry it forward.
Otherwise, for a list of size \(k\), choose a sample that hits every
possible open child meeting at least \(k/\rho\) rims. An actual child
contains no point of a sampled rim, so it has fewer than \(k/\rho\) conflicts.

This bounds list lengths, not the number of cells. For the latter,
charge each new feature to a sampled rim or a pair of rims meeting in
the parent. Two original rims meet at most twice across \emph{all}
disjoint open parents of a level. Lemma~\ProofRef{lem:split} controls
the charge using the fraction of pairs meeting in each parent;
Appendix~\ProofRef{sec:cut-budget} sums it to obtain \(O(\rho^{2j})\)
cells at level \(j\).
\else
Short lists and few cells are separate requirements. A parent at or
below the next threshold is carried forward. Otherwise, if its list
has size \(k\), the sample hits every possible open child meeting at
least \(k/\rho\) rims. An actual open child contains no point of a
sampled rim: those rims are cuts. Its list must therefore have fewer
than \(k/\rho\) conflicts. Even constant branching could
give more than \(\rho^{2j}\) cells at level \(j\).

For the cell count, charge the new features of a split to sampled
circles and pairs that meet inside the parent. A pair of original
circles has at most two contacts across \emph{all} disjoint open
parents of a level, including tangencies. Let \(P_T\) count the
circle pairs meeting inside parent \(T\). The local split bound
(Lemma~\ProofRef{lem:split}) charges \(O_\rho(P_T/k^2)\) for pairs:
it is their frequency among the \(k\) parent circles that matters. Since \(k>m/\rho^j\) for a
split leading to level \(j\) and \(\sum_T P_T\le 2\binom m2\),
the level's pair charge is \(O_\rho(\rho^{2j})\). The remaining term per parent is
\(O(\rho\log^2(2\rho))\) with an absolute constant, at most
\(\rho^2/2\) for our choice of \(\rho\). Thus if \(C_j\) is the
number of cells, \(C_j/\rho^{2j}\) is at most a fixed constant plus
half the preceding normalized count. This proves the level bound;
Appendix~\ProofRef{sec:cut-budget} gives the recurrence. Only one
branch and its pending work are needed at a time.
\fi

\ifIllustrated
\ifdefined\CompletePaper
\paragraph{Why test more patterns than a circle can realize?}
The \emph{trace} of a query records which entries of the input list it
hits. Here a hit means that the query circle meets a listed rim inside the cell.
In Figure~\ref{fig:lifted-traces}, take a cell containing both fixed
rims. A query centred at \((u,0)\) with radius \(1/2\) meets the unit
rim centred at \((a,0)\) exactly when \(1/4\le (u-a)^2\le9/4\).
The \emph{lift} records \(u\) as \((u,v)=(u,u^2)\); each hit test becomes
a strip. Allowing \(v\) to vary independently then adds patterns but
keeps every genuine trace. See Lemma~\ProofRef{lem:trace-catalogue}.

\begin{figure}[!htb]
\centering
\includegraphics[width=\ProofPlateWidth]{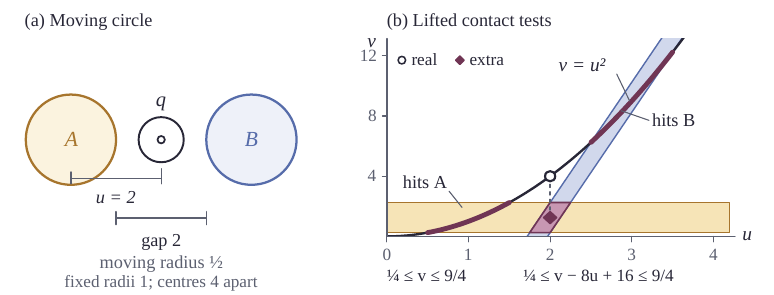}
\caption{Which rims does the moving circle \(q\) meet?
(a) The fixed unit circles \(A,B\) have centres four units apart.
The diameter of \(q\) is one, smaller than the gap of two between
their rims. (b) A real placement has \(v=u^2\). The two heavy arcs
mark separate hits on \(A\) and on \(B\); at the open point \((2,4)\)
neither is hit. Allowing \(v\) to vary independently admits the
diamond \((2,5/4)\) in both strips, a pair of hits no placement of
\(q\) realizes}\label{fig:lifted-traces}
\end{figure}

\FloatBarrier
\fi
\fi

\subsection{Scoring cells, edges and vertices}\label{sec:plane-scoring}
In Algorithm~\ref{alg:plane}, \textsc{Visit}\((T,F,b,j)\) searches an open
cell \(T\) at level \(j\). The list \(F\) contains the rims meeting \(T\).
The \emph{offset} \(b\) sums the constant contributions of all circles
outside \(F\), including those set aside at earlier splits. Thus
\[
   \operatorname{depth}(x)=
   \underbrace{b}_{\text{constant in }T}
   +\underbrace{\sum_{C\in F}w_C(x)}_{\text{still tested at }x},
    \qquad x\in T .
\]
Nonnegativity gives \(w_C(x)\le w_C^{\rm in}+w_C^{\rm out}\) at every
point. If \(M\) is the best score so far and
\(b+\sum_{C\in F}(w_C^{\rm in}+w_C^{\rm out})\le M\), the algorithm may
skip \(T\): even a point on every remaining rim cannot improve on \(M\).

At the root \(b=0\). Initialize the best pair with a corner of \(W\) and
its score \(\sum_C w_C^{\rm out}\), the score everywhere on or outside
the rectangle.

\ifIllustrated
Figure~\ref{fig:child-offset} follows a split with parent offset one.
Keep the tangent unit circles \(C,D\) of Figure~\ref{fig:open-child},
centred at \((0,0)\) and \((2,0)\). Here \(E\) is smaller: its centre
is \((0,0)\) and its radius is \(1/2\). Other listed circles miss
the closed disk of \(C\) and contribute zero there. Only \(C\) is
selected in this window.

\begin{figure}[!htb]
\centering
\begin{tabular}{@{}c@{\hspace{.46cm}}c@{}}
\begin{tikzpicture}[x=1.13cm,y=1.13cm,font=\small,baseline=(current bounding box.center)]
  \fill[depth3!64!white] (0,0) circle[radius=1];
  \fill[depth4!72!white] (0,0) circle[radius=.5];
  \draw[black!42,thin] (-1.34,-1.24) rectangle (3.30,1.24);
  \foreach \x in {-1,1} \draw[densely dotted,draw=figTrace] (\x,-1.24) -- (\x,1.24);
  \draw[thick,draw=figRim] (0,0) circle[radius=1];
  \draw[semithick,dashed] (2,0) circle[radius=1];
  \draw[semithick,dash dot] (0,0) circle[radius=.5];
  \node[inner sep=1pt] at (-.49,-.63) {\(3\)};
  \fill (0,0) circle[radius=1.45pt];
  \node[anchor=north,inner sep=1pt] at (0,-.04) {\(z:4\)};
  \filldraw[fill=depth5,draw=black] (1,0) circle[radius=2pt];
  \draw[thin] (1.05,.02) -- (1.36,.21);
  \node[anchor=west,inner sep=1pt] at (1.36,.21) {\(v:5\)};
  \draw[thin] (-.52,.855) -- (-.60,1.04) node[left,inner sep=1pt] {\(C\)};
  \draw[thin] (2.61,.79) -- (2.84,1.03) node[right,inner sep=1pt] {\(D\)};
  \draw[thin] (.05,.4975) -- (.14,.70) node[right,inner sep=1pt] {\(E\)};
\end{tikzpicture}
&
{\small\setlength{\tabcolsep}{3pt}%
\begin{tabular}{@{}lcl@{}}
  \multicolumn{3}{@{}l}{Parent offset \(b=1\)}\\[3pt]
  Rim & (in, out) & In the child \(T'\)\\
  \hline
  \(C\) & \((1,0)\) & inside: add \(1\)\\
  \(D\) & \((2,1)\) & outside: add \(1\)\\
  \(E\) & \((1,0)\) & keep on the list\\
  \hline
  \multicolumn{3}{@{}l}{Child offset \(b'=3\), list \(F'=\{E\}\)}
\end{tabular}}
\end{tabular}%
\caption{Both shades belong to one open child \(T'\), the interior of
\(C\). The solid rim and dotted walls define the split. The dashed
\(D\) and dash-dotted \(E\) are present only for scoring. The table
gives the child's offset; the numbers on the drawing are full scores. The boundary
vertex \(v\) is scored by the parent}\label{fig:child-offset}
\end{figure}

\Needspace{4\baselineskip}
\ifdefined\ArxivPreprint
The parent scores \(v\) first: its offset contributes one, the rims
of \(C,D\) contribute \(1+0\) and \(2+1\), and \(E\) contributes
zero. The total is five. Inside \(C\), the offset becomes
\(b'=1+1+1=3\): add the inside weight of \(C\) and the outside weight
of \(D\). Only \(E\) remains to test, so \(z\) scores four. Applying
this child's formula at the excluded \(v\) would give only three.
\else
First score the new boundary with the parent list. A wall starts at
the rightmost point of \(C\), so the contact \(v=(1,0)\) is a new
vertex. Its depth is five: one from the parent offset, \(1+0\) on the
rim of \(C\), \(2+1\) on the rim of \(D\), and zero outside \(E\).

Now form the record for the child inside \(C\). Throughout this open
child, \(C\) contributes its inside weight and \(D\) its outside
weight. Only the rim of \(E\) enters the child, so \(F'=\{E\}\) and
\(b'=1+1+1=3\). The probe \(z=(0,0)\) scores four because it also
lies in the disk of \(E\); the outer band scores three. Using the
child formula at \(v\) would likewise give three and lose the score
of five: the formula is valid only in the open child.
\fi
\fi

\begin{algorithm}[htb]
\caption{\PlaneDepth: score a cutting one open cell at a time}\label{alg:plane}
\small
\begin{algorithmic}
\Require Grouped weighted circles \(\mathcal C\); fixed \(\rho\) from Lemma~\ref{lem:cut}
\If{\(\mathcal C\) is empty} \State \Return the origin and zero. \EndIf
\State \(m\gets|\mathcal C|\); take the largest integer \(J\) with \(\rho^J\le m\).
\State Choose an open rectangle \(W\) containing the circles strictly.
\State Set the best pair to a corner of \(W\) and \(\sum_{C\in\mathcal C}w_C^{\rm out}\).
\State Whenever a tested point has greater depth, keep that point and depth as the best pair.
\State \(\Call{Visit}{W,\mathcal C,0,0}\);\quad \Return the best pair.
\Procedure{Visit}{$T,F,b,j$}
  \If{\(j=J\)}
    \State Form the origin, the rightmost points of the rims in \(F\), and their pair contacts.
    \State \hyperref[par:plane-leaf]{Score the candidates strictly inside \(T\)} using \(b,F\); \textbf{return}.
  \EndIf
  \If{\(|F|\le m/\rho^{j+1}\)} \State \(\Call{Visit}{T,F,b,j+1}\); \textbf{return}. \EndIf
  \State Select circles by Lemma~\ProofRef{lem:split}; draw their arcs and stopped walls in \(T\).
  \State With \(b,F\), score each new vertex and \hyperref[lem:edge]{sweep each new open edge}.
  \ForAll{open children \(T'\), one at a time}
    \State \(F'\gets\{C\in F:\text{the rim of }C\text{ meets }T'\}\); choose \(z\in T'\).
    \State \(b'\gets b+\sum_{C\in F\setminus F'}w_C(z)\).
    \State \(\Call{Visit}{T',F',b',j+1}\).
  \EndFor
\EndProcedure
\end{algorithmic}
\end{algorithm}

\FloatBarrier
\Needspace{6\baselineskip}
The first split that puts a point on a boundary determines which list
scores it. All other points of \(W\) reach an open leaf.
\begin{lemma}[Ownership]\label{lem:owner}
Each point of \(W\) is owned exactly once: it belongs either to a new
open edge or vertex at the first split that puts it on a boundary, or to one open leaf.
At a split of an open parent, its open children, new open edges, and new
vertices partition that parent.
\end{lemma}
\begin{proof}
At a split every meeting of drawn arcs and every internal endpoint of a
drawn arc or wall is a single vertex. Its incident edges exclude it. These boundary features
are disjoint. Removing them from the open parent leaves precisely the
disjoint open child cells. Thus a
point not on a new edge or vertex lies in a unique open child; a carried
cell leaves it unchanged. Following this unique sequence, the point is
first owned by a new boundary feature, or reaches a unique open leaf.
These alternatives are disjoint. The boundary of \(W\) is outside this
ownership partition and is handled separately.
\end{proof}

\ifIllustrated
Follow the open edge upward in Figure~\ref{fig:edge-scores}. At height
zero it crosses \(C\) and just touches \(C'\). For ordinary disks
the counts just below, at and just above zero are \(0,2,1\). With the
weights shown, the contact earns \(2+1\) from \(C\) and one from \(C'\).
Comparing only the open gaps, with scores one and two, would miss the maximum.

\begin{figure}[!htb]
\centering
\includegraphics[width=\ProofPlateWidth]{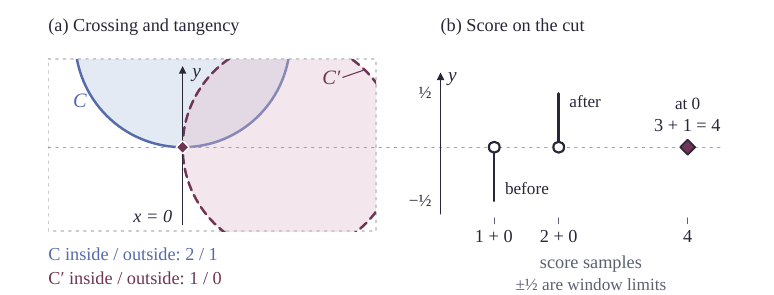}
\caption{An isolated maximum on an open edge. The two unsampled unit rims
centred at \((0,1)\) and \((1,0)\) meet the cut \(x=0\) at the same
height. The hollow endpoints in (b) exclude that height; the diamond
gives its score \(3+1=4\). No other circle or offset contributes}\label{fig:edge-scores}
\end{figure}
\else
Even an open cut edge can contain a winning tangency with an unsampled
rim. We must score the tied event as well as the adjacent intervals.
\fi

\begin{lemma}[Edge sweep with ties]\label{lem:edge}
Given an open edge, its parent offset \(b\) and the parent's conflict list
of size \(k\), the sweep finds the maximum depth on the edge and a point
attaining it in
\(O(k\log k)\) time for \(k\ge2\), and constant time otherwise, using
\(O(k+1)\) extra memory.
\end{lemma}
\begin{proof}
Use the \(x\)-coordinate on a circular edge (a single branch) and
the \(y\)-coordinate on a vertical edge. A circular edge lies entirely
on its supporting rim, so both of that circle's weights count throughout.
Every other circle has at most two contacts on the open edge.
Sort their parameters and group equal ones. Score any point in the
first open interval using \(b\) and the parent list. At a contact \(x_0\), take \(x_-\) and
\(x_+\) at parameter midpoints of the preceding and following open
intervals. Only circles meeting the edge at \(x_0\) can change their
contribution. Let \(G\) consist of these circles, excluding the supporting
circle: its constant contribution is already in the incoming score.
From the score \(S_-\) at \(x_-\), compute
\[
     S_\star=S_-+\sum_{C\in G}\bigl(w_C(x_\star)-w_C(x_-)\bigr),
                  \qquad \star\in\{0,+\}.
\]
Compute both corrections from \(S_-\). Compare the scored points
\((x_0,S_0)\) and \((x_+,S_+)\) with the running best; \(S_+\) is
the incoming score for the \emph{next} event. This evaluates a tangency
or multiway contact at equality even when neither adjacent interval has
that score. Neither endpoint belongs to the open edge: a new vertex
has its own score, and points on older boundaries were scored earlier.
With no contacts the midpoint gives the score on the whole edge. Sorting dominates
the work; all event groups together have \(O(k)\) incidences.
\end{proof}

\Needspace{5\baselineskip}
\phantomsection\label{proof:planar}
\begin{proof}[Proof of Theorem~\ref{thm:planar}]
The root satisfies the offset identity. A circle missing a connected
open child has constant contribution throughout it, so the update in
Algorithm~\ref{alg:plane} establishes the offset identity for that
child. To choose its interior point \(z\), take the midpoint
\(x_m\) between the child's left and right sides. At \(x_m\) take
the height halfway between the bottom and top curves,
\((g_-(x_m)+g_+(x_m))/2\).
The new offset \(b'\) applies only inside \(T'\).

A new boundary feature lies inside its open parent, so the parent
offset and list give its full depth. One scan scores a vertex, and
Lemma~\ref{lem:edge} finds an attaining point of maximum depth on each
open edge. By Lemma~\ref{lem:owner}, it remains to account for a
winning candidate that reaches an open leaf.

\Needspace{5\baselineskip}
\paragraph{Finishing a leaf.}\label{par:plane-leaf}
If the rightmost point of a circle lies in the open leaf \(T\), its
rim meets \(T\) and is on the leaf's list \(F\). If a contact lies
in \(T\), \emph{both} of its defining rims are on that list. The
origin is a candidate for every list, including the empty list. Hence
\[
   \operatorname{cand}(\mathcal C)\cap T
   = \operatorname{cand}(F)\cap T .
\]
Generate the set on the right and score its points with the leaf's
offset and list. If the set is empty, return without a test. We need
one global winner, not the maximum of each open leaf:
Lemma~\ref{lem:plane-candidates}, applied to the weights as in
Section~\ref{sec:plane-contract}, puts a winner among the global
candidates. A winner on or outside the enclosing rectangle is matched by the initial corner,
a winner on a cut by the boundary search, and a winner inside a
leaf by the identity above. With \(k=|F|<\rho\), there are at most
\(1+k+2\binom{k}{2}=1+k^2\) local candidates. Generating, filtering
and scoring them therefore take constant time per leaf.

\ifIllustrated
\ifdefined\ArxivPreprint
For the two projected disks in Figure~\ref{fig:pole-sides}(d), the root
is already a leaf. Their contact at the origin scores two; lifting it
back to the sphere and counting the anchor recovers all three sites.
\else
\begin{figure}[!htb]
\centering

\caption{Candidates for two tangent disks. In (a) the origin \(o\),
the rightmost point \(R(A)\) of the left circle and the contact \(v\)
are the same point. In (b) the centres are \((0,1/2)\) and
\((0,3/2)\): the contact \(v=(0,1)\) is distinct from all three
other candidates. Numbers are disk counts; the enclosing rectangle's
corner, not shown, scores zero in both runs}\label{fig:leaf-samples}
\end{figure}

Figure~\ref{fig:leaf-samples}(a) uses the two disks of
Figure~\ref{fig:pole-sides}(d), here calling the left disk \(A\).
Both have radius \(1/2\) and weights \((1,0)\). The root is a leaf since \(m=2<\rho\).
Its only distinct candidates are the contact \(o=(0,0)\), scoring
two, and \((1,0)\), scoring one. Lifting \(o\) back to the sphere
and counting the anchor recovers all three sites of Section~\ref{sec:caps}.
After rotating and translating the disks in~(b), only their contact
scores two; the origin and both rightmost marks score one.
\fi
\fi

The algorithm needs no global sort of pair contacts. Each edge sweep
sorts only events from the parent list. At a split with
\(k\ge2\) parent conflicts there are \(O_\rho(1)\)
new features by Lemma~\ref{lem:cut}, so scoring \emph{all} of them
costs \(O(k\log k)\); for \(k<2\) the cost is constant.
Large lists occur only near the root, where there are few cells. Count
levels from the bottom, with \(i=1\) at the last level \(J\). Level
\(j=J-i+1\) has \(O(m^2/\rho^{2i})\) cells because \(\rho\) is fixed,
and each list has \(k<\rho^i\) conflicts. Local work of
\(O(k\log^\beta k)\) for \(k\ge2\), constant otherwise, thus costs
\(O(\rho^i i^\beta)\) \emph{per cell} on that level. The level
total is \(O(m^2 i^\beta/\rho^i)\). Summing over levels
gives, for fixed \(\beta\ge0\),
\begin{equation}\label{eq:sum}
 O\!\left(m^2\sum_{i\ge1}\frac{i^\beta}{\rho^i}\right)=O(m^2).
\end{equation}
For the memory bound, sort one edge's events at a time and release
the array before descending. The parent list remains available for
the next sibling; Lemma~\ref{lem:cut} already accounts for these lists. Scoring
adds only one temporary event array of size \(O(m)\) and the running
best point and score.
\end{proof}

\section{Lower bounds via hyperplane incidence}\label{sec:lower}
A \emph{hyperplane} is flat, has dimension \(D-1\), and need not pass
through the origin. An \emph{incidence} means that a point lies on it.
Integer \emph{affine degeneracy} asks whether some \(D+1\) of the
integer input points lie on one hyperplane; in the plane this is three-point
collinearity~\citep{gajentaan1995class,erickson1995better}.
We reduce this question to coverage in the same dimension.

\subsection{One pair of target sites per source point}\label{sec:lens}
\ifIllustrated
Let \(H(P)\) be the most sites on one hyperplane \(b\cdot p=\gamma\),
with normal \(b\ne0\). Give each site two targets. One ball hits every pair and
doubles the pairs on a maximizing hyperplane. Conversely, all pairs
doubled by any ball come from sites on one hyperplane.
\else
Which line contains the most points of a planar set \(P\)? In \(\R^D\)
let \(H(P)\) be the largest number of its points on an affine hyperplane
\(b\cdot p=\gamma\); the nonzero normal \(b\) sets the orientation and
\(\gamma\) the position.

We replace each source point by a pair of labelled targets. The proof
shows that one unit ball hits every pair and covers both targets in
\(H(P)\) pairs. No ball can cover both targets in more pairs: their
source points must lie on one hyperplane. Thus an optimal ball counts
one target per source point and one extra for each source point on a
maximizing hyperplane. The labels tell us which points these are.
\fi

\begin{theorem}[Exact encoding of hyperplane incidence]\label{thm:lens}
For each fixed \(D\ge2\), a set \(P\) of \(s\) distinct points in
\(\Z^D\) can be transformed in \(O(s)\) rational arithmetic
operations into a set \(Q\) of \(2s\) distinct rational points in
\(\Q^D\) whose maximum unit-ball coverage is
\[
                              M(Q)=s+H(P).
\]
Each target is labelled by its source index and a sign. From the labels
covered by an optimal ball one can recover a maximizing source hyperplane
in \(O(s)\) further rational operations. For nonempty input set
\(U=\max(2,1+\max_{i,j}|p_{ij}|)\). Target numerators and denominators
have \(O((D+1)^6\log(2DU))\) bits, hence \(O(\log U)\) for fixed \(D\).
\end{theorem}

\begin{figure}[!htb]
\centering
\ifIllustrated
\includegraphics[width=\ProofPlateWidth]{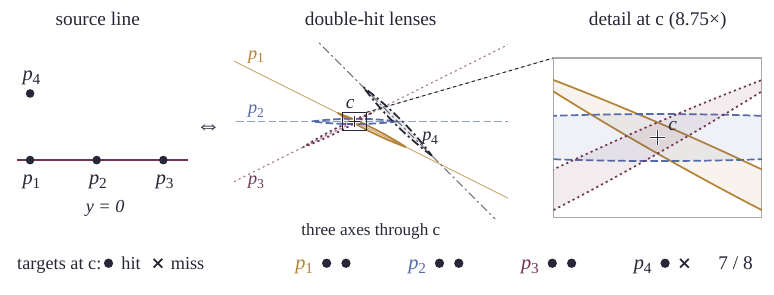}
\caption{Reading the example in both directions. Each lens consists of
the centres covering both sites of a pair. The first three bisectors meet at
\(c=(0,1/2)\); the inset shows \(c\) in all three lenses. Read the
doubled pairs below to recover \(p_1,p_2,p_3\) on \(y=0\).
The drawing uses \(o=(1,2)\), \(R=4\) before rescaling, and pair
spacings chosen for these four sites}
\else

\caption{Recovering a source line from the covered labels. Each row on the
right represents an unordered pair of target labels; filled dots are
covered. The doubly filled rows select the source points on the line
at left}
\fi
\label{fig:incidence-labels}
\end{figure}

\begin{example}\label{ex:lens}
Take \(p_1=(0,0)\), \(p_2=(1,0)\), \(p_3=(2,0)\) and \(p_4=(0,1)\).
The line \(y=0\) contains the first three points, and no line contains
all four, so \(H(P)=3\). An optimal disk covers seven of the eight
targets: one for each of the four source points, and a second for
each of the three on \(y=0\) (Figure~\ref{fig:incidence-labels}).
\end{example}

\ifIllustrated\LayoutFloatBarrier\fi

\begin{proof}[Proof of Theorem~\ref{thm:lens}]
If \(s=0\), return an empty target and the hyperplane \(x_1=0\);
both optima are zero. Otherwise let \(k_2\) count the pairs hit twice
by a proposed ball and \(k_0\) the pairs it misses. We count labelled
targets here; Lemma~\ProofRef{lem:lensparameters} ensures their coordinates
are all distinct. The total score is
\begin{equation}\label{eq:doubled}
                       2k_2+(s-k_2-k_0)=s+k_2-k_0 .
\end{equation}
It suffices to achieve \(k_2\ge H(P)\) with no missed pairs, and to show
that \(k_2\le H(P)\) for \emph{every} possible centre.

All target pairs use the same integer translation \(o\), chosen in
Appendix~\ProofRef{sec:lensparam} so that every \(a_i=o-p_i\) is a
nonzero integer vector.
We want the perpendicular bisectors of some target pairs to share a
point exactly when their sources lie on a hyperplane. Use
\(\Pi_i=\{x:a_i\cdot x=1\}\) as the bisector for \(p_i\).
If \(b\cdot p_i=\gamma\) on an index set and \(b\cdot o\ne\gamma\),
then \(x=b/(b\cdot o-\gamma)\) lies on every corresponding \(\Pi_i\):
\(a_i\cdot x=(b\cdot o-b\cdot p_i)/(b\cdot o-\gamma)=1\).
For each incident set the appendix supplies a containing hyperplane
with bounded integer coefficients and chooses \(o\) off all such
hyperplanes. This gives a bounded witness without constructing it.

The foot of the perpendicular from the origin to \(\Pi_i\) is
\(u_i=a_i/\norm{a_i}^2\). Place the two targets at
\(q_i^\pm=u_i\pm\lambda_i a_i\), equally far along the normal on
opposite sides, with \(\lambda_i>0\) rational. Here \(R\) is the
common rational target-ball radius before rescaling. The centres covering both
targets form a lens, the intersection of their two closed radius-\(R\)
balls (Figure~\ref{fig:lens}). Dividing all coordinates by \(R\)
preserves the covered labels and gives unit balls.

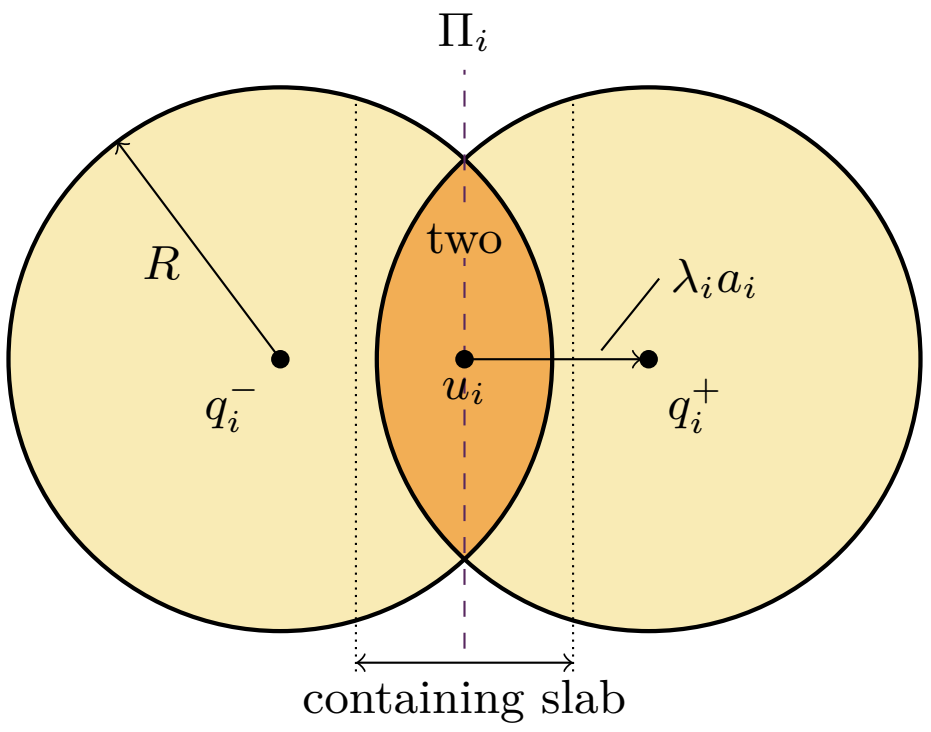
\begin{figure}[ht]
\centering
\begin{tikzpicture}[scale=1.16,font=\small]
  \fill[depth1!62!white] (-1.05,0) circle (1.55) (1.05,0) circle (1.55);
  \begin{scope}
    \clip (-1.05,0) circle (1.55);
    \fill[depth2] (1.05,0) circle (1.55);
  \end{scope}
  \draw[thick] (-1.05,0) circle (1.55);
  \draw[thick] (1.05,0) circle (1.55);
  \draw[dashed,draw=figTrace] (0,-1.65) -- (0,1.65) node[above] {\(\Pi_i\)};
  \draw[densely dotted] (-.619,-1.78) -- (-.619,1.50);
  \draw[densely dotted] (.619,-1.78) -- (.619,1.50);
  \fill (-1.05,0) circle (1.5pt) node[below left] {\(q_i^-\)};
  \fill (1.05,0) circle (1.5pt) node[below right] {\(q_i^+\)};
  \fill (0,0) circle (1.5pt) node[below] {\(u_i\)};
  \draw[->] (0,0) -- (1.01,0);
  \draw[thin] (.78,.05) -- (1.11,.46)
    node[anchor=west,inner sep=2pt] {\(\lambda_i a_i\)};
  \draw[->] (-1.05,0) -- (-1.98,1.24)
    node[pos=.56,below left,inner sep=1.5pt] {\(R\)};
  \node[fill=depth2,inner sep=1pt] at (0,.69) {two};
  \draw[<->] (-.619,-1.73) -- (.619,-1.73);
  \node[below,inner sep=2pt] at (0,-1.76) {containing slab};
\end{tikzpicture}
\caption{A schematic cross-section before division by \(R\). The two
outer dots are target sites; \(u_i\) is
their midpoint. Each circle bounds
the positions of a centre that covers the corresponding site. In the
lens marked ``two'' both sites count; outside the two disks neither counts.
The dashed line bisects the pair; the dotted lines bound the containing
slab at distance \(\delta_i/(2\lambda_i\norm{a_i})\) from it.
They are not the lens endpoints.}\label{fig:lens}
\end{figure}

The quantity \(\delta_i=R^2-\lambda_i^2\norm{a_i}^2\) is the squared
radius of the common rim of the two spheres when \(\delta_i>0\).
Since \(a_i\cdot u_i=1\), the squared distances to the targets are
\[
 \|x-q_i^\pm\|^2=\|x-u_i\|^2+\lambda_i^2\|a_i\|^2
                     \mp2\lambda_i(a_i\cdot x-1).
\]
For one hit compare the smaller distance with \(R\); for two hits
compare the larger. Taking the minimum or maximum gives the absolute value:
\begin{align}
 \text{at least one hit}\quad&\Longleftrightarrow\quad
  \norm{x-u_i}^2-2\lambda_i|a_i\cdot x-1|\le\delta_i ,
           \label{eq:near}\\
 \text{two hits}\quad&\Longleftrightarrow\quad
  \norm{x-u_i}^2+2\lambda_i|a_i\cdot x-1|\le\delta_i .
           \label{eq:far}
\end{align}

\paragraph{One witness attains the lower estimate.}
Let \(I\) be all \(H(P)\) indices on a maximizing source hyperplane. The reach estimate
in Lemma~\ProofRef{lem:lensparameters} supplies one witness \(x\) with
\(a_i\cdot x=1\) on \(I\) and \(\norm{x-u_j}^2\le\delta_j\) for
every \(j\). Thus the nearer-site test~\eqref{eq:near} holds for every pair. On
\(I\) the absolute value is zero, so~\eqref{eq:far} holds as well.
No pair is missed and at least \(H(P)\) are doubled.

\paragraph{No centre can do better.}
Use the coordinate bound \(\alpha\) supplied with the target parameters
by Lemma~\ProofRef{lem:lensparameters}. Its width estimate gives
\(\delta_i/(2\lambda_i)<1/((D+1)!\,\alpha^D)\).
By~\eqref{eq:far}, a double hit forces
\(|a_i\cdot x-1|\le\delta_i/(2\lambda_i)\), wherever \(x\) lies.
For any \(D+1\) doubled indices, form the square integer matrix \(C\)
with rows \((a_i,1)\). Subtract \(\sum_{j=1}^D x_j C_{\cdot,j}\)
from its last column. The determinant is unchanged; the new entries
are \(1-a_i\cdot x\). Each cofactor
\(C_{r,D+1}^{*}\) is a sum of \(D!\) signed products of \(D\)
entries of magnitude at most \(\alpha\). Thus it has magnitude at most
\(D!\alpha^D\). The strict width bound makes each of the \(D+1\)
summands smaller than \(1/(D+1)\) in magnitude:
\[
 |\det C|=\left|\sum_{r=1}^{D+1}(1-a_{i_r}\cdot x)\,C_{r,D+1}^{*}\right|<1.
\]
Since \(C\) is integral, \(\det C=0\).

Thus the doubled rows have rank at most \(D\), also when there are
fewer than \(D+1\). The source rows \((p_i,1)\) have the same rank:
the linear map \((a,t)\mapsto(to-a,t)\) is its own inverse.
A nonzero nullvector \((b,-\gamma)\) of these rows gives \(b\cdot p_i=\gamma\);
on a nonempty doubled set \(b\ne0\) since the rows end in one.
Thus \(k_2\le H(P)\), also for an empty doubled set. By~\eqref{eq:doubled} we
have both bounds on \(M(Q)\). At equality \(k_2=H(P)\) and \(k_0=0\).

To recover a maximizing hyperplane, take the source indices whose two
labels were covered. There is at least one, since \(s>0\) and \(H(P)\ge1\).
Scan their homogeneous rows for a basis and solve for a nonzero
nullvector as above. This recovers a hyperplane through all \(H(P)\)
indices in \(O(s)\) operations. The appendix verifies that the target
points are distinct and gives the cost and coordinate lengths of the
construction.
\end{proof}

\subsection{Conditional lower bound}\label{sec:conditional}
\emph{Integer affine-degeneracy hypothesis in dimension \(D\).}
For fixed \(D\ge2\), no deterministic real RAM with the operations of
Section~\ref{sec:problem} decides in \(O(s^{D-\varepsilon})\) time,
for any constant \(\varepsilon>0\), whether \(D+1\) of \(s\) integer
points lie on an affine hyperplane. This is our model-specific version
of the integer conjecture of Munteanu, Omlor and
Phillips~\citep[Conjecture~2]{munteanu2026hardness}. We use the same model
for both problems so the reduction can call a placement algorithm directly.
On real inputs the
known arrangement algorithms decide affine degeneracy in \(O(s^D)\)
time for fixed \(D\)~\citep[pp.~41--42]{erickson1995better}; this is
also an upper bound for integer inputs in our model. For related formulations
and uses see~\citep{harpeled2018regression,chanzheng2022hopcroft,cardinal2025degeneracy}.

\begin{corollary}\label{cor:exponent}
Under this hypothesis there is no deterministic \(O(N^{D-\varepsilon})\)
algorithm, for any fixed \(\varepsilon>0\), even for the optimum count
on distinct rational sites in \(\R^D\). Thus the exponent of
Theorem~\ref{thm:main} is conditionally optimal.
\end{corollary}
\begin{proof}
For \(s\ge D+1\), Theorem~\ref{thm:lens} gives \(H(P)\ge D+1\) if and
only if \(M(Q)\ge s+D+1\), with \(N=2s\). If the source permits
repeated coordinates, first sort in \(O(s\log s)\) time: two equal
entries and any \(D-1\) other input positions form a dependent
\((D+1)\)-tuple. Otherwise the distinct-site reduction applies.
The sorting and construction costs are both truly sub-\(s^D\).
\end{proof}
The hypothesis allows \(N^D/\log N\) time and therefore does not imply
an \(\Omega(N^D)\) lower bound.
Polynomially bounded source coordinates give \(O_D(\log s)\)-bit
target numerators and denominators. The quoted conjecture does not
require such a bound; nor does this representation result simulate
arbitrary real-RAM computations on a word RAM. For rational sources
given by explicit fractions, a common positive scaling clears all
denominators in \(O_D(s)\) arithmetic operations. For arbitrary reals
the determinant proof has no integer gap.

For the status of the older planar 3SUM evidence and the restricted
sidedness-query result, see Section~\ref{sec:prior-lower}.

\paragraph{A weaker bound from even \(k\)-SUM.}
Take a fixed even \(k\ge4\) and a set of distinct integers. Map
each value \(y\) to \((y,y^2,\ldots,y^{k-2},y^k)\in\Z^{k-1}\).
For any \(k\) values the affine determinant, with the column of ones
first, equals
\[
  \det[\,1,y_i,\ldots,y_i^{k-2},y_i^k\,]_{i=1}^k
  =\prod_{i<j}(y_j-y_i)\,\sum_i y_i.
\]
Dividing the alternating determinant by the Vandermonde product
leaves a symmetric homogeneous form of degree one, a multiple of
\(\sum_i y_i\). The coefficient of \(y_k^k\) on both sides is the
Vandermonde on the first \(k-1\) values, so the multiplier is one.
Distinctness makes the product nonzero, so
affine dependence detects a zero sum. Apply Theorem~\ref{thm:lens}
in dimension \(k-1\) and embed the \emph{target sites} into any
\(D\ge k-1\). A centre in the smaller space is still available;
projecting any other centre back cannot increase its distances to the
targets. Hence the optimum is unchanged.
If no deterministic algorithm in the same model improves the exponent
\(k/2\) by a positive constant for this distinct-integer problem,
none does so for coverage. Taking the largest even \(k\le D+1\)
gives exponent \(\lfloor(D+1)/2\rfloor\) for \(D\ge3\), independently
of affine-degeneracy hardness. Parity enters in the hypothesis, not
the determinant identity~\citep[for the odd case, see][]{chan20263sum}.

\subsection{What can be proved without the hardness hypothesis?}
Even a collinear input gives an unconditional obstruction.

\begin{proposition}\label{prop:sorting}
Even for distinct points on a line, deciding whether \(M(P)\ge2\)
takes \(\Omega(N\log N)\) time in the algebraic computation-tree model.
\end{proposition}
\begin{proof}
Restrict the sites to \((x_i,0,\ldots,0)\). A pair fits precisely when
its separation is at most two. In a no-instance every gap in sorted
order exceeds two. For each of the \(N!\) strict orders, these gap
inequalities define a nonempty convex region. Truncate a tree for distinct
inputs at its promised worst-case height. Extend it to all inputs by
guarding each division and square root against a zero divisor or
negative radicand, respectively, and assigning an arbitrary answer on
a failed guard or truncated branch. This costs only a constant factor
in height and leaves the answers on the promise unchanged. The no-output
set of a finite guarded tree is semialgebraic, so its connected components
are path-connected. Along a path between different orders, two
coordinates must eventually coincide. The minimum pairwise separation
is continuous: it starts above two and reaches zero at the first
coincidence. It must first equal two, when all sites are still distinct
and one pair fits a closed unit ball. A no-output path cannot contain
that promised yes-instance. Thus there are at least \(N!\)
no-output components. Ben-Or's component bound~\citep{benor1983lower}
gives height \(\Omega(\log(N!))-O(N)=\Omega(N\log N)\), charging
arithmetic operations and square roots. Sorting and a sliding window
attain this order in one dimension.
\end{proof}

\section{Related work and conclusions}\label{sec:context}
Quadratic time in the plane is not new. The question here is whether
one can return a point of maximum \emph{closed} depth in that time
without storing the arrangement.

\subsection{Exact algorithms, cuttings and approximations}
Counting candidate balls does not give the cost of finding the best one.
De Figueiredo and da Fonseca~\citep[Section~1.1]{defigueiredo2009enclosing}
describe an exact weighted approach that scores \(O(N^D)\) candidate
balls by halfspace range searching, in roughly \(O(N^{D+1/D})\) total
time.

For static ball MaxRS, Gusain, Rahul and Subramanian
\citep[Section~7, p.~19]{gusain2025bouquet} point to a
\(\widetilde O(N^D)\) time bound by computing the entire arrangement of
the site spheres. The covered labels are constant on each relatively
open face, but a vertex may cover more sites than any incident
full-dimensional cell. The notation hides polylogarithmic factors;
that passage specifies neither their exponent nor the workspace.

A different construction gives an explicit single logarithm. The
whole-rim sweep of Section~\ref{sec:depth} sorts \(O(m)\) contacts on
each of \(m\) planar circles in \(O(m^2\log(2+m))\) total time and
\(O(m)\) extra words. In \(D\ge3\), Section~\ref{sec:high} uses
\(O_D(N^{D-2})\) sections, each needing at most two planar calls on
\(m\le2N\) circles. Substituting the sweep gives
\(O_D(N^D\log(2+N))\) time and \(O_D(N)\) extra words; in \(D=2\)
we apply the sweep directly to the site circles. Theorem~\ref{thm:main}
removes this sorting logarithm.

The planar exponent for congruent disks is classical
\citep{bajaj1983duality,chazelle1986circle}. Chazelle and
Lee~\citep[Fig.~1 and Theorems~2 and~3]{chazelle1986circle} maintain
the graph of intersections and arcs; this can take quadratic space.
For a reverse-nearest-neighbor facility-location problem, Cabello et
al.~\citep[Theorem~1]{cabello2010facility} construct the arrangement
of the client disks and its dual. In \(O(n^2)\) time they obtain the
whole set of optimal facility locations, rather than a single witness.
For arbitrary radii in general position, another baseline follows
from Mulmuley's randomized incremental construction
\citep[Theorem~2]{mulmuley1991fast}. Splitting at vertical extrema gives
\(2m\) monotone arcs and \(K=O(m^2)\) intersections. The vertical
partition takes \(O(m\log m+K)\) expected time to construct. Starting
with the exterior weights on the unbounded face, crossing an arc
exchanges its two side weights; crossing an artificial wall changes
nothing. At an arc or crossing add the missing side weights to the
label of an incident face. This gives a deepest point in linear time
in the map size, but the map itself can occupy \(\Theta(m^2)\) space.

De, Nandy and Roy~\citep[Theorem~4]{de2014place} use \(O(1)\) extra
words by permuting the disk records; for read-only input a writable
copy costs \(O(m)\) words.
The edge parameter in Table~\ref{tab:prior} counts even containment,
so it bounds the number of proper crossings up to a constant. For
\(e=\Theta(N^2)\) their time bound retains a logarithmic factor.
Moreover, a closed exterior is the complement of an \emph{open} disk:
subtracting a closed-disk weight loses the rim. Cazals and
Loriot~\citep[Theorems~1 and~4]{cazals2009computing} also describe the
stereographic correspondence between spherical circles and planar
circles or lines. Their direct spherical sweep reports intersection
points in linear storage; their tree for recovering the balls covering
each face uses space proportional to the number of faces plus balls.
A contact can have greater closed depth than every incident open face.

\paragraph{What a sweep supplies.}
Edelsbrunner and Guibas~\citep{edelsbrunner1989topologically} visit a
\emph{line} arrangement in quadratic time and linear space; Rafalin,
Souvaine and Streinu~\citep{rafalin2002topological} handle parallel
and multiply concurrent lines. Two circles, in contrast, can cross
twice. Snoeyink and Hershberger~\citep[Theorem~3.1 and Section~6]{snoeyink1989sweeping}
prove sweep existence for pseudocircles whose intersections are
crossings. Their local moves can use an already constructed
arrangement; they left a space-efficient implementation open.
Dalal et al.~\citep{dalal2024sweeping} also prove sweep existence for
non-piercing regions, assuming transverse crossings and no three curves
through a point.

Balaban~\citep[Theorems~8 and~9, p.~219]{balaban1995optimal} reports
intersections in \(O(n\log n+K)\) time and \(O(n)\) workspace. His
extension to suitable \(x\)-monotone curves assumes distinct intersection
points; \(K\) counts those points. Reporting a pair of intersecting
arcs does not give the number of other disks containing their contact.
At a tangency, neither adjacent open gap need have the contact's score.

Wang~\citep[Appendix~A, Theorem~12]{wang2023unit} adapts Chazelle's
hierarchical hyperplane cutting~\citep{chazelle1993cutting} to arcs of
different radii. For conflict bound \(m/r\), the construction has
\(O(r^2)\) cells and takes \(O(mr)\) time; see also~\citep{wang2026counting}.
At \(r=m\) this already gives quadratic construction time, including
the conflict lists; the appendix assumes no three arcs meet at a point.
Our additional tasks are to account for the space live while a sample
is being built, to keep rims wholly inside a child on its list, and to
score tangencies and multiple contacts on the original input. We visit
one branch at a time (Appendix~\ProofRef{sec:construction}).

For approximate depth, Aronov and Har-Peled~\citep[Theorem~4.6]{aronov2008approximating}
obtain a factor \(1-\varepsilon\) for planar pseudodisks with high
probability in \(O(N\varepsilon^{-2}\log N)\) expected time. Their
regions are \(x\)-monotone, have constant algebraic complexity and
constant-time geometric primitives, and their boundaries cross at most
twice per pair.

There is also an exact formulation with a prescribed count: \(M(P)\ge t\)
exactly when the smallest ball enclosing at least \(t\) sites has radius
at most one. In the plane this leads to algorithms sensitive to \(t\)
\citep{efrat1994computing,matousek1995enclosing}; Har-Peled and
Mazumdar~\citep{harpeledmazumdar} find the smallest such disk in
\(O(Nt)\) expected time. The complementary parameter is the number of
sites one may miss. Matou{\v s}ek~\citep{matousek1995geometric} studied
geometric optimisation with a prescribed number of violated constraints,
including the smallest enclosing circle with outliers. Cavaleiro and
Alizadeh~\citep{cavaleiro2022branch} gave an exact branch-and-bound
method for the smallest ball enclosing a prescribed number of sites.
For a given miss budget \(k\), such a minimum radius is at most one
exactly when a unit ball covers at least \(N-k\) sites. In contrast,
Theorem~\ref{thm:missed} needs no budget: its deterministic time depends
on the number actually missed, and its workspace stays linear.

\subsection{Prior lower bounds}\label{sec:prior-lower}
The 3SUM problem asks whether three input numbers sum to zero.
De Berg, Cabello and Har-Peled~\citep[p.~448]{deberg2008covering}
report 3SUM-hardness for planar unit-disk placement; the lemma they
cite from Aronov and Har-Peled~\citep[Lemma~4.5]{aronov2008approximating}
is stated for disk arrangements. In that construction the two unweighted
disks forming each thin lens have the same radius, but a common radius
across pairs and rational target centres are not specified.
Keil, McLeod and Mondal~\citep[Section~1]{keil2024quantum} note an
adaptation using disks of one common radius. Our reduction builds those properties explicitly
in every fixed dimension. De Figueiredo and da Fonseca
\citep[Theorem~2.2]{defigueiredo2009enclosing} prove an unconditional
\(\Omega(N^D)\) bound for \emph{weight-oblivious} weighted-placement
algorithms, whose arithmetic on weights is chosen from the coordinates
alone. Section~\ref{sec:weighted-sites} attains this order within their
class. Their bound does not apply to an algorithm specialized to unit weights.

The classical reduction is \emph{from} 3SUM \emph{to} three-point
collinearity~\citep{gajentaan1995class}. Alman and Vassilevska
Williams~\citep[Theorems~22 and~35]{alman2026truly} give a deterministic
\(O(n^{1.9992})\) algorithm for polynomially bounded integer 3SUM
and a Las Vegas \(O(n^{1.998})\) expected-time algorithm for real
3SUM. Chan et al.~\citep{chan20263sum} subsequently derandomized real
3SUM. A faster source algorithm does not solve collinearity, but the old
3SUM conjecture no longer supports its quadratic exponent. Fischer's
\(k\)-term arithmetic-progression hypothesis restores several geometric
lower bounds; his paper explicitly excludes three-point
collinearity~\citep{fischer2026beyond}. The
\(\Omega(s^D)\) lower bound of Erickson and Seidel
\citep{erickson1995better,ericksonseidel1997erratum} concerns sidedness
queries: a decision tests the sign of the orientation determinant of
\(D+1\) input points. It neither proves integer real-RAM hardness nor
transfers that query bound to ball coverage.

\paragraph{Computational evidence.}
The companion software~\citep{petrucena2026coverage} includes the reference
implementation, independent exact tests and Lean proofs. The tests check
closed-depth witnesses and compare the two-disk encoding of
singleton tips with direct scoring at the original tips. The Lean
development covers the direct finite-candidate search and its use in
unweighted 3D placement with direct singleton tests. The two-call
transfer, the weighted placement corollary and the resource bounds of
the faster planar solver are mathematical proofs in this paper.

\subsection{Concluding remarks}\label{sec:checks}
Can one place the ball in \(O(N^D/\log N)\)
time on \emph{every} input while retaining linear workspace?
By Section~\ref{sec:high} it would suffice to save that factor for
closed planar disk and exterior depth. The 3D saving in
Theorem~\ref{thm:missed} depends on how many sites the optimal ball
misses. A saving on every input is compatible
with the hypothesis, and logarithmic savings are known for other
geometric 3SUM-hard problems~\citep{chan2020logshaving}. Conversely,
is there a natural query model for unweighted ball placement in which
one can prove an unconditional \(\Omega(N^D)\) lower bound?

\ifdefined\CompletePaper\else
\appendix
\section{Coordinates and correctness of the sphere searches}\label{sec:geometry}
The sphere search needs a projection that sends every regular
cap to a disk or its exterior. We first show how to find a suitable
pole. We then give the coordinates for the sections searched by
Algorithm~\ref{alg:place} and prove that one of them yields an optimum.
We then show how to search fewer spheres and circles in 3D when an
optimal ball misses few sites, and how to transfer weighted depth in
both directions between a two-sphere and the plane.

\subsection{Projecting caps without boundary lines}\label{sec:chart-details}
On the unit sphere \(a\cdot u\) ranges from \(-\norm a\) to \(\norm a\).
For \(a\ne0\) the cap \(a\cdot u\ge\beta\) is therefore empty if
\(\beta>\norm a\), full if \(\beta\le-\norm a\), and the single tip
\(a/\norm a\) if \(\beta=\norm a\). Only \(|\beta|<\norm a\) gives a
regular rim. The sign of \(\beta\) and a comparison of squares suffice;
for \(a=0\) just test \(0\ge\beta\).

Figure~\ref{fig:pole-sides} projects from south. Here we use north,
\((0,0,1)\), as the pole. Reflecting the sphere across its equatorial
plane interchanges these conventions: \(X,Y\) stay the same and the
third coordinate of the inverse map changes sign. The chart and its inverse are
\[
 x=(u_1,u_2)/(1-u_3),\qquad
 u=(2X,2Y,\|x\|^2-1)/(1+\|x\|^2).
\]
Substitution and multiplication by the positive denominator turn the
cap inequality into
\begin{equation}\label{eq:chart}
 (a_3-\beta)(X^2+Y^2)+2a_1X+2a_2Y-(a_3+\beta)\ge0 .
\end{equation}
For a regular cap with \(\kappa=a_3-\beta\ne0\), completing the square
rewrites the test as
\[
 \kappa\bigl(\|x-z\|^2-r^2\bigr)\ge0,\qquad
 z=-(a_1,a_2)/\kappa,\quad r^2=(\norm a^2-\beta^2)/\kappa^2>0 .
\]
Thus \(\kappa<0\) keeps the closed disk. When \(\kappa>0\), north
satisfies \(a_3>\beta\) and the image keeps the closed exterior.
If \(\kappa=0\), the rim passes through
north and appears as a line: regularity gives \((a_1,a_2)\ne(0,0)\).

If no rim projects to a line, keep north. Otherwise find a planar
point off every projected rim in two scans. First choose \(X_0\) strictly
to the right of every vertical boundary line. Then choose \(Y_0\)
above the top of every circle and above the height of every nonvertical
line at \(X_0\). A circle of centre height \(y_0\) and radius \(r_C\)
has top below \(y_0+1+r_C^2\), since \(1+r_C^2>r_C\); we need no
square roots. In each scan take the maximum with zero and add one.
The inverse image \(v\) of \((X_0,Y_0)\) is off all rims, so it is a
valid pole. Score \(v\) in the original coordinates.
To reuse the north-pole formula let \(H\) be
reflection in the plane through the origin with normal \(v-(0,0,1)\),
so \(Hv=(0,0,1)\). Apply~\eqref{eq:chart} with normal \(Ha\) and
unchanged threshold \(\beta\): the reflected coordinates are \(Hu\) and
their north pole is off every regular rim. Reflect the inverse chart
image back by \(H^{-1}=H\) to recover the winning sphere point.
Pole selection takes linear time and uses only rational arithmetic
on rational cap data.

\subsection{Computing a section and finding the next tight site}\label{sec:section-coordinates}
Fix \(p\in I\) and write \(c=p+u\). For each other site \(q\in I\)
put \(q-p\) in a row of \(M_I\) and \(\norm{q-p}^2/2\) in the
corresponding entry of \(b_I\). By equality in~\eqref{eq:anchor},
the section is given by \(M_Iu=b_I,\ \norm u=1\). The minimum-norm
solution of the linear system is \(h=0\) for \(|I|=1\), and otherwise
\(h=M_I^{\mathsf T}(M_IM_I^{\mathsf T})^{-1}b_I\). The rows are
independent, so the inverse exists. This \(h\) solves \(M_Ih=b_I\)
in the row space. Every solution is \(h+v\) with \(v\in\ker M_I\);
orthogonality gives \(\|h+v\|^2=\|h\|^2+\|v\|^2\), minimized at \(v=0\).
Thus the section is
\[
  S_I=\{p+h+v:v\in\ker M_I,\ \norm v^2=1-\norm h^2\}.
\]

Test \(1-\norm h^2\) before taking a square root. If it is negative,
the section is empty; if zero, its only centre is \(p+h\). Otherwise
let \(r=\sqrt{1-\norm h^2}>0\) and choose the columns of \(V\) as
an orthonormal basis of \(\ker M_I\). Project the ambient coordinate
vectors onto \(\ker M_I\), then apply Gram--Schmidt, omitting zero residuals; normalization
uses a square root. The section and the constraint for a site \(p+w\) are
\begin{align}
  c&=L_I(z)=p+h+rVz,\qquad z\in\R^{D-|I|+1},\quad\norm z=1,\label{eq:flat}\\
  (rV^{\mathsf T}w)\cdot z&\ge\norm w^2/2-w\cdot h.\label{eq:cap}
\end{align}
The kernel has dimension \(D-(|I|-1)\); its unit sphere has one
dimension less. When \(D-|I|>2\) take \(z=e_1\) as the representative.
If the section contains an optimum but the representative is not optimal,
the following argument preserves an optimum on a smaller section.

\begin{proof}[Proof of Lemma~\ref{lem:sections}]
Maintain an affinely independent set \(I\) of at most \(D-2\) sites
whose section contains a globally optimal centre \(c^*\).
Lemma~\ref{lem:tight} starts the induction with one site. An empty
section cannot satisfy the invariant. A point section is scored
directly; at size \(D-2\) a positive-radius section is searched by
Theorem~\ref{thm:sphere}. Otherwise the positive-radius sphere has
dimension greater than two. If its representative is optimal we are
done. If not, consider the closed subset of \(S_I\) covering every
site covered by \(c^*\). It contains \(c^*\) but not the representative.

The sphere is connected, so this nonempty proper closed subset has
a relative boundary point. The point retains every required site.
It lies on the boundary of at least one of their caps: otherwise it
would be interior to each cap, hence to the finite intersection.
A constraint with \(V^{\mathsf T}w=0\) is constant on the section by
\eqref{eq:cap}, so it has no boundary relative to that sphere. A cap
supplying a boundary point, even a singleton cap, must therefore have
\(V^{\mathsf T}w\ne0\). Thus \(w\) is outside the row span of \(M_I\).
At the boundary~\eqref{eq:cap} holds with equality, so \(p+w\)
is tight and \(I\cup\{p+w\}\) is affinely independent. The enlarged set still has
an optimal centre in its section. Each step increases \(|I|\) until
we reach an optimal representative, a point, or a two-sphere. The
section depends only on the set, not the order of additions; the
enumeration tries every set required by the induction.
\end{proof}

\subsection{Finding a second anchor on a sphere}\label{app:missed-anchors}
We prove Theorem~\ref{thm:missed}. Both stages ask the same question:
does any allowed centre cover the entire list? We need either such a
centre or at most four sites that no allowed centre covers together.

\paragraph{How to find a small infeasible group.}
In the first stage a centre can be anywhere in \(\R^3\). Compute the
smallest enclosing ball of the remaining sites, which takes deterministic
linear time in fixed dimension~\citep{dyer1986multidimensional,megiddo1989ball}.
If its radius is at most one, its centre works. Otherwise we need a
small group to set aside, not just a report that the sites do not fit.
Helly's theorem gives such a group of at most four labels.

Work on a copy of the infeasible list. While it has more than four
labels, divide it into five blocks whose sizes differ by at most one.
An infeasible group of at most four labels can meet at most four of
the blocks, so omitting one of the five preserves infeasibility.
Try the five complements and keep the first infeasible one. Each round costs linear time in the current size
and shrinks it by a constant fraction. The costs sum to \(O(m)\) on a
list of size \(m\). Only the final group of at most four is set aside
from the original list. This procedure will work on an anchor sphere
too, once we have a linear-time feasibility test and the same four-label
guarantee there.

\paragraph{Can the sites be covered with \(p\) tight?}
Write the centre as \(p+u\), with \(\norm u=1\). The anchor \(p\)
always counts. For any other site \(q\), expansion of its squared
distance gives
\[
    \norm{p+u-q}\le1
    \quad\Longleftrightarrow\quad a_q\cdot u\ge1,
    \qquad a_q=\frac{2(q-p)}{\norm{q-p}^2}.
\]
Distinct sites make the denominator nonzero. Temporarily allow
\(\norm u\le1\): any vector satisfying a nonempty list of these
inequalities is nonzero, and scaling it to unit length preserves them
because their right side is positive. Each constraint in the solid
unit ball is convex. By Helly's theorem, if there is no common
direction, at most four sites already have no common direction.
An empty list admits any unit direction.

To test the dot products in linear time, we turn them into distances.
If all the vectors have the same length, the squared distance from a
unit vector depends only on the dot product. One extra coordinate
makes the lengths equal. Choose \(H=\max(1,\max_q\norm{a_q}^2)\) for
the list being tested and give each \(a_q\) two opposite heights:
\[
       \bigl(a_q,\ \pm\sqrt{H-\norm{a_q}^2}\bigr)\ \in\R^4 .
\]
They all have squared norm \(H\); our choice \(H\ge1\) makes \(H-1\)
a valid squared radius. A feasible unit direction \(u\)
puts \((u,0)\) within squared distance \(H-1\) of every lifted point,
since that squared distance is \(1+H-2a_q\cdot u\).
Conversely, suppose the smallest enclosing ball of the lifted points
has squared radius at most \(H-1\) and centre \((x,t)\). For a pair
at heights \(\pm h\), the greater vertical separation is \(h+|t|\);
setting \(t=0\) reduces it to \(h\). Thus \((x,0)\) still satisfies
the bound for every point:
\[
       \norm{x}^2+H-2a_q\cdot x\le H-1
       \quad\Longrightarrow\quad
       2a_q\cdot x\ge \norm{x}^2+1\ge2\norm{x}.
\]
Here \(x\ne0\), so \(x/\norm{x}\) is a feasible direction. This test
takes \(O(m)\) time on \(m\ge1\) constraints, including lifts and square
roots. Applying the five-block procedure to the original site labels
now finds an infeasible group within the same bound.

\paragraph{Which second sites must be tried?}
Recall that \(c_0\) misses \(b\) sites globally. On the sphere of a
first anchor \(p\), repeatedly set aside an infeasible group of caps.
If \(b+1\) disjoint groups are found, every centre on this sphere
misses at least \(b+1\) sites. Discard the sphere: \(c_0\) is better.
Otherwise we obtain a direction \(u_0\) covering the remaining caps.
Score \(p+u_0\) on the original input. It misses at most \(4b\) sites.

If \(u_0\) is not best on the sphere, take a best direction \(u_*\).
We need a path to \(u_0\) that stays on the sphere and preserves every
cap containing both ends. Unless the directions are opposite, take the
straight segment between them and normalise each of its vectors to unit
length. A vector \(v\) on this segment has \(0<\norm v\le1\) and
satisfies every common inequality; hence
\(a_q\cdot(v/\norm v)\ge1/\norm v\ge1\).
Opposite directions cannot both satisfy \(a_q\cdot u\ge1\): adding
their inequalities would give \(0\ge2\). A semicircle in any plane through their
axis works. Consider the sites covered at \(u_*\) but missed at \(u_0\);
there is at least one because \(u_*\) is better. Along the path each of
their dot products starts at or above one and ends below one. Stop the
first time any of them equals one, even if that is the starting point.
None has yet dropped below one, and all common caps still count. Thus
this centre is best on the sphere and has a second tight site among the
misses at \(u_0\).

The references alone can miss the optimum. With sites \(A=(0,0,0)\),
\(B=(2,0,0)\), \(C=(1,3/2,0)\), covering \(A,B\) forces their midpoint,
which misses \(C\). We may set aside the triple and take \(c_0=A\);
on either missed site's sphere the other two caps are disjoint.
After setting them aside, a reference one unit above the anchor covers
only that site, but the pair search on \(B\)'s sphere with \(A\) tight
recovers depth two.

\paragraph{Search one circle and add the costs.}
Subtracting the two tight-site equations gives
\((c-z)\cdot(q-p)=0\), where \(z=(p+q)/2\). Thus
\(\|c-p\|^2=\|c-z\|^2+\|p-q\|^2/4=1\): at distance greater
than two there is no centre, at two only \(z\), and otherwise a circle
of radius \(r=\sqrt{1-\|p-q\|^2/4}\) in the perpendicular bisector plane.
Write \(c=z+rv\) with \(v\) a unit vector parallel to that plane.
Expanding \(\|rv-(s-z)\|^2\le1\) gives the test for site \(s\):
\[
        (s-z)\cdot v\ \ge\ (\norm{s-z}^2+r^2-1)/(2r).
\]
In an orthonormal frame of the plane this is a halfplane test on a unit
circle. It gives the whole circle, nothing, a one-point arc, or a proper closed
arc. For a proper arc intersect the boundary line with the circle and
keep the arc containing the unit direction of the inward normal; it can
be longer than a semicircle. Order the rays counterclockwise from the
positive horizontal ray. Sort by half-circle (the upper half includes
the positive ray, the lower half the negative ray), then by determinant.
Just before the first endpoint, count the whole-circle sites and the
proper arcs whose end precedes their start: exactly those arcs wrap
around the start of the order.
At each tied direction the preceding count already includes ending arcs:
add starts and one-point arcs, score the point, then remove ends and
one-point arcs. Closed arcs give each endpoint at least the score of the
preceding gap. If there are no events, score any point. The sweep
returns a best centre in \(O(N\log(2+N))\) time and \(O(N)\) extra words.

There are \(b\) first anchors, at most \(b+1\) linear-time rounds on
each sphere, and at most \(4b^2\) pair searches in total. Compare
\((b+1)\lceil\log_2(2+N)\rceil\) with \(N\), computing the integer
ceiling by doubling. Use the pair searches if the first quantity is
smaller; otherwise search each whole sphere by Theorem~\ref{thm:sphere}.
Since \(k\le b\le4k\), these costs, including construction of \(c_0\),
give the minimum in Theorem~\ref{thm:missed} up to a constant factor.
Process one sphere and one circle at a time, retaining the best centre
and count. A final scan returns its original site indices.

\subsection{Reducing cap depth to planar depth and back}\label{app:cap-transfer}
\begin{proof}[Proof of Theorem~\ref{thm:sphere}]
Translate and scale the given two-sphere to the unit sphere in its
three-dimensional affine span.
Classify its caps as in Appendix~\ref{sec:chart-details}. Full caps
contribute a constant \(F\); empty caps contribute nothing. Choose a
pole off all regular rims and score it against every original cap.
Project the regular caps, sorting equal circles together and summing
their weights on each closed side separately. Let \(R(x)\) be the
resulting regular-cap weight at a finite planar point \(x\).
The first planar call returns a point maximizing \(R\); if there are
no regular caps, use any finite point.

\emph{Force the second search to choose a singleton.}
Project each singleton tip other than the pole. Group equal projected
tips, writing \(t_i\) for each distinct point and \(g_i\) for the sum
of its singleton weights. If there are no such tips, skip this call.
Sort the tips by first coordinate and, within each equal-first-coordinate
group, by second coordinate. Take the least of the gaps between successive
first coordinates and the gaps between successive second coordinates
within a group; call it \(\delta\), using \(1\) for a lone tip. Distinct
tips differ by at least \(\delta\) in the first coordinate if their
groups differ, and in the second otherwise. Thus their Euclidean
distance is at least \(\delta\).

At each \(t_i\) add two closed disks of radius \(r=\delta/8\), centred at
\(t_i\pm(r,0)\). Their centres are \(2r\) apart, so their only common
point is \(t_i\). Each pair lies within \(2r\) of its tip, and \(4r<\delta\);
different pairs are disjoint. Let \(W\) be the
sum of all original cap weights and put \(A=W+1\). Give \emph{each}
new disk inside weight \(A+g_i/2\) and outside weight zero. Keep the
regular records in this call. A tip earns both disk weights; any other
point earns at most one. The two scores are separated by a strict gap:
\[
 \begin{aligned}
 \text{at }t_i:\quad &R(t_i)+2(A+g_i/2)=2A+R(t_i)+g_i\ \ge 2A,\\
 \text{in pair }i\text{ off }t_i:\quad &R(x)+A+g_i/2\ \le A+W<2A .
 \end{aligned}
\]
Indeed \(R(x)+g_i\le W\): the regular and singleton labels are
disjoint and all weights are nonnegative. Outside every pair the score
is just \(R(x)\le W\). Thus \emph{every} winner of the second call is
a tip, and among tips the common bonus \(2A\) leaves their order unchanged.

Lift both returned points, when present, and compare them and the pole
using the original caps. Away from the pole and all tips the true score
is \(F+R(x)\), which the first winner matches or exceeds. At a finite tip
it is \(F+R(t_i)+g_i\), whose maximum the second call attains.
A final scan reports every containing label, including repeats of the
same cap and labels of weight zero. Each call has at most \(2m\)
records before grouping. If an added circle has the same rim as a
regular circle, add the weights on each side separately as before.
All sorting and scanning use
\(O(m\log(2+m))\) operations and \(O(m)\) words beyond the calls.

\emph{From the plane back to the sphere.}
For a planar circle of centre \(c\) and radius \(r>0\), put
\(h=\|c\|^2-r^2\). In the north-pole chart \(x=(u_1,u_2)/(1-u_3)\),
we have \(\|x\|^2=(1-u_3^2)/(1-u_3)^2=(1+u_3)/(1-u_3)\) off north.
Expanding the planar test gives an affine function of \(u\):
\[
 \begin{aligned}
 (1-u_3)(\|x-c\|^2-r^2)
 &=1+u_3-2c\cdot(u_1,u_2)+h(1-u_3)\\
 &=1+h+(1-h)u_3-2c\cdot(u_1,u_2)=:L(u).
 \end{aligned}
\]
Off north, \(1-u_3>0\). Use the two closed halfspace caps \(L(u)\le0\)
and \(L(u)\ge0\) with the disk and exterior weights, respectively.
Both count on the rim. At north, \(L(0,0,1)=1+h+(1-h)=2\) for
every record. If the spherical call returns north, choose a finite
point strictly to the right of every disk: it too earns every exterior
weight and no disk weight. Otherwise project the spherical winner. Forming
the caps and recovering the planar point and its score take \(O(m)\)
time and extra words.
\end{proof}

\section{Constructing the dense cutting without general position}
\label{sec:construction}
Lemma~\ref{lem:cut} promises both short conflict lists and few cells.
The first property alone is easy to arrange: draw every circle and the
open children have no conflicts. But constructing that subdivision
inside every parent would not give the time bound. We follow the
hierarchical cuttings of Chazelle~\citep{chazelle1993cutting} and
Wang's adaptation to circular arcs~\citep[Appendix~A]{wang2023unit}.
When sampled rims meet inside a parent, even at a tangency, we keep
their contact as a vertex: it could be the only winning point. We
also need to visit one branch at a time. We first describe how to draw a split
once its sample of circles is known. To choose that sample, we
compress the parent's circle list while approximating the fraction of
circles that meet each possible child. We can then select circles
from the short list using the sampling techniques of Matou{\v s}ek
and Chazelle--Matou{\v s}ek
\citep{matousek1995approximations,chazellematousek1996derandomization}.

\subsection{What a split must achieve}\label{sec:localcut}
Let \(m\) be the total number of distinct circle boundaries, and let
\(F_\sigma\) be the list of the \(k\) circles meeting an open parent
\(\sigma\). For a subset \(S\) of that list, write \(P_\sigma(S)\) for
the number of unordered pairs in \(S\) that meet \emph{inside} \(\sigma\).
Two crossings of the same pair count once. A contact only on the parent
boundary does not count.

Suppose the sample \(\mathcal H\subseteq F_\sigma\) has already been
chosen. Clip each of its circles to the
open parent, retaining the endpoints in its closure. Draw vertical
walls from contacts among the sampled arcs, their clipped endpoints
and their vertical tangencies. Mark these points; each vertical line
through one of them is an event column. Unsampled rims do not stop a
wall. On each event column list the heights of the drawn arcs and of
the top and bottom of the parent, grouping ties. From a marked point extend a vertical segment
in each direction that enters the parent. Stop at the next distinct
listed height, even if it too is marked; a zero-length segment creates
no wall. Neither vertical direction enters the open parent from its
vertical end side. Merge overlapping walls and cut every incident
arc at each internal wall endpoint. In particular, the
ends of a clipped arc on a curved parent side can send walls inward
(Figure~\ref{fig:wall}(b)); the endpoints themselves are not scored
again inside this open parent. The vertical cuts separate places where
the bounding graphs change. Stopping at the next arc leaves bands beyond
it joined whenever their two bounds continue unchanged.

\begin{lemma}[Shape and size of a stopped-wall split]\label{lem:stopped-shape}
Suppose the parent has the form
\(\{(x,y):a<x<b,\ g_-(x)<y<g_+(x)\}\), with each bounding graph
horizontal or a single upper or lower circle branch. Every open child has
this form. The children, new open edges and new vertices together number
\(O(1+|\mathcal H|+P_\sigma(\mathcal H))\), with an absolute constant.
\end{lemma}
\begin{proof}
The empty sample leaves the parent unchanged. Otherwise put
\(s=|\mathcal H|,\ P=P_\sigma(\mathcal H)\). Each sampled circle has
at most eight intersections with the four parent sides and two vertical
extrema. If it supports a side, discard the coincident portion; only
its ends can matter there. At an interior meeting of \(v\ge2\) circles,
the number of circle incidences is \(v\le2\binom v2\), and a pair has
at most two contacts. Thus at most \(K=10s+4P\) marked incidences suffice.
Cutting the circles at these marks leaves at most \(s+K\) retained arcs.
There are at most \(2K\) stopped walls; their interiors meet no arc,
and walls on one column occupy disjoint gaps after merging. Starting
with one face, inserting an arc or a wall creates at most one more.
Consequently there are at most \(1+s+3K\le32(1+s+P)\) open children.
Wall stops split at most \(2K\) additional arcs. All new edges and
vertices together therefore also number \(O(1+s+P)\).

For the shape assertion, temporarily slice at all event columns. Mark
every retained vertical tangency (both on a full circle) and every
clipping endpoint. Between columns the branches present have a fixed
vertical order, so each band has one lower and one upper graph.
On an internal event column consider the interval between two
consecutive distinct heights. If either end is marked, a stopped wall
fills the interval. Otherwise both bounds continue unchanged to either
side: a branch starting, ending or meeting another there would give a
mark. No intervening branch can limit to the interior of this interval,
since its height would have been listed. Thus an unblocked interval
joins exactly one band on each side with the same two bounds. A gap
that closes at a tangency joins none, even if it reopens to the right.

A band can consequently continue at most once to the left and once
to the right: an interior mark in its limiting vertical interval
blocks the adjacent gaps, and any unmarked internal height would
already separate bands in the slab. The joined bands form chains in
increasing \(x\), not branching components. Each chain and its
unblocked fibres give exactly one open child with fixed bounding
graphs; an end fibre in its closure may collapse to a point. The
rectangle starts the induction. For bounded \(|\mathcal H|\), even
constructing all temporary slabs and then joining the bands costs
constant time by circle and line intersections and exact comparisons.
\end{proof}

We can now say what to look for when choosing \(\mathcal H\). First,
if many rims meet a possible child, at least one of those rims should
be selected. No selected rim meets an actual open child, so every
actual child will have a short conflict list. Second, the size of
the drawing depends on \(P_\sigma(\mathcal H)\): where few parent pairs
meet, few selected pairs should meet. The next two tests correspond to
these two requirements. Neither uses the coverage weights.

\subsection{Two geometric tests and their finite descriptions}
For any cell \(T\) as in Section~\ref{sec:depth} and any circle \(q\),
define two subsets of the input boundaries:
\[
 \mathcal U_T=\{C:C\cap T\ne\varnothing\},\qquad
 \mathcal V_{q,T}=\{C:C\ne q,\ C\cap q\cap T\ne\varnothing\}.
\]
The first test counts a circle wholly inside \(T\); the second counts a
tangency in \(T\), but not a contact confined to its boundary. The
parameters of \(T\) and \(q\) need not come from the input.

To test whether \(C\in\mathcal U_T\), let \(z_C\) and \(r_C>0\) be
the centre and radius of \(C\), and put
\(f(x)=\norm{x-z_C}^2\). Unlike the closed coverage test, the
conflict test asks whether the \emph{rim} reaches the \emph{open}
cell. The range of squared distances on its closure must straddle
the squared radius strictly:
\begin{equation}\label{eq:extrema}
 C\in\mathcal U_T\quad\Longleftrightarrow\quad
       \min_{\overline T} f <r_C^2<\max_{\overline T} f .
\end{equation}
Indeed, at a point of \(C\) in the open cell one can move a little
towards and away from \(z_C\). Conversely, strict inequalities on the
closure give points on both sides in \(T\) by density, and \(T\) is
connected. Thus~\eqref{eq:extrema} includes the internal circle in
Figure~\ref{fig:wall}(a) and excludes the exterior tangency.

There is a constant list of candidates for these extrema. Use the
corners; the perpendicular foot from \(z_C\) on each straight side if it
belongs to that segment; and \(z_C\) itself if it is in the closure.
On a curved side supported by a circle of centre \(z_S\ne z_C\), also test
the intersections with the line \(z_Sz_C\) that lie on the side, if
any. These are its only possible interior stationary points, since
there the tangent is perpendicular to \(x-z_C\). When \(z_S=z_C\) the
distance is constant on the arc. This covers collapsed vertical sides
as well. For \(\mathcal V_{q,T}\), equal or concentric circles contribute
nothing; otherwise test each of the at most two circle intersections
against the strict cell bounds.

On a list \(L=(C_1,\ldots,C_\ell)\), the \emph{trace} of a query is
the set of list positions it selects. A sample that works for a larger
family of traces still works for every geometric query. We can therefore
list sign patterns rather than geometric objects, provided a query on
a merged list restricts to the same query on either sublist.

\begin{lemma}[Hereditary trace enumeration]\label{lem:trace-catalogue}
Let the ground set consist of positive-radius circle boundaries, and
let the ranges be all \(\mathcal U_T,\mathcal V_{q,T}\) above. There
is a fixed enlarged family of ranges with the following properties.
It contains both geometric families and has shatter function at most
\(A(1+\ell)^h\): at most this many distinct traces occur on any list
of \(\ell\) slots, including repeats. In deterministic
\((1+\ell)^{O(1)}\) real-RAM operations one can list all its traces.
Deleting slots restricts each listed trace to a listed trace on the
remaining slots. The constants \(A,h\) are independent of the input,
the cell and the splitting parameter \(\rho\).
\end{lemma}
\begin{proof}
Fix the test and the types of the lower and upper sides (horizontal,
upper branch or lower branch). The query has at most eleven scalars:
the two end \(x\)-coordinates, three parameters per supporting curve and, for
\(\mathcal V\), three for \(q\). Store squared radii. There are infinitely
many queries, but this fixed number of parameters will bound the number
of distinct sign patterns on a list, independently of tree depth.

Here are the sign tests explicitly. Write
\(\operatorname{pow}_s(z)=\|z-s\|^2-R_s\) for the power with respect to the circle
with centre \(s\) and squared radius \(R_s\).
At an \(x\)-coordinate where the branch is defined, a point \(z\) is strictly
below the upper branch if \(z_y<s_y\) or \(\operatorname{pow}_s(z)<0\);
it is strictly below the lower branch if \(z_y<s_y\) and \(\operatorname{pow}_s(z)>0\).
Reflection gives the tests for ``above''; replacing strict comparisons
by weak ones gives the closure tests. Together with the interval in
\(x\) these decide membership without a second square root.

For \(\mathcal U\), test whether powers of both strict signs occur among
the admissible extremum candidates above. Each candidate has the form
\(z=(P+Q\sqrt K)/S\), where \(P,Q\) are two-vectors and \(S>0\).
Corners on a circle use \(K=R_s-(x-s_x)^2\); straight-side feet and
the input centre \(c\) require no root. The radial candidates on a
curved side are \(s\pm(c-s)\sqrt{R_s d}/d\) with \(d=\|c-s\|^2>0\).
For \(\mathcal V\), set \(w=c_q-c,\ d=\|w\|^2\). If \(d=0\) return
false, including when the circles coincide. Otherwise put
\(n=d+R_c-R_q,\ K=4dR_c-n^2\). Return true exactly when \(K\ge0\)
and one of the two points
\[
             c+(n w\pm\sqrt K\,(-w_y,w_x))/(2d)
\]
belongs to the open cell. This includes the single contact at \(K=0\).

In all cases the degrees of \(P,Q,K,S\) are at most \(3,1,4,2\),
respectively. Thus a coordinate comparison or a power test, after
clearing \(S\) or \(S^2\), asks for the sign of \(A_0+B_0\sqrt K\)
with degrees at most \(6,4\) for \(A_0,B_0\). Use the signs of the
two summands; if opposed, compare \(A_0^2\) with \(B_0^2K\).
Zero summands and a zero radicand are tested first. The resulting
polynomial signs have degree at most twelve. There are at most eleven
candidates for either test: four corners, two vertical feet, at most two
stationary points on each nonvertical side, and \(c\) for \(\mathcal U\);
just two points for \(\mathcal V\). Per candidate there are at most seven
coordinate or power comparisons (two bounds on \(x\), two comparisons
per graph and one input power). Each needs at most three polynomial
signs after the common root guard; even allowing four additional guards
per candidate, \(s_0=512\) signs per slot suffice.

We \emph{lift} the query by recording each nonconstant monomial in its
parameters as a coordinate of \(z\in\R^h\), where
\(h=\binom{11+12}{12}-1\) suffices. After an input
circle is fixed, each polynomial sign is now the sign of an affine
form in \(z\)~\citep{agarwalmatousek1994ranges}. The enlarged query
is the type and \emph{one} vector \(z\) for all slots. The lift itself
preserves the geometric trace. We enlarge the family by allowing the
coordinates of \(z\) to vary independently. Restricting slots changes
neither the type nor \(z\).

Put \(n=s_0\ell\). The number of
feasible ternary sign patterns is at most
\(\sum_{j=0}^h 2^j\binom{n}{j}\), including lower-dimensional faces.
Indeed, inserting a hyperplane splits each face it cuts into three.
The cut faces inject into the faces on the inserted hyperplane, giving
the recurrence \(F(n,h)\le F(n-1,h)+2F(n-1,h-1)\); constant or
repeated forms cannot increase the bound. Summing over finitely many
types gives the claimed polynomial, for example with
\(A=18(1+2s_0)^h\), even if one emits one mask per face without removing
duplicate masks. These deliberately loose constants are for the bound.

Enumerate feasible sign prefixes, trying \(-,0,+\) for each next
form. Test a prefix by Fourier--Motzkin elimination: express equality
as two weak inequalities and mark a combined lower/upper bound strict
if either parent bound is strict. In fixed dimension at most \(h\)
eliminations are needed, with at most \(O_h((1+n)^{2^h})\) inequalities
in any one feasibility test. The face bound holds for every prefix
length, so the total work is polynomial. Evaluate the Boolean test at the signs of each complete
pattern. No witness realizing an extra mask as a geometric query is needed.
\end{proof}

\ifIllustrated
Figure~\ref{fig:lifted-traces} illustrates the enlargement on just two
input rims. Take a cell containing both rims. For a query circle of
radius \(1/2\) centred at \((u,0)\) to meet a unit rim centred at
\((a,0)\), the centre distance must be between the difference and the
sum of the radii:
\[
                 \tfrac14\le (u-a)^2\le\tfrac94 .
\]
Use \(a=0,4\) for the two rims. Replacing \(u^2\) by one shared
coordinate \(v\) turns the two quadratic tests into strips.
Their overlap adds the trace \(\{A,B\}\). The two slots must use
the \emph{same} \((u,v)\), even when \(v\ne u^2\).

\begin{figure}[!htb]
\centering
\includegraphics[width=\ProofPlateWidth]{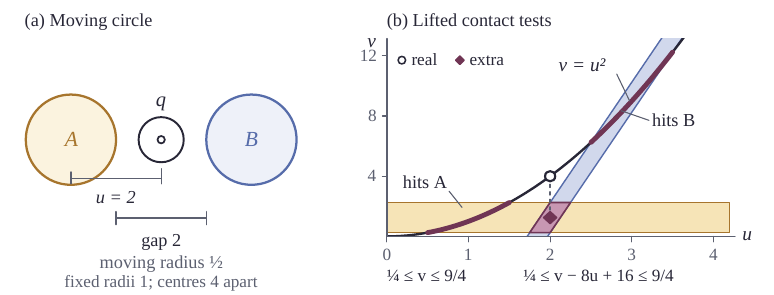}
\caption{Which rims does the moving circle \(q\) meet?
(a) The fixed unit circles \(A,B\) have centres four units apart.
The diameter of \(q\) is one, smaller than the gap of two between
their rims. (b) A real placement has \(v=u^2\). The two heavy arcs
mark separate hits on \(A\) and on \(B\); at the open point \((2,4)\)
neither is hit. Allowing \(v\) to vary independently admits the
diamond \((2,5/4)\) in both strips, a pair of hits no placement of
\(q\) realizes}\label{fig:lifted-traces}
\end{figure}

\FloatBarrier
\else
For example, let \(q\) have centre \((u,0)\) and radius one, and choose
\(T\) to contain every input rim. For an input circle with centre
\((a,0)\) and radius \(r\), contact is equivalent, apart from identical
circles, to
\((r-1)^2\le u^2-2au+a^2\le(r+1)^2\).
Replacing \(u^2\) by \(v\) gives two affine comparisons in \((u,v)\).
Points off the parabola may add traces. Giving each circle its own
\((u,v)\), in contrast, would destroy the polynomial bound.
\fi

\subsection{An \texorpdfstring{\(\eta\)}{eta}-approximation by merge-reduce}
Listing all traces on a large parent could already cost more than the
entire search. Instead, make a small list \(\mathcal A\) that preserves the
\emph{fractions} of positions selected by the tests. Only small
intermediate lists need trace enumeration. Pad to the next power of
two by duplicating distinct entries once as needed; this changes
sampling mass, not coverage weight. The proof budgets half the error
for the merge tree and half for shortening its root list. At a merge,
equal-sized siblings' frequency errors average; only halving adds error.

\begin{lemma}[Local approximation]\label{lem:sample}
For \(k\ge1\) distinct circle boundaries and \(0<\eta<1/2\), one can
find a multiset \(\mathcal A\) of \(O(\eta^{-4})\) entries such that the
following holds. Let \(\mu\) be uniform on the padded list and \(\nu\)
uniform on \(\mathcal A\). For every range \(E\) of either linearized family,
\(\lvert\nu(E)-\mu(E)\rvert\le\eta\), where the measures denote the
fractions of slots hit. Each original circle has \(\mu\)-mass between
\(1/(2k)\) and \(2/k\). Thus \(\mathcal A\) is an \(\eta\)-approximation
of the padded multiset. For fixed \(\eta\) this takes \(O(k\log^\beta k)\)
time for \(k\ge2\), and constant time for \(k=1\), for an absolute exponent \(\beta\).
\end{lemma}
\begin{proof}
A singleton already has the desired distribution. Otherwise the padded
list has fewer than \(2k\) entries and each boundary occurs once or
twice, giving the mass bounds.

The basic operation is to halve a list. Above a fixed cutoff
\(\ell_0\), one can halve an even list of size \(\ell\) while changing every range
frequency by at most \(\ell^{-1/4}\). Pair the positions and retain
one from each pair. For one trace suppose
\(M\le\ell/2\) pairs are mixed,
meaning exactly one member is marked. If \(X\) of those mixed pairs
retain their marked member, the change in hit fraction is
\((2X-M)/\ell\). Under independent fair choices the probability that its
absolute value exceeds \(\ell^{-1/4}\) is at most \(2\exp(-\sqrt{\ell})\)
by Hoeffding's bound (\(M=0\) never fails). Lemma~\ref{lem:trace-catalogue}
gives at most \(A(1+\ell)^h\) masks, with repetitions allowed.
Thus above a fixed cutoff \(\ell_0\) the expected number of failed
masks is less than one. For an effective choice on the dyadic lengths
used here, take \(\ell_0=4^v\) with \(2^v\ge4(h+1)\) and
\(2A(1+\ell_0)^h<2^{2^v}\), found by trying successive powers of
four. At each subsequent
doubling the polynomial grows by at most \(2^h\), whereas
\(2^{\sqrt{\ell}}\) grows by at least that factor.
Make the half deterministic
by choosing from each pair to minimize the remaining conditional
expectation of the number of failed traces. The two choices average to
the old expectation; after all choices, the failure count is an integer
less than one. With \(a_0\) hits already fixed in mixed pairs and \(b_0\)
mixed pairs unset, the exact conditional failure probability is
\[
    2^{-b_0}\sum_{\substack{0\le i\le b_0\\ |2(a_0+i)-M|^4>\ell^3}}\binom{b_0}i .
\]
Pascal's recurrence computes these counts in polynomial time in \(\ell\);
the exponential tail estimate is never evaluated by the algorithm.

We apply this halving step in a balanced tree of height \(H\) to
summarize the padded list with total error at most \(\eta/2\).
Use one padded entry per leaf and concatenate the two summaries at
each merge. Choose a capacity \(B_{\rm cap}=O((H/\eta)^4)\), a power of
two at least \(\ell_0\) and \((2H/\eta)^4\); halve a list when its
size reaches \(2B_{\rm cap}\). Siblings represent equal numbers of original
slots and have retained lists of equal length. For one lifted query
restricted to the two lists, merging averages their frequency errors;
halving adds at
most \(B_{\rm cap}^{-1/4}\). There are at most \(H\) such increments on any path.

Finally choose a power of two \(L_*=O(\eta^{-4})\) with
\(L_*\ge\ell_0\) and \(\sum_{i\ge1}(2^i L_*)^{-1/4}\le\eta/2\);
choose \(B_{\rm cap}\ge L_*\) above. Halve the root list while its size exceeds
\(L_*\). This last geometric sum explains the fourth power of \(1/\eta\);
we need no optimal dependence on this fixed parameter. We obtain the
claimed list \(\mathcal A\). If trace enumeration and halving on
\(\ell\) slots take \(O((1+\ell)^b)\) operations, the \(O(k)\) merges
cost \(O_\eta(k(1+\log k)^{4b})\): no trace catalogue is built on the
whole parent unless the parent itself is small. In postorder at most
\(H\) waiting summaries and one polynomial-size halving computation are
live. For fixed \(\eta\) this takes \(\operatorname{polylog}(k+1)=O(k)\)
extra words: the exponent from trace enumeration is large, but independent
of \(k\). Each binomial count uses one exact real-RAM word (at most
\(\ell+1\) bits in binary); the geometric bound is in words.
\end{proof}

\subsection{Turning the compressed list into a split}
Fix an integer \(\rho>1\), put \(\eta=1/(16\rho)\), and choose the draw
budget as follows. If \(L_*\) is the final capacity in Lemma~\ref{lem:sample},
take \(t=4\rho j\), where \(j\) is the first integer with
\(2^j>4A(1+L_*)^h\). Doubling computes \(j\) without logarithms or
floor. Since \(L_*=O(\rho^4)\), \(t=O(\rho\log(2\rho))\) with an
absolute constant. Write \(P_\sigma=P_\sigma(F_\sigma)\).
\begin{lemma}[Local split]\label{lem:split}
For a parent with \(k\ge2\) conflicts, one can select at most \(t\)
circles whose subdivision leaves fewer than \(k/\rho\) conflicts in
each open child. The total number of children, new edges and vertices is
\[
 O(1+t+t^2\eta+t^2P_\sigma/k^2)
\]
with an \emph{absolute} constant. Construction takes \(O(k\log^\beta k)\)
time and \(O(k)\) extra memory for fixed \(\rho\); the constants in these
resource bounds may depend on \(\rho\). An empty list is carried unchanged;
for a singleton, draw its circle in constant time.
\end{lemma}
\begin{proof}
Let \(\chi_\sigma(u,v)=1\)
if distinct boundaries \(u,v\) meet in the open parent, and zero
otherwise, in particular if the same circle is drawn twice. For a
distribution \(\pi\) on circles, put
\(\Delta(\pi)=\tfrac12\mathbb E_{\pi\times\pi}\chi_\sigma\).
The half compensates for the two orders of each pair.

Use Lemma~\ref{lem:sample} with \(\eta=1/(16\rho)\). It estimates
one-circle queries, yet we need information about \emph{pairs}. Fix
the first draw at \(u\): the possible second draws with \(\chi_\sigma(u,v)=1\)
form precisely \(\mathcal V_{u,\sigma}\). Replacing the second draw from \(\mu\)
by one from \(\nu\) thus changes its hit probability by at most \(\eta\).
This replaces \(\mu\times\mu\) by \(\mu\times\nu\) after averaging
over \(u\). By symmetry, replacing \(\mu\times\nu\) by \(\nu\times\nu\)
changes the contact probability by at most another \(\eta\).
The factor of a half in \(\Delta\) gives
\(\lvert\Delta(\nu)-\Delta(\mu)\rvert\le\eta\), so
\[
                  \Delta(\nu)\le4P_\sigma/k^2+\eta .
\]
Here \(\Delta(\mu)\le4P_\sigma/k^2\) since neither member of a pair has
mass above \(2/k\). We can compute \(\Delta(\nu)\) on the small list
\(\mathcal A\); the algorithm never computes \(P_\sigma\) on the parent.

Choose \(t\) slots of \(\mathcal A\) with
replacement. We want a \(1/(4\rho)\)-net~\citep{hausslerwelzl1987nets}:
hit every listed cell trace of \(\nu\)-mass greater than \(1/(4\rho)\).
At the same time we must keep the
number \(Z\) of contacting pairs of \emph{draw indices} small:
two draws of the same boundary contribute zero.
Measure both failures by the single quantity
\[
          \#\{\text{unhit heavy traces}\}+\frac{Z}{4t^2\Delta(\nu)},
\]
omitting the fraction when \(\Delta(\nu)=0\): then no pair of entries
of \(\mathcal A\) meets inside \(\sigma\), so \(Z=0\). Initially imagine
independent draws. The expected first term is at most the number of
listed traces times \((1-1/(4\rho))^t\), below \(1/4\) by the choice of
\(j\), since \((1-1/(4\rho))^{4\rho}<1/2\). The second has expectation
at most \(1/4\) since \(\mathbb E Z=t(t-1)\Delta(\nu)\).

The draws describe how to average, not a source of randomness for
the algorithm. At each step try the possible next entries and choose
one minimizing the expected final penalty given that choice and the
prefix already fixed. For
a prefix \(u_1,\ldots,u_i\), put \(n_*=t-i\). An unhit trace of mass \(p\)
has final miss probability \((1-p)^{n_*}\). The conditional mean of \(Z\)
counts pairs within the prefix, between the prefix and the remaining
draws, and among the remaining draws:
\[
 Z_{\rm prefix}
  + n_*\sum_{j\le i}\mathbb E_{v\sim\nu} \chi_\sigma(u_j,v)
  + n_*(n_*-1)\Delta(\nu).
\]
These are rational counts on \(\mathcal A\). The next choices average to the
current conditional value, so the terminal value is below one. No heavy
trace is missed and \(Z\le4t^2\Delta(\nu)\). This selection takes polynomial
work in the listed traces, \(|\mathcal A|\) and \(t\), without enumerating tuples.
Drop duplicate selected circles and call the resulting set \(\mathcal H\). This
preserves every hit, and \(P_\sigma(\mathcal H)\le Z\).

Now use \(\mathcal H\) for the geometric split in Section~\ref{sec:localcut}.
An open child \(T\) meets no member of \(\mathcal H\). Its trace \(\mathcal U_T\) cannot
be heavy, since every heavy trace was hit. Returning first to the
padded measure and then to the actual, unpadded conflict count gives
\[
 |F_T|\ \le\ 2k\,\mu(\mathcal U_T)
      \ \le\ 2k\bigl(\nu(\mathcal U_T)+\eta\bigr)
     \ \le\ \frac{5k}{8\rho}<\frac{k}{\rho}.
\]
The first inequality uses the minimum mass \(1/(2k)\) of an original
circle. Construct \(F_T\) explicitly by applying~\eqref{eq:extrema}
to each parent conflict. The drawing bound of Section~\ref{sec:localcut}
and \(P_\sigma(\mathcal H)\le Z\le4t^2(4P_\sigma/k^2+\eta)\) give the
claimed feature count. Only Lemma~\ref{lem:sample} processes growing
lists during sample selection; the remaining work uses the bounded
sample and one scan of the parent list per child.
\end{proof}

\subsection{Counting cells and bounding memory}\label{sec:cut-budget}
For running time we count every cell generated. For memory we count
only what the traversal retains along its current path, including
children waiting to be visited.
\begin{proof}[Proof of Lemma~\ref{lem:cut}]
Write \(b_j=m/\rho^j\) for the threshold at the \emph{destination}
level \(j\). A parent at level \(j-1\) has \(k\le b_{j-1}\). If
\(k>b_j\), the split gives each child fewer than \(k/\rho\le b_j\)
conflicts; otherwise carry the parent forward with its list unchanged.
At each level the parents have disjoint open
interiors, so one pair belongs to at most two parent pair counts:
\(\sum_\sigma P_\sigma\le2\binom m2\). By Lemma~\ref{lem:split}, the
pair contribution summed over refined parents is at most
 \(O(t^2 b_j^{-2}\sum_\sigma P_\sigma)=O(t^2\rho^{2j})\).
Summing the remaining terms and the carried cells gives,
for the number \(C_j\) of cells,
\[
 C_j\le 512t^2\rho^{2j}
          +32(1+t+t^2/(4\rho))C_{j-1}.
\]
Here the constants follow by substituting
\(P_\sigma(\mathcal H)\le4t^2(4P_\sigma/k^2+\eta)\) into the
\(32(1+t+P_\sigma(\mathcal H))\) child bound, and using
\(\sum_\sigma P_\sigma\le m^2\). Starting at four, double \(\rho\)
until \(32(1+t+t^2/(4\rho))\le\rho^2/2\), recomputing its budget \(t\)
each time. This terminates since the left side is
\(O(\rho\log^2(2\rho))\). Both this \(\rho\) and the exponent for
constructing the small catalogues are independent of the input size.
Dividing by \(\rho^{2j}\)
gives \(C_j/\rho^{2j}\le512t^2+\tfrac12 C_{j-1}/\rho^{2(j-1)}\).
Since \(C_0=1\), the normalized count stays bounded, which proves
\(C_j=O(\rho^{2j})\).
The potentially large \emph{time} constant of finding a sample is not
part of this cell-count coefficient. For \(m\ge1\), stop when the
next power of \(\rho\) exceeds \(m\); this is level
\(J=\lfloor\log_\rho m\rfloor\), with lists of size less than \(\rho\).
No real logarithm or floor is needed. The time per parent
is \(O(k\log^\beta k)\) for \(k\ge2\) and constant otherwise;
equation~\eqref{eq:sum} bounds the total time.

Now visit the geometric hierarchy depth first. The level thresholds
\(b_j=m/\rho^j\) are known at the root; \(C_j\) and \(P_\sigma\) are
used in the proof, not to choose the next parent. Score
the new edges and vertices before releasing their parent's list. Each
child needs only the \(x\)-coordinates of its sides, its bottom and top curves, its
conflicts and its offset. Finish one child before visiting the next. A parent with
\(k\) conflicts retains its list and at most \(O_\rho(1)\) waiting child
records; even if all child lists are formed immediately they use
\(O_\rho(k+1)\) words. The local sampling computation fits the same
budget by Lemma~\ref{lem:sample} and is released before descent.
A carried cell advances without copying its list. At level \(j\),
\(k+1=O(m/\rho^j)\), even when the list is empty, because
\(m/\rho^j\ge1\) on every visited level. The total live storage is therefore
\[
       O\!\left(\sum_{j=0}^{J}m/\rho^j\right)=O(m)
\]
words. This visits the same cells as the level count above. The empty
family needs only constant space and time.
\end{proof}

\section{Choosing rational parameters for the lens reduction}
\label{sec:lensparam}
The two tests in Section~\ref{sec:lens} must work for the same targets.
For the first, take an index set \(I\) of source sites on one hyperplane.
We need a centre that hits at least one target in every pair and both
targets for \(i\in I\). The \emph{reach} bound~\eqref{eq:reach}
supplies that centre; it may depend on \(I\). For the converse,
\emph{any} centre that hits both targets of pair \(i\) must be close
to \(\Pi_i\). The \emph{width} bound~\eqref{eq:width} controls
\(\lvert a_i\cdot x-1\rvert\) in the two-hit test~\eqref{eq:far};
divide by \(\norm{a_i}\) to get the distance to \(\Pi_i\).
Table~\ref{tab:lens-bits} follows the construction order. First the
minor bound \(K\) and base \(\tau\) give a translation \(o\) with bounded
witnesses. Choose \(B\) large enough to reach them, then a much larger
displacement scale \(L=B^4\) to make the lenses thin. Finally, rational
Newton iterates \(g_i\) control each displacement's length; its direction
is exactly \(a_i\) throughout.

\begin{lemma}[Lens parameters]\label{lem:lensparameters}
For fixed \(D\ge2\) and nonempty distinct integer source points, rational
parameters \(o,\lambda_i,R\), with \(o\) integer and \(\lambda_i,R>0\),
can be chosen as follows. Put \(a_i=o-p_i\ne0\),
\(u_i=a_i/\norm{a_i}^2\), and \(\delta_i=R^2-\lambda_i^2\norm{a_i}^2\).
There is an integer \(\alpha\ge1\) bounding all \(\norm{a_i}_\infty\),
and every nonempty index set \(I\) whose source points lie on a
hyperplane has a witness \(x\) satisfying
\begin{equation}\label{eq:reach}
 a_i\cdot x=1\quad(i\in I),\qquad
 \norm{x-u_j}^2\le\delta_j\quad(1\le j\le s).
\end{equation}
Moreover, for every \(i\),
\begin{equation}\label{eq:width}
 \frac{\delta_i}{2\lambda_i}<\frac1{(D+1)!\,\alpha^D}.
\end{equation}
The scaled sites \((u_i\pm\lambda_i a_i)/R\) are distinct.
\end{lemma}

\begin{table}[htb]
\centering
\caption{Parameters in order of construction. Put
\(\Lambda=\lceil\log_2(2DU)\rceil\). For a rational number the last
column bounds each of its numerator and denominator lengths}\label{tab:lens-bits}
\small
\begin{tabular}{@{}lll@{}}
\toprule
Parameter & Role & Bits, up to constants\\
\midrule
\(U\) & Input coordinate height & \(\Lambda\)\\
\(K,\tau\) & Minor bound and translation base & \(D\Lambda\)\\
\(o,a_i,\alpha\) & Shift, normals and coordinate bound & \(D^2\Lambda\) per coordinate\\
\(B,L,R\) & Witness reach, offset length and radius & \(D^3\Lambda\)\\
\(j_{\rm N}\) & Newton steps, \(O(\log D)\) & \(O(\log(2+\log D))\)\\
\(g_i,\lambda_i\) & Normalization and rational displacement & \((D+1)^6\Lambda\)\\
\bottomrule
\end{tabular}
\end{table}

\begin{proof}[Proof of Lemma~\ref{lem:lensparameters}]
We first choose \(o\) so that every incident set has a bounded
witness. Set \(U=\max(2,1+\max_{i,j}|p_{ij}|)\), and define
\[
 K=D!U^D,\qquad \tau=32(D+1)K,\qquad o=(\tau,\tau^2,\ldots,\tau^D).
\]
Every nonempty set \(I\) of source indices on a proper affine
hyperplane is contained in one with integer coefficients
\(b\cdot p=\gamma\), all of magnitude at most \(K\), with \(b\ne0\).
To see this, choose independent homogeneous rows \((p_i,1)\) spanning
the rows from \(I\). If there are \(j\) basis rows, then \(j\le D\).
In the matrix of these \(j\) basis rows choose \(j\) independent columns
and one further column. The signed
\(j\)-by-\(j\) minors of this submatrix give a nullvector, nonzero
because the first \(j\) columns are independent. Extend it by zeros to
all columns. Its entries are integers of magnitude at most \(j!U^j\le K\);
its spatial part cannot vanish because the rows end in one.

No hyperplane with such coefficients contains \(o\). In the integer
polynomial \(b\cdot o-\gamma\), viewed as a polynomial in \(\tau\),
the leading coefficient has magnitude at least one. If its degree is
\(j_*\ge1\), the lower terms have total magnitude at most
\[
 K\sum_{r=0}^{j_*-1}\tau^r=K(\tau^{j_*}-1)/(\tau-1)<\tau^{j_*}.
\]
In particular the denominator
of \(x=b/(b\cdot o-\gamma)\) has magnitude at least one. Thus
\(\norm{x}_\infty\le K\) and \((o-p_i)\cdot x=1\) for every \(i\in I\).
The reduction computes only the common \(o\) at this stage; it does
not enumerate the sets \(I\) or their witnesses.

We next put rational target sites on the normal line of each plane.
Write \(a_i=o-p_i\) and \(\alpha=2\tau^D\). The last coordinate satisfies
\(a_{iD}\ge \tau^D/2>0\), and
\(\norm{a_i}_\infty\le \alpha\). The choice of translation also makes each
\(a_i\) nearly parallel to the last coordinate axis:
\[
 \xi_i=\sum_{j<D}(a_{ij}/a_{iD})^2\le 16(D-1)/\tau^2\le 1/\tau .
\]
Here \(|a_{ij}|\le2\tau^{D-1}\) for \(j<D\). Set
\(B=\tau^{(D+1)^2}\) and \(L=B^4\). We want the displacement from a
midpoint to either site to have length close to \(L\), while remaining
\emph{exactly parallel} to \(a_i\). Rather than divide by the possibly
irrational \(\norm{a_i}\), use the rational vector \(g_i a_i/a_{iD}\),
whose direction is exact and whose squared length is \((1+\xi_i)g_i^2\).
The displacement is \(L\) times this vector, so it suffices to find a rational
\(g_i\ge1/2\) such that
\begin{equation}\label{eq:uniterror}
                     0\le1-(1+\xi_i)g_i^2\le L^{-2}.
\end{equation}
Take \(j_{\rm N}=\lceil\log_2(8(D+1)^2)\rceil\). Starting at \(g=1\),
use Newton's iteration \(g\gets g(3-(1+\xi_i)g^2)/2\) for the inverse
square root, for at most \(j_{\rm N}\) steps. Stop once
\eqref{eq:uniterror} holds and set \(g_i=g\).
One step replaces the residual \(E=1-(1+\xi_i)g^2\) by
\(E^2(3+E)/4\). Initially \(E=-\xi_i\); after one step
\(0\le E\le \tau^{-2}\) and \(g=1-\xi_i/2\ge1/2\). Thereafter
the error is at most its previous square and \(g\) does not decrease.
Hence after \(j_{\rm N}\) steps the residual is at most
\(\tau^{-2^{j_{\rm N}}}\le L^{-2}\), proving~\eqref{eq:uniterror}.

With \(g_i\) fixed we can check reach and width. Take
\(\lambda_i=Lg_i/a_{iD}\) and \(R=L+B^2/L\). Since
\(\norm{a_i}^2=a_{iD}^2(1+\xi_i)\), the remaining squared radius is
\[
 \delta_i=R^2-\lambda_i^2\norm{a_i}^2
          =2B^2+B^{-4}+L^2\bigl(1-(1+\xi_i)g_i^2\bigr),
 \qquad 2B^2\le\delta_i\le4B^2 .
\]
For the incidence witness \(x=b/(b\cdot o-\gamma)\) constructed above,
\(\norm{x-u_j}\le\sqrt D\,K+1\le B\) for every \(j\):
\(\norm{u_j}=1/\norm{a_j}\le1\). This proves reach~\eqref{eq:reach}.
For the width,
\[
       \frac{\delta_i}{2\lambda_i}
          =\frac{a_{iD}\delta_i}{2Lg_i}\le\frac{4\alpha}{B^2}.
\]
Now \(\alpha\le \tau^{D+1}\) gives \(\alpha^{D+1}\le B\), and \(B>4(D+1)!\).
Consequently \((D+1)! \alpha^D(4\alpha/B^2)<1\), which is~\eqref{eq:width}.

To see that the target points are distinct, first note that the two
points of a pair lie on the line \(\R a_i\). Neither is zero:
\(\lambda_i\norm{a_i}\ge L/2>1\ge\norm{u_i}\), and they are
different since \(\lambda_i>0\). For \(i\ne j\), the two distinct
source points lie on some hyperplane because \(D\ge2\). Its witness
in~\eqref{eq:reach} has \(a_i\cdot x=a_j\cdot x=1\). If \(a_i\)
and \(a_j\) were proportional this would force \(a_i=a_j\), contrary
to \(p_i\ne p_j\). Thus different pairs lie on different lines
through the origin and share no target point. Scaling by \(1/R\)
preserves this property.

It remains to bound the coordinate lengths. Since
\(\log\tau=O(D\Lambda)\), we have \(\log\alpha=O(D^2\Lambda)\)
and \(\log B,\log L=O(D^3\Lambda)\). Form \(1+\xi_i=u/v\) over the common denominator
\(v=a_{iD}^2\), with \(u=v+\sum_{j<D}a_{ij}^2\), rather than
adding independent unreduced fractions. If \(g=P/Q\), one exact
Newton step is represented without reduction by
\[
 P'=P(3vQ^2-uP^2),\qquad Q'=2vQ^3.
\]
For numerator and denominator lengths at most \(b\), the next lengths
are at most \(3b+O(\log v+1)\), since \(\xi_i\le1\) gives \(u\le2v\).
Over \(j_{\rm N}\) Newton steps the multiplier on representation length
is \(3^{j_{\rm N}}=O((D+1)^{2\log_2 3})\), and
\(\log v=O(D^2\Lambda)\). The squared norm \(\norm{a_i}^2\) is an
integer of \(O(D^2\Lambda)\) bits. Each target coordinate uses only a
constant number of remaining products, sums and quotients with \(g_i\),
\(R\) and these integers. Since \(2+2\log_2 3<6\), the lengths are
\(O((D+1)^6\Lambda)\). In particular, for fixed \(D\)
they are \(O(\log U)\), even without fraction reduction.

After one scan to find \(U\), each target
\((u_i\pm\lambda_i a_i)/R\) takes \(O(1)\) rational operations for
fixed \(D\). This is an arithmetic-operation count, not a bit-time bound.
This also gives the construction bounds in Theorem~\ref{thm:lens}.
\end{proof}

\fi

\begingroup
\NextLinkTarget{geometry-glossary}
\section*{Glossary of geometric terms}\label{sec:glossary}
\addcontentsline{toc}{section}{Glossary of geometric terms}
\small
\newcommand{\geomterm}[2]{\par\addvspace{.32\baselineskip}%
  {\interlinepenalty=10000\noindent\textbf{#1.}\enspace #2\par}}

\geomterm{Affine hull; affine independence}{The affine hull is the smallest
flat containing the sites. They are affinely independent if, from one of
them, the vectors to the others are linearly independent
(Section~\ref{sec:high}).}

\geomterm{Anchor; tight site}{A site \(p\) is tight at a centre \(c\) if
\(\|c-p\|=1\). An anchor is a site we \emph{choose} to keep tight while
searching. Other sites can be tight at the winner too
(Lemma~\ref{lem:tight}).}

\geomterm{Arrangement; cutting}{The arrangement records all circle
boundaries, their contacts and the regions between them. A cutting splits a
connected open region (a cell) using selected circle arcs and vertical
segments (walls). The open pieces are its children; the parent scores the
new edges and vertices. A leaf is never split (Section~\ref{sec:cell-tree}).}

\geomterm{Ball; sphere}{The closed ball \(B(p,1)\) about a site \(p\)
contains exactly the centres that cover \(p\). Its sphere is just its
boundary. In three dimensions this surface is called a \emph{two-sphere}
(Section~\ref{sec:three}).}

\geomterm{Cap; rim; tip}{A cap is the part of a sphere in a closed
halfspace. On a two-sphere a regular cap has a circular boundary, its
rim. A singleton cap consists of just one point, its tip; a cap may
also be empty or the whole sphere (Section~\ref{sec:caps}).}

\geomterm{Conflict list; offset}{List a circle if its rim meets the open
cell. A rim wholly inside belongs on the list; one touching only the
boundary does not (Figure~\ref{fig:wall}a). The unlisted circles contribute
a constant total, the offset. Only the listed circles remain to be tested.}

\geomterm{Depth; closed exterior}{For site coverage, depth counts the sites
covered at a centre or sums their weights. Each planar circle carries a
weight for its disk (on or inside the circle) and one for its closed exterior
(on or outside). A rim point earns both (Section~\ref{sec:plane-contract}).}

\geomterm{Hyperplane incidence}{A hyperplane is a line in 2D, a plane in
3D, and a flat of dimension \(D-1\) in \(\R^D\). An incidence is a point
lying on the hyperplane. \emph{Affine degeneracy} asks whether some \(D+1\)
input points share one hyperplane (Section~\ref{sec:lower}).}

\geomterm{Lens; double hit}{The lens of a pair of target sites is the
intersection of their two closed balls of possible centres. A ball placed
with its centre in the lens covers both targets: a double hit
(Figure~\ref{fig:lens}).}

\geomterm{Pole; stereographic projection}{A ray from the pole through any
other sphere point meets the equatorial plane at its image. Choose the pole
off every regular cap rim. A regular cap missing it maps to a closed disk;
for a regular cap containing it, the rest maps to a closed exterior. Score the
omitted pole separately (Figure~\ref{fig:pole-sides}).}

\geomterm{Section; relative boundary}{The section \(S_I\) consists of
centres exactly one unit from \emph{every} chosen site in \(I\). To find the
boundary of a feasible subset relative to \(S_I\), look only within the
section: every neighbourhood there must meet both the subset and its
complement in \(S_I\) (Section~\ref{sec:high}).}

\geomterm{Trace; lift}{A trace records which list entries a test selects.
In Lemma~\ProofRef{lem:trace-catalogue}, lifting introduces extra coordinates
for products of query parameters. In Section~\ref{sec:caps}, lifting a
planar point means applying the inverse projection to return to the sphere.}
\endgroup

\ifIllustrated\else
\section*{Data and code availability}
The reference implementation, scripts for generating the exact test instances
and Lean sources are archived on Zenodo as version~1.0.0 of the companion
software~\citep{petrucena2026coverage}. No external dataset is required.
\fi
\NextLinkTarget{coverage-ai-use}
\section*{Use of generative AI}
\addcontentsline{toc}{section}{Use of generative AI}
The author directed the research on exact unit-ball placement. OpenAI
models used through OpenCode, including GPT-6.2 Astra, helped formulate
algorithms and theorems, develop and review proofs, and formalize results
in Lean. They also helped write the Python implementations and tests,
draft and revise the manuscript, and prepare the code for the figures.
Opus and DeepSeek models were used for additional draft critiques.
Literature searches used DeepSeek Search and Semantic Scholar. The author
takes responsibility for the manuscript, its claims and citations, and
the accompanying code.

\renewcommand{\bibfont}{\small}
\setlength{\bibsep}{4pt plus 1pt minus 1pt}
\bibliographystyle{plainnat}
\bibliography{cuttings,refs}
\end{document}